\documentclass[useAMS,usenatbib,fleqn,english,a4paper]{mnras}
\pdfoutput=1
\usepackage{newtxtext,newtxmath}
\usepackage[T1]{fontenc}
\usepackage{ae,aecompl}

\usepackage{hyperref}	
\hypersetup{colorlinks=true,linkcolor=blue,citecolor=blue,filecolor=blue,urlcolor=blue}

\usepackage{graphicx}	
\usepackage{amsmath}	
\usepackage{amssymb}	
\usepackage{enumitem}
\usepackage{subfig}
\usepackage{natbib}
\usepackage{booktabs}
\usepackage{cleveref}
\crefname{figure}{Fig.}{Figs.}
\crefname{table}{Table}{Tables}
\crefname{chapter}{Chapter}{Chapters}

\newcommand{\perc}[1]{$#1$ per cent}
\newcommand{\units}[1]{\, \mathrm{#1}}
\newcommand{\unitstx}[1]{\mathrm{#1}}
\newcommand{\diff}{\mathop{}\!\mathrm{d}}

\newcommand{\rfsec}[1]{\mbox{\S\ref{sec:#1}}}
\newcommand{\equnp}[1]{eq.\@ \ref{eq:#1}}
\newcommand{\equnpTwo}[2]{eqs.\@ \ref{eq:#1} and \ref{eq:#2}}

\newcommand{\subrfig}[1]{\protect\subref{fig:#1}}

\newcommand{\kboltz}{k_{\mathrm{B}}}
\newcommand{\zeq}[1]{\mbox{$z=#1$}}
\newcommand{\nbody}{$N$-body~}
\newcommand{\delvir}{\Delta_{\mathrm{vir}}}
\newcommand{\msun}{\units{M_\odot}}
\newcommand{\Rv}{R_{\mathrm{vir}}}
\newcommand{\Mv}{M_{\mathrm{vir}}}
\newcommand{\Vv}{V_{\mathrm{vir}}}
\newcommand{\Tv}{T_{\mathrm{vir}}}
\newcommand{\cvir}{c_{\mathrm{v}}}
\newcommand{\Vc}{V_{\mathrm{c}}}

\newcommand{\Rsat}{R_{\mathrm{sat}}}
\newcommand{\Msat}{M_{\mathrm{v,sat}}}
\newcommand{\cvs}{c_{\mathrm{v,sat}}}
\newcommand{\Rc}{R_{\mathrm{c}}}
\newcommand{\Mc}{M_{\mathrm{v,clust}}}
\newcommand{\cvc}{c_{\mathrm{v,clust}}}
\newcommand{\Rd}{R_{\mathrm{s}}}
\newcommand{\zd}{z_{\mathrm{s}}}
\newcommand{\Rb}{R_{\mathrm{b}}}

\newcommand{\Ms}{M_{\mathrm{s}}}
\newcommand{\Mb}{M_{\mathrm{b}}}
\newcommand{\Mg}{M_{\mathrm{g}}}
\newcommand{\Rg}{R_{\mathrm{g}}}
\newcommand{\fgs}{f_{\mathrm{gs}}}
\newcommand{\fbs}{f_{\mathrm{bs}}}
\newcommand{\md}{m_{\mathrm{d}}}
\newcommand{\jd}{j_{\mathrm{d}}}
\newcommand{\tR}{\widetilde{R}}
\newcommand{\tl}{\tilde{\ell}}
\newcommand{\bet}{\beta}
\newcommand{\xii}{\xi}

\newcommand{\fg}{f_{\rm g}}
\newcommand{\fc}{f_{\rm c}}
\newcommand{\sigstar}{\Sigma_{\rm s}}
\newcommand{\sigclust}{\Sigma_{\rm c}}

\newcommand{\ftilde}{\mathcal{F}} 
\newcommand{\ptilde}{\mathcal{P}} 
\newcommand{\bfunc}[2]{B_{#1}(#2)}

\newcommand{\afunc}[2]{\mathcal{A}\left(#1;#2\right)}
\newcommand{\ttrav}{t_{\mathrm{travel}}}
\newcommand{\tdepl}{t_{\mathrm{depl}}}
\newcommand{\alfP}{\varepsilon^\prime}
\newcommand{\alf}{\varepsilon}
\newcommand{\olSig}{\overline{\Sigma}}

\graphicspath{ {figs/} }
\DeclareGraphicsExtensions{.pdf,.eps}

\title[Galaxy Quenching in Outskirts of Clusters]{Quenching of Satellite Galaxies at the Outskirts of Galaxy Clusters}

\author[Zinger et al.]{ Elad Zinger$^1$\thanks{E-mail: elad.zinger@mail.huji.ac.il}, Avishai Dekel$^1$, Andrey V.\@ Kravtsov$^2$ \& Daisuke Nagai$^3$ \\
$^1$Center for Astrophysics and Planetary Science, Racah Institute of Physics, The Hebrew University, Jerusalem 91904, Israel\\
$^2$Department of Astronomy \& Astrophysics, The University of Chicago, Chicago, IL 60637 USA \\
$^3$Department of Physics, Yale University, New Haven, CT 06520, USA\\}

\date{Accepted 2017 December 22. Received 2017 December 21; in original form 2016 October 8}

\pubyear{2018}

\begin{document}
\pagerange{\pageref{firstpage}--\pageref{lastpage}}
\maketitle
\label{firstpage}

\begin{abstract}
  We find, using cosmological simulations of galaxy clusters, that the hot
  X-ray emitting intra-cluster medium (ICM) enclosed within the outer
  accretion shock extends out to $R_{\rm shock}\sim(2\textrm{--}3)\Rv$, where
  $\Rv$ is the standard virial radius of the halo. Using a simple analytic
  model for satellite galaxies in the cluster, we evaluate the effect of
  ram pressure stripping on the gas in the inner discs and in the haloes at
  different distances from the cluster centre. We find that significant
  removal of star-forming disc gas occurs only at $r\lesssim0.5\Rv$, while gas
  removal from the satellite halo is more effective and can occur when the
  satellite is found between $\Rv$ and $R_{\rm shock}$.  Removal of halo gas
  sets the stage for quenching of the star formation by starvation over
  $2\textrm{--}3\units{Gyr}$, prior to the satellite entry to the inner
  cluster halo. This scenario explains the presence of quenched
  galaxies, preferentially discs, at the outskirts of galaxy clusters, and the
  delayed quenching of satellites compared to central galaxies.
\end{abstract}


\section{Introduction}\label{sec:intro}
The link between the properties of galaxies and their environment has long
been known, and nowhere is this link clearer than in galaxy clusters. Galaxies
which reside in groups and clusters are more likely to be `quenched', i.e.\@
characterized by quiescent star formation, and possess less atomic and
molecular gas than similar `field' galaxies
\citep{Oemler1974,Butcher1978,Dressler1980}.

In recent years, numerous observations have established that the effects of
the cluster environment on nearby galaxies in terms of star formation,
morphology, color, gas content etc.\@ extends farther than previously assumed,
out to $\sim (2\textrm{--}3)\Rv$
(\citealt{Solanes2002,Braglia2009,Park2009,Hansen2009,vonderLinden2010} and
see \citealt{Boselli2006a} for an extensive review).  In particular, an
elevated fraction of quenched galaxies was detected beyond the virial radius
of the clusters compared with similar populations of `field' galaxies
\citep{Balogh2000,Verdugo2008,Wetzel2012}. A related phenomenon is that of
`galactic conformity' \citep{Weinmann2006,Ann2008,Kauffmann2010} in which the
star formation properties of satellites are found to correlate with the
central galaxy in the halo - if the central is quenched then the satellites
are more likely to be quiescent as well, even when the satellites are found
well beyond the virial radius of the host halo.

One possible explanation for the extended effect of the environment is that
the quenched, gas-poor galaxies observed are `pre-processed' galaxies which
were already subjected to quenching mechanisms in smaller haloes prior to
becoming satellites in the cluster \citep{Mihos2004,Fujita2004}. Another
explanation is that these are actually `ejected' or `splashback' galaxies,
i.e.\@ galaxies which entered the virial radius on a highly eccentric orbit at
a much earlier time, passed through the central regions of the cluster and are
now found beyond the virial radius once again \citep{Mamon2004,Gill2005}. In
their passage through the inner regions of the cluster myriad processes such
as tidal stripping \citep{Zwicky1951,Gnedin2003,Villalobos2014} ram pressure
stripping \citep{Gunn1972}, thermal evaporation \citep{Cowie1977}, and
encounters with other satellites (`harassment')
\citep{Moore1996,Moore1999,Gnedin2003a} can lead to gas depletion and star
formation quenching \citep{Mamon2004,Wetzel2014}. However, some studies have
indicated that a substantial fraction are infalling into the system for the
first time \citep{Cen2014a,Fang2016}. \citet{Bahe2013} in their analysis of
satellites in simulated clusters find that neither of these explanations is
sufficient to account for the quenched fraction at the cluster outskirts.

Another, more obvious explanation is that the cluster environment actually
extends to beyond the virial radius. It is well established that as structures
form from the initial density perturbations the dark matter and gas converge
towards the centre of the potential well. As the gas falls towards the centre
and compresses, the growing pressure prevents the gas elements from passing
through each other, limiting further compression. Under these conditions, the
centre of the potential well, is characterized by vanishing infall
velocity. However, since the velocity of the infalling cold gas is typically
super-sonic, information of the zero-velocity boundary condition cannot reach
the gas and thus an accretion shock is
formed\citep{Bertschinger1985,Furlanetto2004,Keres2005}.

The virial accretion shock, though unstable in low mass haloes, is a robust
feature in massive haloes of $> 10^{12} \msun$ \citep{Birnboim2003,Dekel2006},
and found consistently in numerical simulations \citep[e.g.\@][]{Keshet2003}.
Analytic studies \citep{Voit2003,Book2010} have shown that in cluster sized
systems, the virial shock should be found at $\sim 1.5\Rv$, and, as we show
below, numerical simulations have also found that in clusters the virial shock
extends much farther than the virial radius \citep{Molnar2009,Lau2015}.

Galaxies crossing the accretion shock, which is found at $\sim
(2\textrm{--}3)\Rv$, enter the `Intra-Cluster Medium' (ICM) -- gas which has
been shock heated to $10^7\textrm{--}10^8\units{K}$ and is in near
hydro-static equilibrium within the potential well of the dark matter halo,
radiating primarily in the X-ray \citep{Sarazin1988}. Within the ICM the
satellites are subjected to the aforementioned environmental processes which
can lead to gas depletion and star-formation quenching.

In this study we focus on the effect of gas removal from the galaxy due to the
ram pressure exerted by the ICM as a result of the relative motion of galaxy
in the medium. This process is commonly called ram pressure stripping
(hereafter RPS). The effects of RPS on galaxies close to the centres of
clusters, where the ambient densities are high $\sim
10^{-3}\textrm{--}10^{-1}\units{cm^{-3}}$, is potent and can lead to removal
of significant amounts of the ISM from within the galaxy. This process has
been studied extensively since the pioneering work of \citet{Gunn1972} who
derived an estimate for the effect \citep[see also][]{Gisler1976}.  Direct
evidence of RPS in action can be found in detailed analyses of specific
galaxies
\citep{Boselli2006,Abramson2011,Ehlert2013,Ebeling2014,Abramson2014,Kenney2014},
while large observational surveys of cluster satellites, coupled with analytic
galaxy evolution models reveal its role in determining satellite properties
\citep{Cayatte1994,Boselli2009,Scott2010}. The issue has also been explored
numerically in idealized, `wind tunnel' simulations
\citep{Gisler1976,Balsara1994,Quilis2000,Tonnesen2009,Weinberg2014,Roediger2015,Roediger2015a},
simple systems
\citep{Roediger2007,Kapferer2008,Kapferer2009,Vijayaraghavan2015} and in
cosmological settings \citep{Tonnesen2007,Vollmer2001}. See also
\citet{Roediger2009} for an extensive review. The insights from analytic and
numerical studies have also been applied in semi-analytic modelling of cluster
systems \citep{Font2008,Book2010}.

A galactic system, consisting of a galaxy surrounded by a gas halo embedded in
a dark matter halo, undergoing RPS will first shed the hot gas in the halo
before the gas within the galaxy is affected
\citep{Larson1980,McCarthy2008,Bekki2009,Bahe2013,Vijayaraghavan2015}. In
regions of intermediate densities,
$\sim10^{-5}\textrm{--}10^{-4}\units{cm^{-3}}$, the ram pressure may remove
the gas halo, but not the cold gas from the galaxy. Eventually, star formation
quenching can occur by `starvation', once the star-forming gas inside the
galaxy is consumed by star formation or removed by feedback-driven outflows
\citep{Kawata2008}. As stars are formed, the dwindling cold gas reservoir
within the galaxy is no longer be replenished from the halo, and the galaxy is
quenched, albeit over longer time-scales of a few giga years.

In this paper we demonstrate, through the use of a suite of zoom-in
simulations of galaxy clusters, that the zone of the cluster influence on
star-formation quenching, i.e.\@ the virial accretion shock, extends well
beyond the standard virial radius derived from the density contrast.  We
explore, via simple analytic models, the effect of RPS on the gas haloes
surrounding galaxies and on the gas within the galaxies themselves, at
different radii in the host cluster and in its outskirts, in an attempt to
ascertain the main quenching channel responsible for the quenched galaxy
population found beyond the virial radius.

The paper is organized as follows: in \rfsec{sims}, we describe the suite of
simulated clusters and in \rfsec{edge} we determine the location of the virial
accretion shock. We construct analytic models to study the effectiveness of
RPS on the halo gas of satellites in \rfsec{rpsHalo} and do the same for RPS
of gas from the galactic discs in \rfsec{rpsDisc}.  In \rfsec{quench}, we
explore the implications of RPS on star formation quenching. In
\rfsec{discuss}, we discuss the strengths and weaknesses of our RPS models and
in \rfsec{summary} we summarize our findings.

\begin{table}
\centering
  	 \begin{tabular}{@{}lccccc@{}}
	 \toprule
	 Cluster & $\Mv$ & $\Rv$ & $\Tv$ &$\Vv$ \\ 
	  &$(10^{14}\msun)$&$(\unitstx{Mpc})$&$(10^7\units{K})$&$(\unitstx{km\,s^{-1}})$   \\ 
        \midrule
CL101 & 22.1 & 3.37 & 10.1 &  1678 \\
CL102 & 13.7 & 2.88 & 7.4  & 1433 \\
CL103 & 15.7 & 3.01 & 8.1  & 1497 \\
CL104 & 11.9 & 2.74 & 6.7  & 1365 \\
CL105 & 12.0 & 2.75 & 6.7  & 1369 \\
CL106 & 9.5  & 2.54 & 5.8  & 1266 \\
CL107 & 6.6  & 2.26 & 4.5  & 1125 \\
CL3   & 6.3  & 2.22 & 4.4  & 1107 \\
CL5   & 3.1  & 1.76 & 2.8  & 875 \\
CL6   & 3.3  & 1.80 & 2.9  & 894 \\
CL7   & 3.3  & 1.78 & 2.8  & 886 \\
CL9   & 1.9  & 1.48 & 1.7  & 739 \\
CL10  & 1.3  & 1.32 & 1.6  & 658 \\
CL11  & 1.8  & 1.45 & 1.9  & 721 \\
CL14  & 1.7  & 1.43 & 1.8  & 709 \\
CL24  & 0.86 & 1.14 & 1.2  & 569 \\
 	\bottomrule 
	\end{tabular}
	\caption{Cluster Properties at \zeq{0}. Virial quantities were
          calculated for an over-density of $\delvir=337$ above the
          mean density of the universe.}
	 \label{tab:clusterProperties}
 \end{table}

\begin{table}
\centering
  	 \begin{tabular}{@{}lccccc@{}}
	 \toprule
	 Cluster&$\Mv$&$\Rv$&$\Tv$&$\Vv$\\
	  &$(10^{14}\msun)$&$(\unitstx{Mpc})$&$(10^7\units{K})$&$(\unitstx{km\,s^{-1}})$  \\ 
        \midrule
CL101 & 3.3  & 1.29 & 4.0 & 1053 \\
CL102 & 3.3  & 1.28 & 4.0 & 1051 \\
CL103 & 2.9  & 1.23 & 3.7 & 1009 \\
CL104 & 7.2  & 1.66 & 6.7 & 1362 \\
CL105 & 6.2  & 1.58 & 6.0 & 1296 \\
CL106 & 2.6  & 1.18 & 3.3 & 965  \\
CL107 & 2.9  & 1.22 & 3.6 & 1002 \\
CL3   & 2.8  & 1.21 & 3.6 & 995  \\
CL5   & 1.8  & 1.04 & 2.6 & 858  \\
CL6   & 2.4  & 1.15 & 3.2 & 946  \\
CL7   & 2.3  & 1.13 & 3.1 & 929  \\
CL9   & 1.2  & 0.91 & 2.0 & 744  \\
CL10  & 1.1  & 0.89 & 1.9 & 730  \\
CL11  & 0.86 & 0.78 & 1.5 & 643  \\
CL14  & 0.99 & 0.86 & 1.8 & 703  \\
CL24  & 0.37 & 0.62 & 0.9 & 505  \\
 	\bottomrule 
	\end{tabular}
	\caption{Cluster Properties at \zeq{0.6}. Virial quantities were
          calculated for an over-density of $\delvir=224$ above the
          mean density of the universe.}
	 \label{tab:clusterProperties06}
 \end{table}

\begin{figure*}
  \subfloat[CL103, entropy]{\label{fig:cl103_S_a1}
    \includegraphics[height=5cm,keepaspectratio,bb=0 0 4.986in 4.375in,trim=0.5in 0.54in 0in 0in, clip]{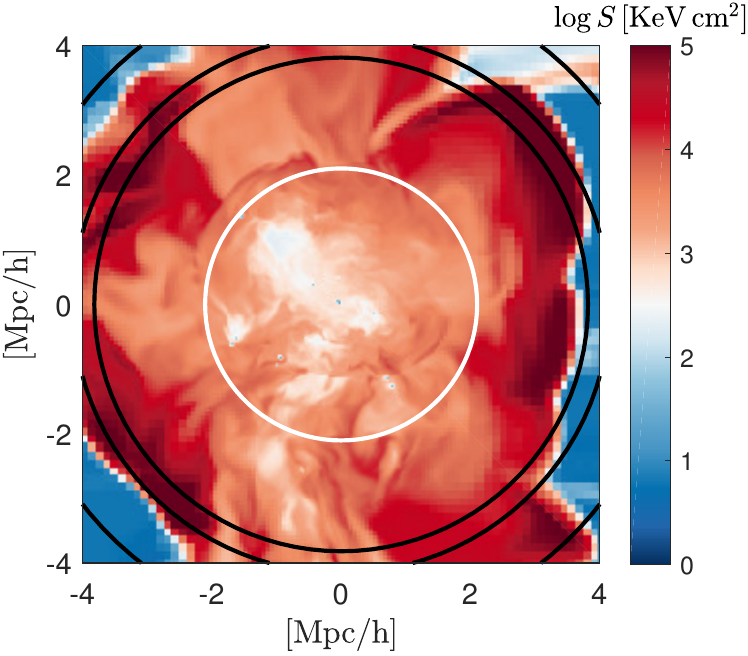}}
  \subfloat[CL103, temperature]{\label{fig:cl103_T_a1}
    \includegraphics[height=5cm,keepaspectratio,bb=0 0 4.792in 4.375in,trim=0.5in 0.54in 0in 0in, clip]{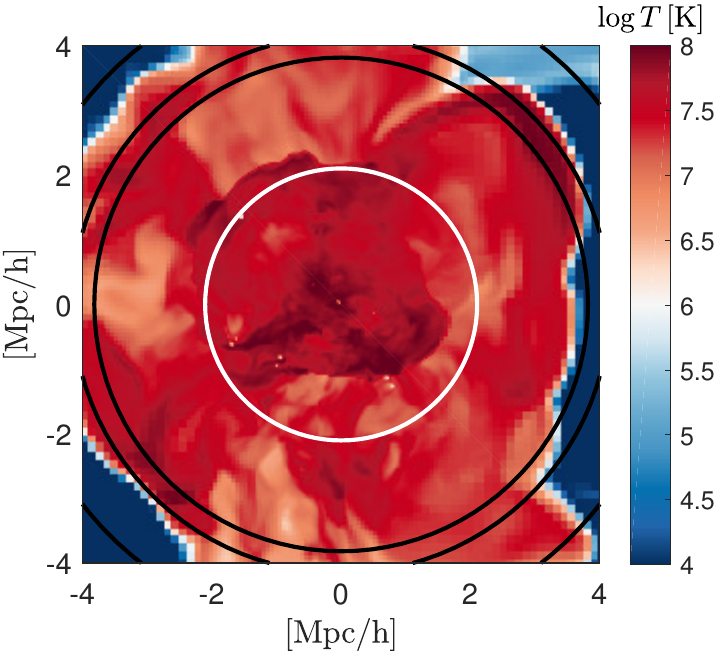}}
  \subfloat[CL103, density]{\label{fig:cl103_N_a1}
    \includegraphics[height=5cm,keepaspectratio,bb=0 0 4.931in 4.375in,trim=0.5in 0.54in 0in 0in, clip]{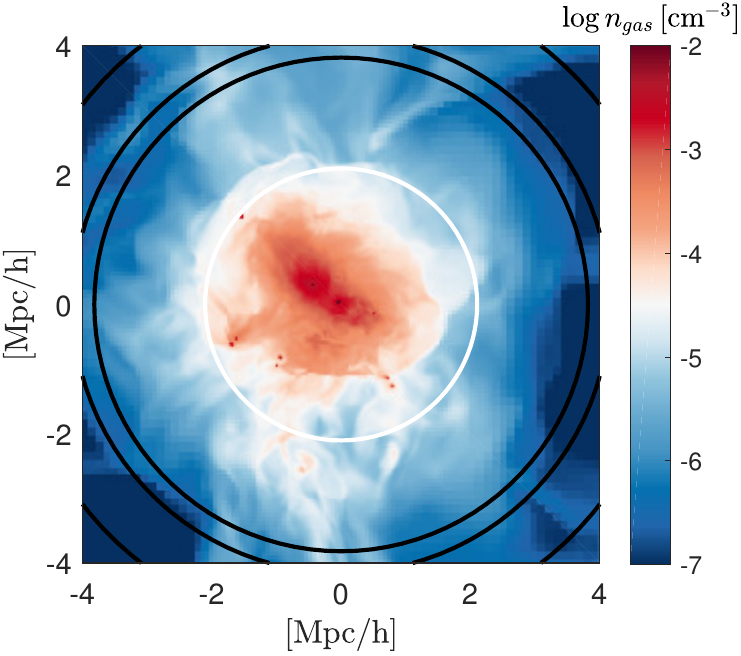}}\\
 
 \subfloat[CL106, entropy]{\label{fig:cl106_S_a1}
    \includegraphics[height=5cm,keepaspectratio,bb=0 0 4.986in 4.374in ,trim=0.5in 0.54in 0in 0in, clip]{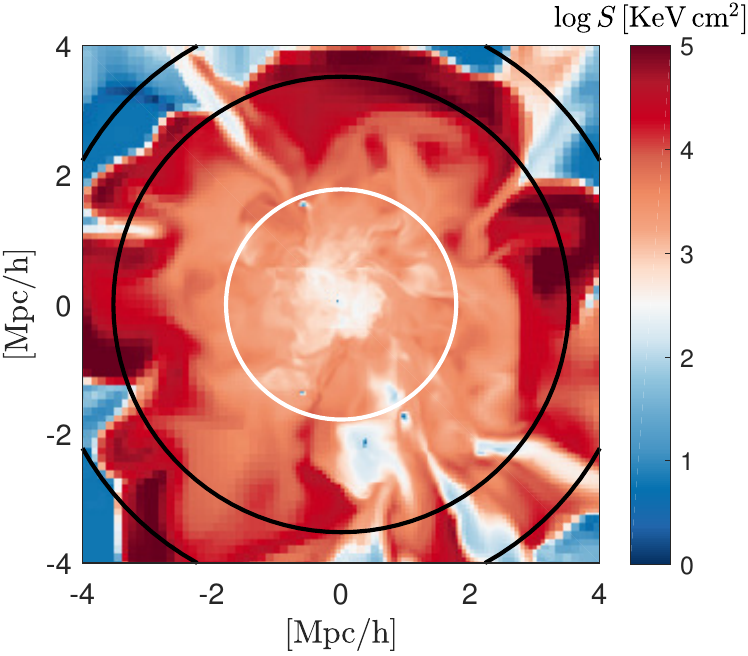}}
\subfloat[CL106, temperature]{\label{fig:cl106_T_a1}
  \includegraphics[height=5cm,keepaspectratio,bb=0 0 4.792in 4.375in,trim=0.5in 0.54in 0in 0in, clip ]{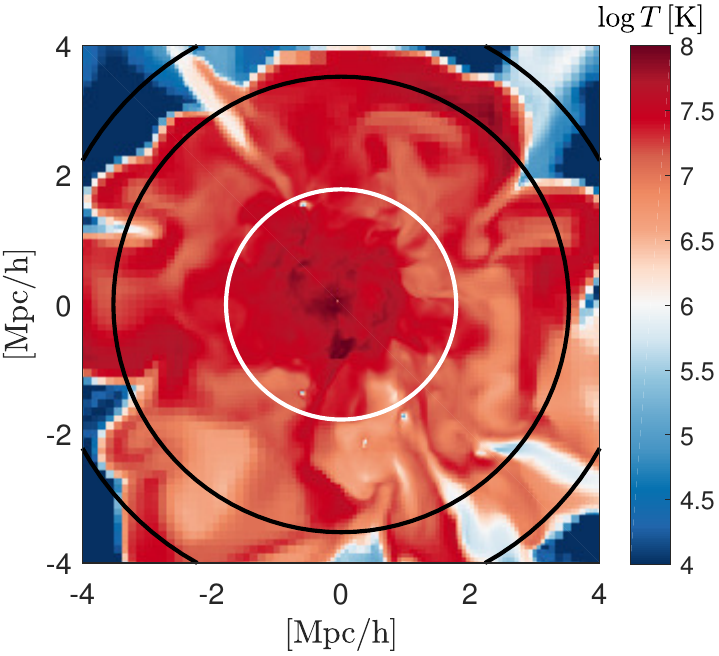}}
\subfloat[CL106, density]{\label{fig:cl106_N_a1}
  \includegraphics[height=5cm,keepaspectratio,bb=0 0 4.931in 4.375in,trim=0.5in 0.54in 0.0in 0in, clip ]{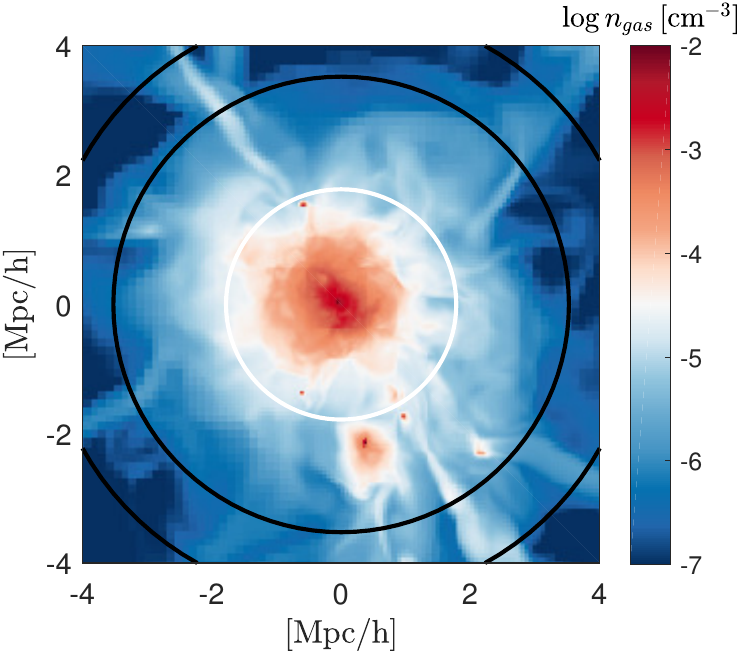}}
\caption{Entropy, temperature and density maps (left, middle and right,
  respectively) of 2 representative clusters at \zeq{0} (top and bottom) are
  plotted in box of size $8 \units{Mpc\,h^{-1}}$ (see
  \cref{tab:clusterProperties} for details). The shock front can be seen to
  extends well beyond the virial radius which is marked by the white
  circles. The entropy is low in the centre of the cluster and rises to a peak
  near the shock front, before dropping sharply, allowing a simple method of
  identifying the shock front. Black circles mark these drops in the entropy
  profile ( see \rfsec{edgeFind}).  The temperature is highest in the centre
  of the cluster and drops gradually outwards, with a sharp drop in
  temperature seen at the shock front. The gas density drops sharply from the
  centre outwards and the density drop at the shock radius is only a factor of
  few. The complex shape of the shock front is evident. Lower entropy gas
  streams flowing towards the centre can also be seen.}
  \label{fig:entTempRhoMaps_a1}
\end{figure*}

\begin{figure*}
  \subfloat[\zeq{0}]{\label{fig:cl14_S_a1}
    \includegraphics[height=6cm,keepaspectratio,bb=0 0 4.97in 4.32in,trim=0.41in 0.44in 0in 0in, clip]{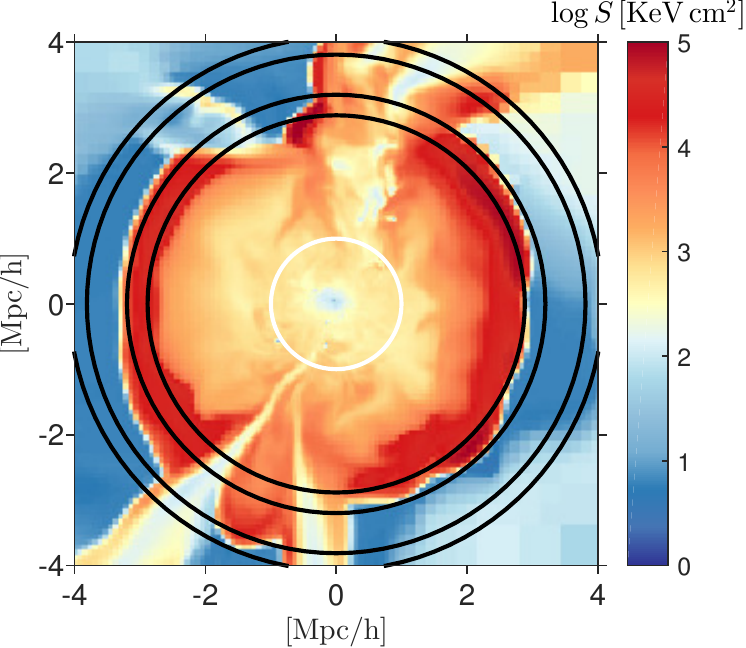}}
  \subfloat[\zeq{0.6}]{\label{fig:cl14_S_a06}
    \includegraphics[height=6cm,keepaspectratio,bb=0 0 5.15in 4.32in,trim=0.62in 0.49in 0in 0in, clip]{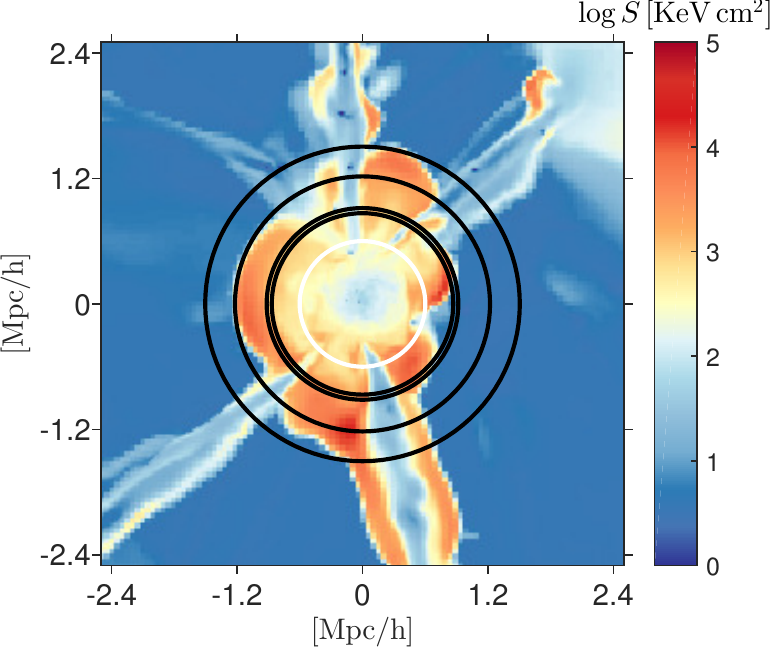}}
  \caption{Entropy maps of CL14 at \zeq{0} and \zeq{0.6} (left and right,
    respectively) are shown. The box at \zeq{0} is of size $8
    \units{Mpc\,h^{-1}}$ and the box at \zeq{0.6} is of size $5
    \units{Mpc\,h^{-1}}$. The cluster dimension and radii are shown in proper
    coordinates.As before, black circles drops in the entropy profile which we
    use to identify the locations of the shock fronts. At this redshift the
    shock front already extends well beyond the virial radius.  The intricate
    shape of the shock front is due in part to the large scale filaments that
    `puncture' the shock front which expands in lobes between the streams.
    Shocks can already be seen surrounding the large scale filaments feeding
    the cluster and merging seamlessly with the virial shock front.}
 \label{fig:cl14Maps}
\end{figure*}

\section{Simulations}\label{sec:sims}
The simulation suite analysed in this study is comprised of $16$ cluster-sized
systems at \zeq{0} spanning a mass range of $8.6\times 10^{13} \textrm{--}
2.2\times 10^{15} \units{M_\odot}$, and their most massive progenitors at
\zeq{0.6}. The systems were extracted from cosmological simulations in a flat
$\Lambda$CDM model: $\Omega_{\mathrm{m}}=1-\Omega_{\Lambda}=0.3$,
$\Omega_{\mathrm{b}}=0.04286$, $h=0.7$, and $\sigma_8=0.9$, where the Hubble
constant is defined as $100h\units{km\,s^{-1}\,Mpc^{-1}}$, and $\sigma_8$ is
the power spectrum normalization on an $8h^{-1} \units{Mpc}$ scale. The
simulations were carried out with the Adaptive Refinement Tree (\textsc{art})
\nbody+gas-dynamics code \citep{Kravtsov1999}, an Eulerian code that uses
adaptive refinement in space and time, and (non-adaptive) refinement in mass
\citep{Klypin2001} to reach the high dynamic range required to resolve cores
of haloes formed in self-consistent cosmological simulations.

The computational boxes of the large-scale cosmological simulations were $120
h^{-1}\units{Mpc}$ and $80 h^{-1}\units{Mpc}$, and the simulation grid was
adaptively refined to achieve a peak spatial resolution of order $\sim 7$ and
$5\, h^{-1}\units{kpc}$ respectively. These simulations are discussed in
detail in \citet{Kravtsov2006}, \citet{Nagai2007} and \citet{Nagai2007a}. The
adaptive mesh refinement technique employed in the simulation is especially
suited to capture discontinuous features such as shock waves and contact
discontinuities which make it especially suitable for our purposes.

Besides the basic dynamical processes of collisionless matter (dark matter and
stars) and gas-dynamics, several physical processes critical for galaxy
formation are incorporated: star formation, metal enrichment and feedback due
to Type II and Type Ia supernov\ae, and self-consistent advection of
metals. The cooling and heating rates take into account Compton heating and
cooling of plasma, UV heating \citep{Haardt1996}, and atomic and molecular
cooling, which is tabulated for the temperature range $10^2 < T <
10^9\units{K}$, a grid of metallicities, and UV intensities using the
\textsc{cloudy} code (version\@ 96b4, \citealt{Ferland1998}). The
\textsc{cloudy} cooling and heating rates take into account metallicity of the
gas, which is calculated self-consistently in the simulation, so that the
local cooling rates depend on the local metallicity of the gas. The star
formation recipe incorporated in these simulations is observationally
motivated \citep[e.g.\@][]{Kennicutt1998} and the code also accounts for the
stellar feedback on the surrounding gas, including injection of energy and
heavy elements (metals) via stellar winds, supernov\ae, and secular mass
loss. The simulations do not include an AGN feedback process.

The main purpose of this paper is to investigate the effect of the
environment at the cluster outskirts on the satellite galaxies within
the cluster, focusing on mass-loss mechanism and the quenching of star
formation. The simulations reliably reproduce the large scale
properties of the ICM. However, the resolution of the simulations is
not high enough to properly model the galaxies within the cluster
since the peak resolution is of order the scale radius of a typical
disc galaxy, and one cannot trust the simulated galaxies to study mass
loss and star formation quenching. Instead, we will rely on analytic
models of the galaxies with which to examine the effects of the
environment.

The virial parameters $\Mv$ and $\Rv$ are defined by the relation
\begin{equation}\label{eq:virialDef}
\frac{3\Mv}{4\mathrm{\pi} \Rv^3}= \delvir \rho_{\mathrm{ref}},
\end{equation}
with various choices in the literature for the over-density parameter
($\delvir = 178, 200, 337$) and reference density
$\rho_{\mathrm{ref}}$ -- either $\rho_{\mathrm{crit}}$ or
$\rho_{\mathrm{mean}} = \rho_{\mathrm{crit}}\Omega_{\mathrm{m}}$. Additional
distance scales for clusters are sometimes set by over-density factors
of $500$ and $1500$.

In this paper, the virial quantities of the mass, radius, velocity and
temperature ($\Mv, \Rv, \Vv, \Tv$) of the clusters are defined for an
over-density $\delvir = 337$ at \zeq{0} and $\delvir = 224$ at
\zeq{0.6} above $\rho_{\mathrm{mean}}$, the mean density of the
universe \citep{Bryan1998}. The properties of the clusters for \zeq{0}
and \zeq{0.6} are summarized in
\cref{tab:clusterProperties,tab:clusterProperties06} respectively.
\begin{figure*}
  \subfloat{\label{fig:cl6_yz_a06}
    \includegraphics[height=5cm,keepaspectratio,bb=0 0 5.15in 4.32in ]{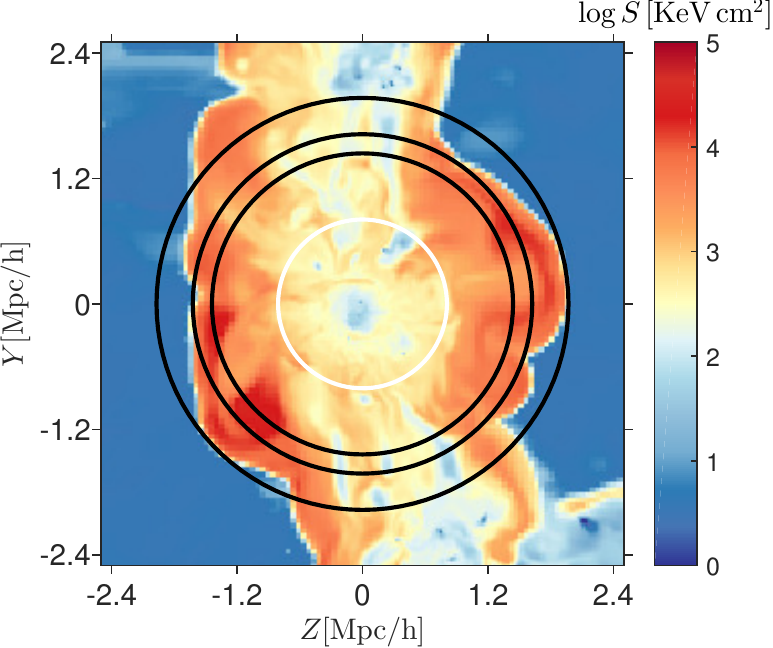}}
  \subfloat{\label{fig:cl6_xy_a06}
    \includegraphics[height=5cm,keepaspectratio,bb=0 0 5.15in 4.32in ]{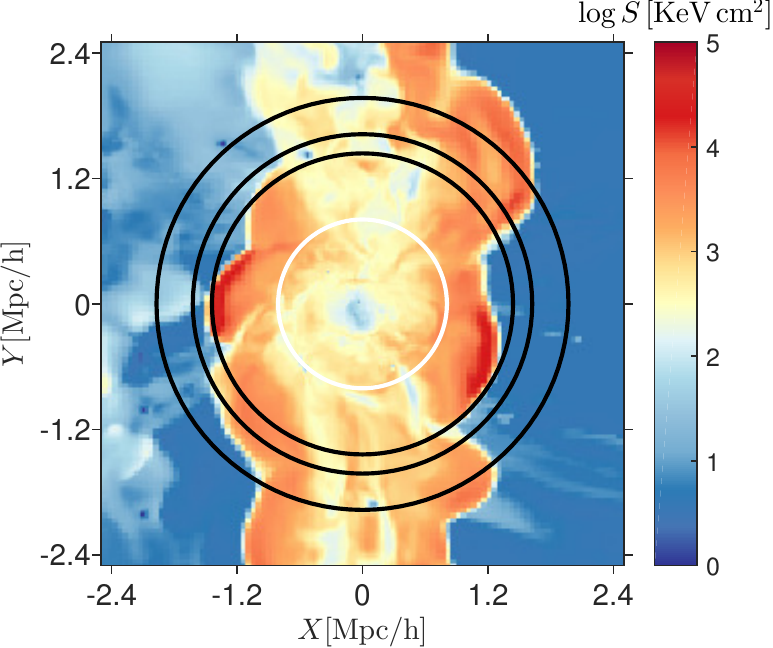}}
  \subfloat{\label{fig:cl6_xz_a06}
    \includegraphics[height=5cm,keepaspectratio,bb=0 0 5.15in 4.32in ]{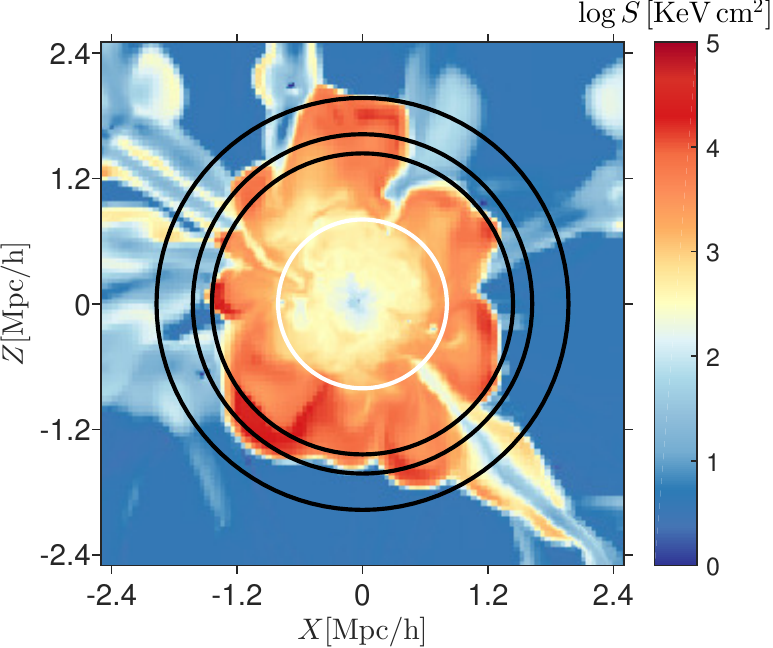}}
  \caption{Entropy maps of the cluster CL6 in three orthogonal
    projections at \zeq{0.6}. White circles mark the virial radius and
    black circles mark the approximate cluster edge as defined in
    \rfsec{edgeFind}. We see that even if an edge estimation does not
    seem to apply in a given projection, it corresponds to a shock
    front best observed from a different projection, due to the
    irregular topography of the shock front.}
  \label{fig:cl6Maps}
\end{figure*}

\begin{figure} 
  \subfloat[\zeq{0}]{\label{fig:rhoProfs_a1}
    \includegraphics[width=8cm,keepaspectratio,bb=0 0 5.32in 4.08in]{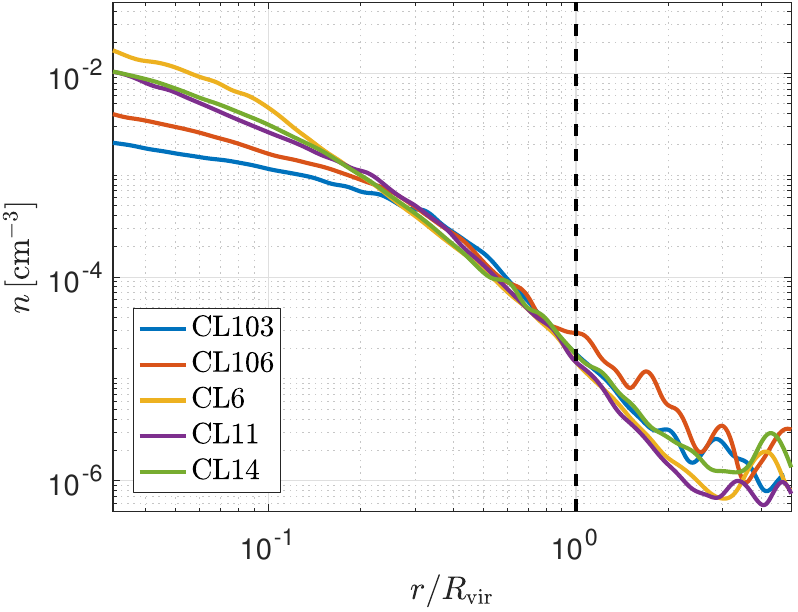}}\\
  \subfloat[\zeq{0.6}]{\label{fig:rhoProfs_a06}
    \includegraphics[width=8cm,keepaspectratio,bb=0 0 5.32in 4.08in]{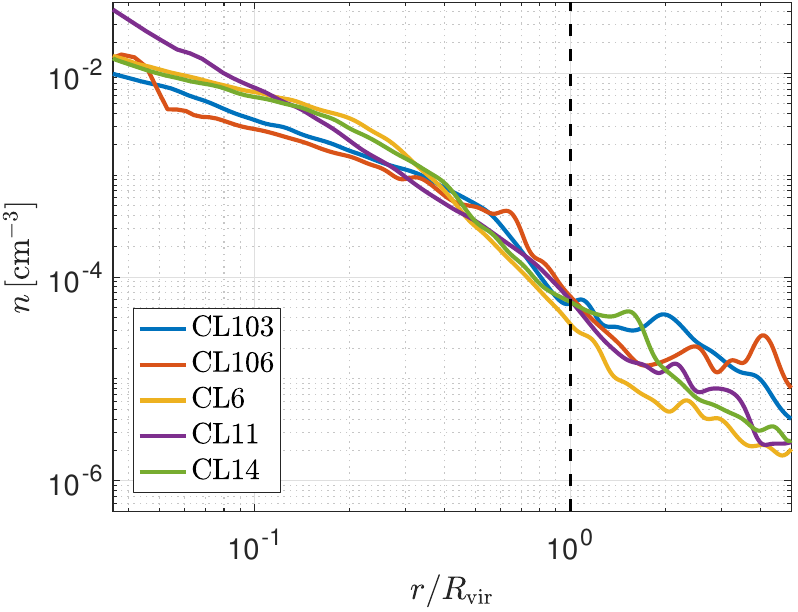}}
  \caption{Density profiles averaged over spherical shells of 5
    representative clusters at \zeq{0} \subrfig{rhoProfs_a1} and
    \zeq{0.6} \subrfig{rhoProfs_a06}. Dashed lines denote $\Rv$.  The
    density drops steadily from the central regions outwards. The
    density jump across a shock is only a factor of a few at
    most. Coupled with the spherical averaging, it is very difficult
    to identify the shock front from the density profiles.}
  \label{fig:rhoProfs}
\end{figure}

\begin{figure} 
  \subfloat[\zeq{0}]{\label{fig:tempProfs_a1}
    \includegraphics[width=8cm,keepaspectratio,bb=0 0 5.24in 4.19in]{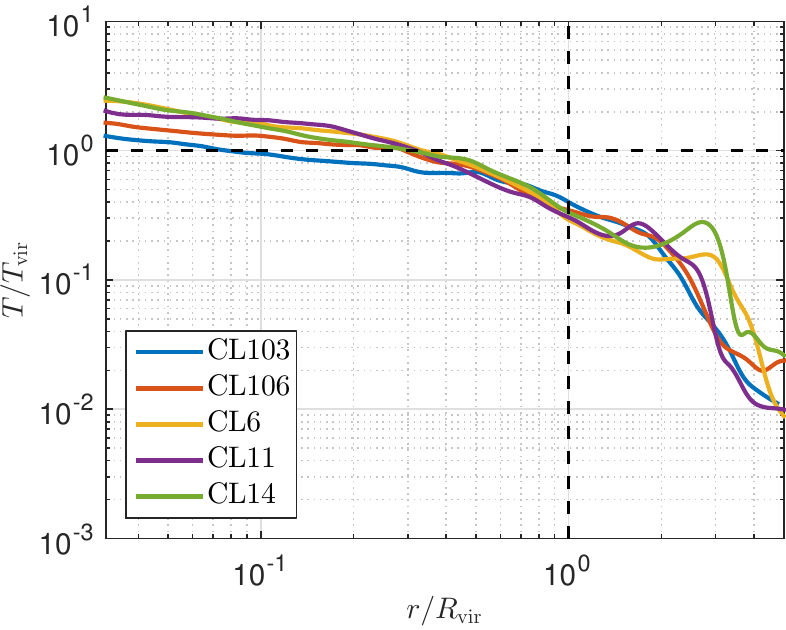}}\\
  \subfloat[\zeq{0.6}]{\label{fig:tempProfs_a06}
    \includegraphics[width=8cm,keepaspectratio,bb=0 0 5.24in 4.19in]{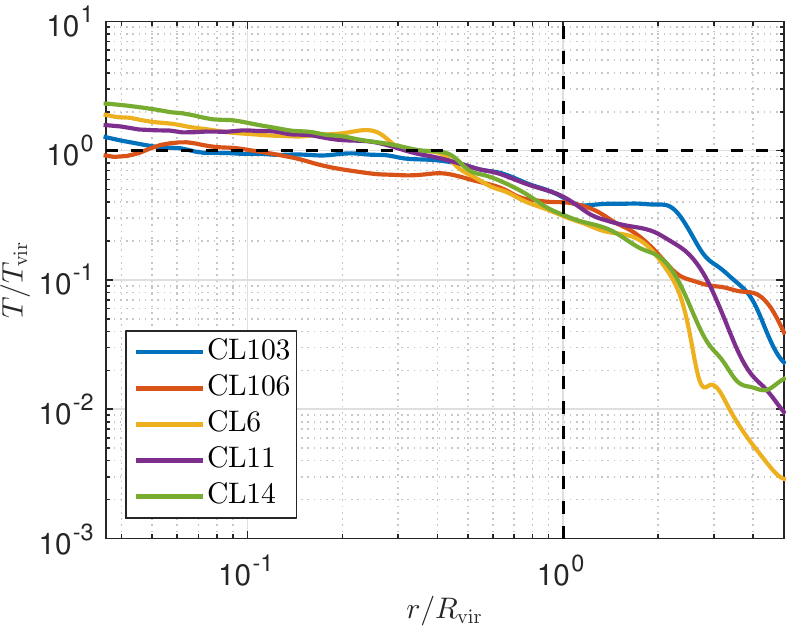}}
  \caption{Temperature profiles averaged over spherical shells of 5
    representative clusters at \zeq{0} \subrfig{tempProfs_a1} and
    \zeq{0.6} \subrfig{tempProfs_a06}. Dashed lines denote $\Rv$. The
    temperature rises sharply across the shock and then continues to
    climb gradually towards the centre where the temperature is
    highest. Due to the averaging, the drop in temperature is smeared,
    making it difficult to identify the position of the shock front.}
  \label{fig:tempProfs}
\end{figure}

\begin{figure} 
  \subfloat[\zeq{0}]{\label{fig:entProfs_a1}
    \includegraphics[width=8cm,keepaspectratio,bb=0 0 5.26in 4.19in ]{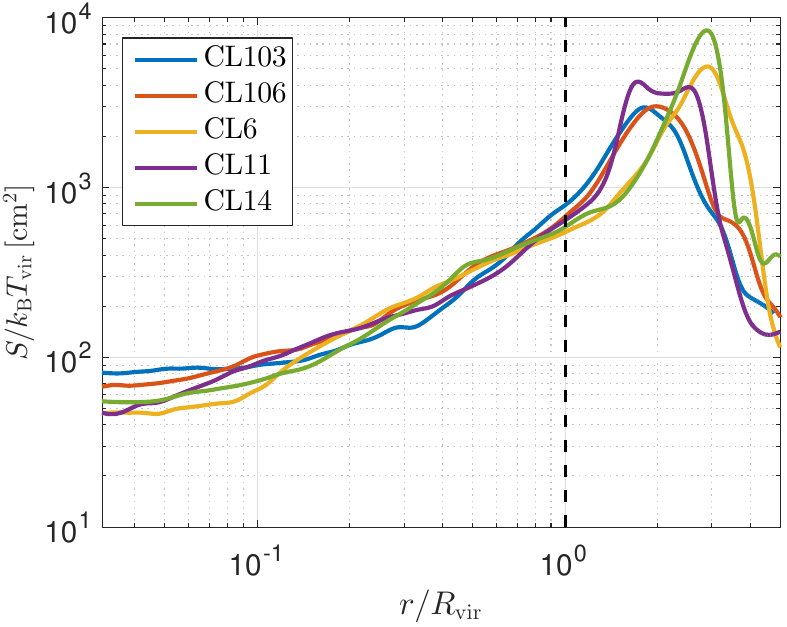}}\\
  \subfloat[\zeq{0.6}]{\label{fig:entProfs_a06}
    \includegraphics[width=8cm,keepaspectratio,bb=0 0 5.26in 4.19in ]{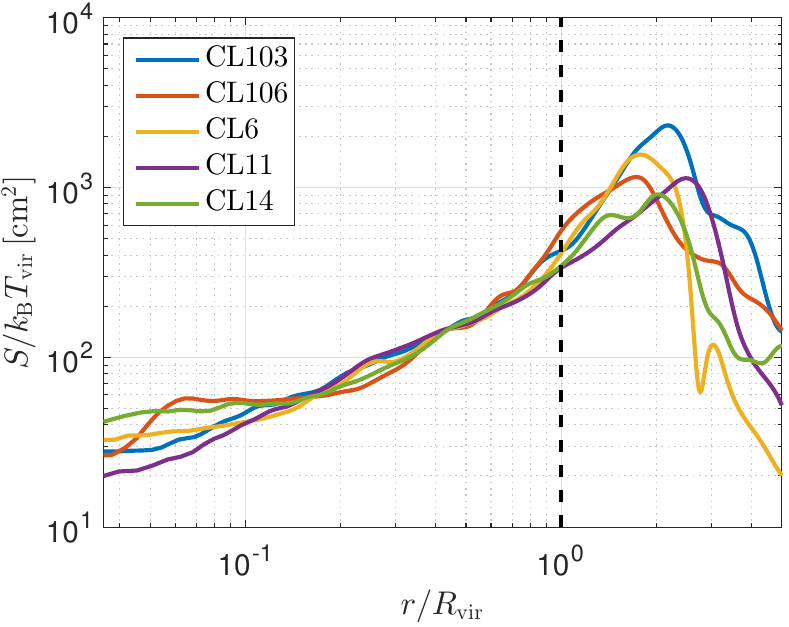}}
  \caption{Entropy profiles averaged over spherical shells of 5
    representative clusters at \zeq{0} \subrfig{entProfs_a1} and
    \zeq{0.6} \subrfig{entProfs_a06}. Dashed lines denote $\Rv$. The
    entropy is lowest in the cluster centre rising to a peak value
    near the shock before dropping sharply across the shock
    front. Since the shock front is not perfectly spherical the
    averaging leads to a widening of the entropy peak or in some cases
    to a double peak (CL11 in \subrfig{entProfs_a1}).}
  \label{fig:entProfs}
\end{figure}

\begin{figure} 
  \subfloat[\zeq{0}]{\label{fig:cl14_entProf_a1}
    \includegraphics[width=8cm,keepaspectratio,bb=0 0 5.26in 4.19in ]{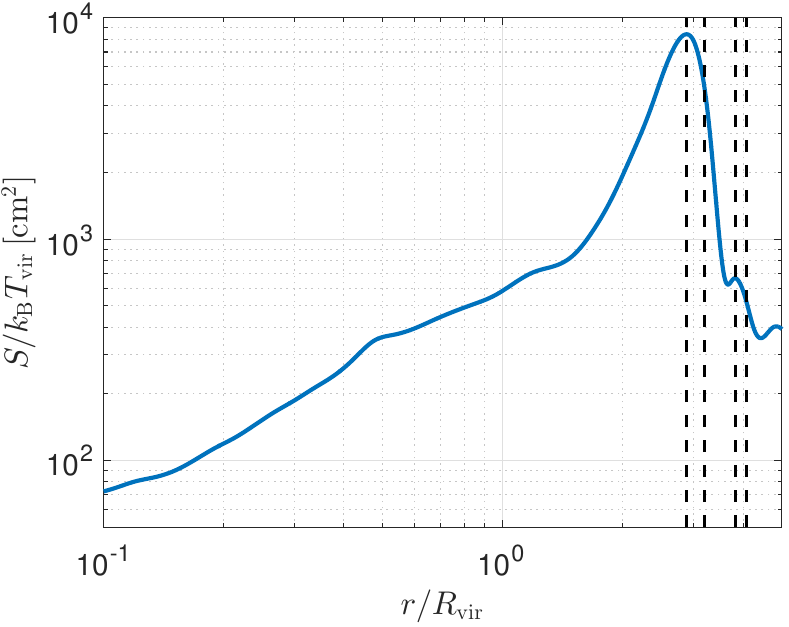}}\\
  \subfloat[\zeq{0.6}]{\label{fig:cl14_entProf_a06}
    \includegraphics[width=8cm,keepaspectratio,bb=0 0 5.26in 4.19in ]{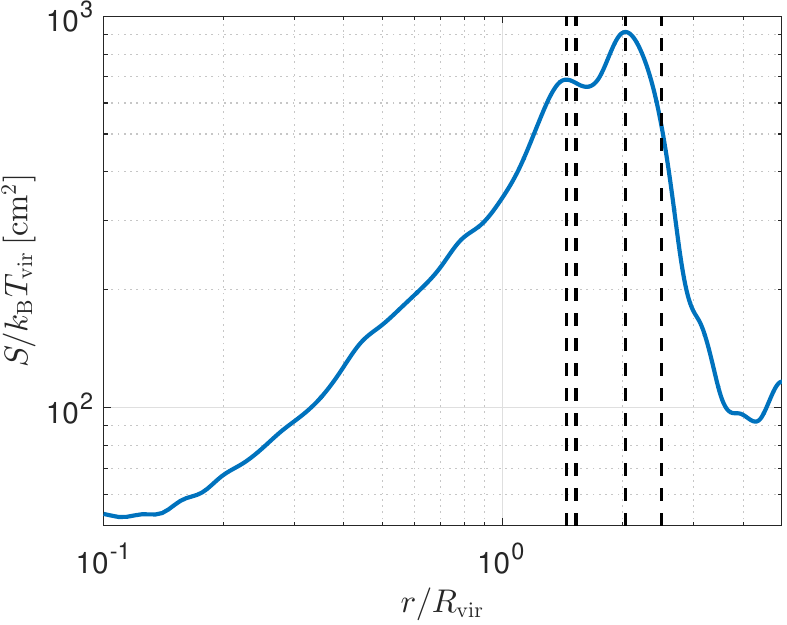}}
  \caption{Entropy profile for cluster CL14 at \zeq{0}
    \subrfig{cl14_entProf_a1} and \zeq{0.6}
    \subrfig{cl14_entProf_a06}. Dashed lines denote the points of
    local entropy maxima and local maxima of negative entropy
    gradient, which we use as a proxy for the shock edge.}
  \label{fig:cl14_entProf}
\end{figure}

\section{Virial Shock versus Virial Radius}\label{sec:edge}
In this section we determine the location of the accretion shock in the
simulated cluster suite, and show it is found well beyond the virial radius of
the cluster. Readers of the paper only interested in the way this affects the
galaxies found beyond the virial radius may safely skip to \rfsec{rpsHalo}.

\subsection{Cluster Maps}\label{sec:edgeMaps}
We examine our cluster systems at \zeq{0.6} and \zeq{0} and find that
the accretion shock extends well beyond the virial radius in all
cases. As representative examples we show the entropy, temperature and
density maps of 2 clusters (CL103 and CL106) at \zeq{0} in
\cref{fig:entTempRhoMaps_a1}. the entropy is defined as $S\propto
T\rho^{-2/3}$. Furthermore, in \cref{fig:cl14Maps} we examine the
cluster CL14 at \zeq{0} and \zeq{0.6} and find that the accretion
shock extends well beyond the virial radius even at that epoch.

Detailed examination of the virial shock in the maps shows it to be
spherical only in the roughest approximation and that in reality it
contains many features and extends to different extents in different
directions. The shock front is sometimes segmented into `lobes'
(e.g.\@ \cref{fig:cl106_T_a1}), by the large scale filaments which
pierce the shock front (e.g.\@ \cref{fig:cl106_S_a1,fig:cl14Maps}). In
some cases it can be seen to merge seamlessly with the virial shock
surrounding the large scale filaments (e.g.\@ \cref{fig:cl14_S_a06}).

The black circles in
\cref{fig:entTempRhoMaps_a1,fig:cl14Maps,fig:cl6Maps} mark the
approximate positions of the accretion shock which we estimate using
the entropy profile in the next section (\rfsec{edgeFind}). At first
glance, some of the estimations may seem erroneous since they do not
appear to correspond to the shock position in a given projection, when
in fact they are indicative of the shock position better viewed from a
different direction. This can be most clearly seen in
\cref{fig:cl6Maps}.
 
In \cref{fig:cl6Maps} we map cluster CL6 from 3 different
projections. The virial shock can be seen to extend out to different
distances in each direction though in all cases it is found well
beyond the virial radius. In this particular system, the cluster can
be seen to be positioned along a cosmic web filament of stretching in
the $Y$ direction. As can be seen, there is a cylindrical virial shock
around the filament as well (this is most evident in the \zeq{0.6}
maps, bottom), which merges seamlessly with the more spherical virial
shock of the cluster. This is expected for massive filaments, of
longitudinal density higher than $10^{12}\msun/\units{Mpc}$, which
feed massive clusters at low redshifts, as shown by
\citet{Birnboim2016}.

In this cluster, and in most other systems as well, the lower entropy
filaments can be seen to penetrate deep into the cluster virial shock,
as far as the virial radius and even, in some cases, into the very
central regions of the cluster. In \citet{Zinger2016} we address the
issue of these gas streams, their origin and structure as well as the
way they carry in energy which heats and stirs up turbulence in the
inner regions of the ICM.

\begin{figure*}
\centering 
\includegraphics[width=15cm,keepaspectratio,bb=0 0 9.75in 4.17in ]{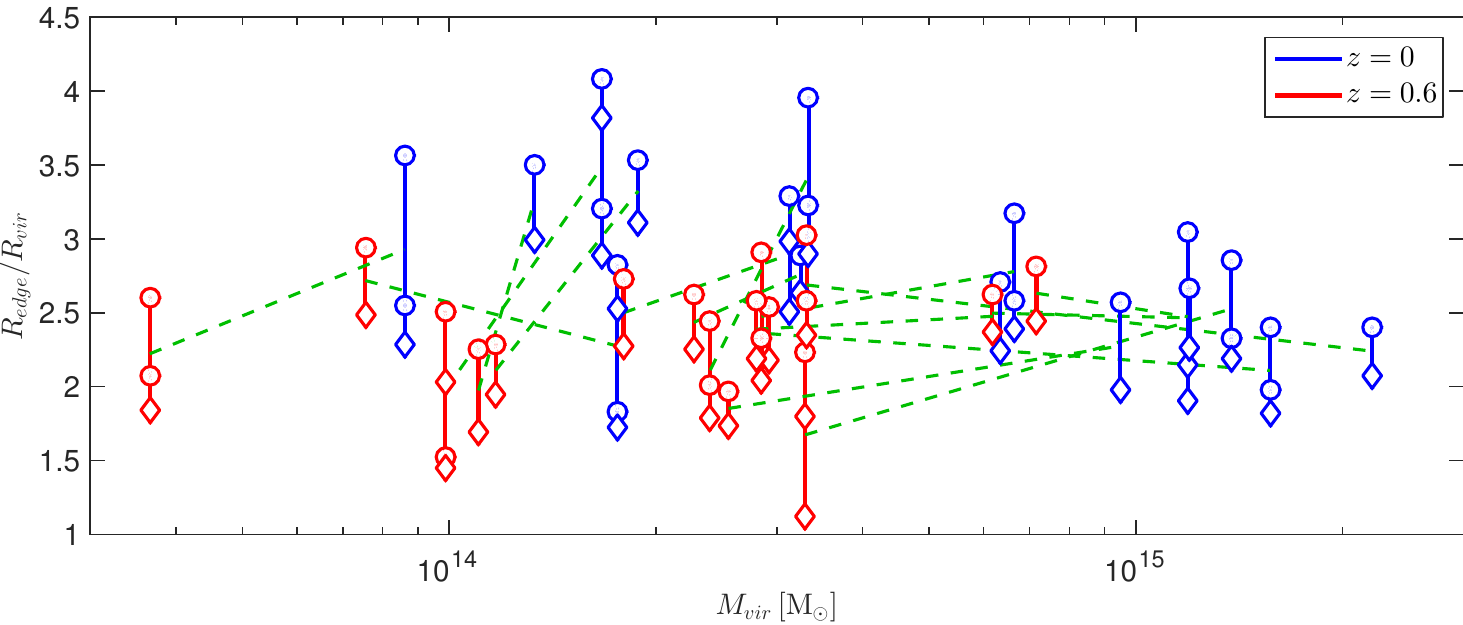}
  \caption{Shock edge estimation for all clusters at \zeq{0}
    ({blue}) and \zeq{0.6} ({red}) plotted versus the virial
    mass of the clusters.  The edge estimations of a given cluster at
    the two redshifts are connected by the {green} lines. Edge
    estimations by entropy gradient are denoted by {circles} and
    entropy maximum by {diamonds}.  All clusters exhibit shock
    edges which are well beyond the virial radius of the systems,
    extending to more than twice the virial radius in all cases.}
  \label{fig:edgeMvir}
\end{figure*}

\subsection{Determining the Shock Edge}\label{sec:edgeFind}
As can be seen in the temperature and entropy maps of the clusters
(\cref{fig:entTempRhoMaps_a1,fig:cl6Maps}), the edge of the virial
shock is characterized by a sharp drop in temperature and entropy of
several orders of magnitude. The drop in density at the shock is only
a factor of a few. Within the shock, the temperature rises gradually
towards the centre of the cluster, whereas the entropy in the cluster
drops, such that the entropy achieves its peak value close to the
shock front.

Spherically averaged density, temperature and entropy profiles were
created for all the cluster systems at both epochs, several
representative cases of which are shown in
\cref{fig:entProfs,fig:tempProfs,fig:rhoProfs}. These profiles are in
general agreement with observations \citep{Simionescu2013} beyond the
central regions of the cluster \mbox{$(r>0.05\Rv)$}.

The density profiles drop from the peak values in the center of the
cluster outwards. Beyond the $0.2\Rv$ region the profiles are well
approximated by a power-law. For an ideal gas equation of state, as is
used in these simulations, the maximal jump in density across a strong
shock is only an order of a few, thus it is very hard to identify the
accretion shock in these profiles.

The temperature profiles show that the virial shock, characterized by
a steep jump in temperature, is found at $\gtrsim 2 \Rv$. The reason
the jump in temperature is not as large nor as sharp as in the maps is
due to the fact that the profile is averaged over a spherical shell
and as we have seen, the shock front is not spherical, leading to a
smearing of the shock front signature. Behind the shock front, as one
moves towards the centre, the temperature rises steadily reaching the
virial temperature value at $\sim (0.3\textrm{--}0.4)\Rv$. The
constant rise makes the identification of the shock front somewhat
tricky, since the exact point in which the temperature drops is hard
to determine. A similar predicament occurs for the pressure profile
$P\propto \rho T$ since both the temperature and gas density rise
towards the centre.

The entropy profiles on the other hand rise steadily from the low
entropy core typical of galaxy clusters, reaching a peak value near
the shock and drops off suddenly at the shock front. The drop in
entropy is roughly one order of magnitude rather than a factor of
$3$--$4$ orders of magnitude, as can be seen in the maps, due to the
smearing of the shock front when averaging over a spherical shell. The
shape of the entropy profile with the distinctive maximum point and
subsequent drop naturally lead to a `quick-and-dirty' method of shock
edge identification, based on the position of the local maxima of the
entropy and the local minima of the entropy gradients, i.e.\@ maximal
negative gradient. While not as precise as state-of-the-art shock
detection methods \citep[e.g.\@][]{Ryu2003,Vazza2011,Schaal2015} which
can identify and trace the intricate three-dimensional structure of
the shock front, our method manages to find a reasonable estimate for
the location of the shock front (or fronts) as can be seen in
\cref{fig:entTempRhoMaps_a1,fig:cl14Maps,fig:cl6Maps}.

In \cref{fig:cl14_entProf} we show the entropy profile the cluster
CL14 at \zeq{0} and \zeq{0.6} and mark the radii of maxima in
the entropy profile as well as the maximal negative gradients. We
designate these radii as the shock edges, which are marked as black
circles in \cref{fig:cl14_S_a1,fig:cl14_S_a06}. As can be seen, in
this particular case (as well as in many others), more than one local
maxima was identified.  This should come as no surprise since we have
already noted that the shock front extends outwards to different
distances in different directions.

As one can see, there is a certain redundancy in identifying both the
local entropy maxima as well as the points of maximal negative
gradient, since they are usually close by. However, when examining all
the simulated systems we found that at times one method failed in
identifying a shock where the other method succeeded, due to the
intricate shape of the shock front. In addition, in this manner we can
treat the maximal negative gradients as the shock front position and
the local maxima as a lower bound to it. We find that employing this
simple method, of identifying both the entropy maxima and gradient
maxima as a rough estimate of the position of the shock front to be
perfectly adequate for determining its position with relation to the
virial radius. For each cluster, more than one `edge' may be defined
due to identification of both the entropy and entropy gradient maxima,
as well as due to multiple instances of these features.

We performed this analysis on all systems, and verified the results by
eye, removing specific cases which were obviously errors such as
entropy maxima found close to the systems centre or identification of
shocks in the large-scale structure very far from the virial shock. In
\cref{fig:edgeMvir} we show the shock edge positions, in units of
$\Rv$, for all the clusters in our sample, both at \zeq{0} and
\zeq{0.6} plotted versus the clusters' virial mass. The green lines
connect the edge estimation of a given system at \zeq{0.6} with the
edge estimation of the same system at \zeq{0}. We find that the shock
front is always found beyond the virial radius and in all systems
extends to well beyond twice the virial radius. One may make a case
for a trend of decreasing $R_{\rm edge}/\Rv$ with increasing system size
for the clusters at \zeq{0}. However, attempts at generating an
acceptable robust fit to these data points resulted in a very weak
relation. In the \zeq{0.6} sample such a relation is non-existent.

Generally, the extent of the shock in units of $\Rv$ is higher at
\zeq{0} than at \zeq{0.6} but we find no special trend in the
evolution of the shock edge for a given cluster, compared to the
growth in the virial radius. Most clusters evolve such that the ratio
is growing, and the growth in some cases is large, compared to a small
decline in other cases.

\section{Ram Pressure Stripping of the Gas Halo}\label{sec:rpsHalo}
Now that it has been established that the virial shocks of clusters
extend well beyond the virial radius, as early as \zeq{0.6}, we can
turn our attention to how this would affect the galaxies within the
shock radius, specifically with a view towards the quenching of star
formation.

The temperature of the area encompassed within the shock front is of
order $10^6\textrm{\textrm{--}}10^8\units{K}$,
(\cref{fig:cl103_T_a1,fig:cl106_T_a1}), and the velocity field is
dominated by random motions, with the exception of the inflowing gas
streams which is characterized by ordered inflow \citep{Zinger2016}.

One hydrodynamical mechanism which is known to affect galaxies in the
ICM is ram pressure stripping of the gas in the galaxy and its gas
halo. The motion of the galaxy within the ICM generates a ram pressure
which, in some cases, can overcome the gravitational binding force of the
satellite. The condition for gas stripping is
\begin{equation}\label{eq:rpsCond}
P_{\rm ram} \ge \frac{F_{\rm grav}}{\diff A},
\end{equation}
where $\diff A$ is the cross-section area over which the ram pressure
affects the gas.

It has long been known that the motion of a satellite galaxy within
the inner regions of the cluster can strip a galaxy of its gas
\citep{Gunn1972}, but we wish to assess the effect RPS has on an
infalling galaxy, and its gas halo, within the accretion shock but
\emph{before} it reaches the virial radius of the cluster.
 
To address this issue we will treat separately the gas confined to the
galactic disc and the gas in the surrounding gas halo. In all that follows the
term `galaxy' refers to a stellar and gaseous disc (possibly with a stellar
bulge), and the terms `gas halo' or `halo gas' refer to the gaseous medium
which surrounds the galactic disc and is embedded within the dark matter
sub-halo. In this section we address only the RPS of the gas halo, while the
fate of the gas found within the galactic disc is examined in \rfsec{rpsDisc}.

The actual process of gas stripping by ram pressure is a complicated
one. Detailed numerical simulation of isolated galaxies, i.e.\@ `wind
tunnel' simulations (e.g.\@ \citealt{Close2013,Roediger2015}) reveal
the highly complex nature of gas stripping. The gas in the hemisphere
facing the ram pressure is compressed and flows towards the edges,
where it is swept back and stripped in turbulent tails. The gas
distribution is deformed into a mushroom-like shape changing the area
affected by the ram pressure.

 While numerical studies are indispensable for the study of RPS, they
 are not without their limitation. As a case in point, most
 simulations do not usually include physical viscosity in the
 hydro-dynamic treatment and as such cannot properly capture the
 structure of the wake which forms around the moving object. The
 pressure in the wake reduces the effect of the ram pressure induced
 by the motion of the object through the medium. A notable exception
 is \citet{Roediger2015a}, where the effects of viscosity are studied
 and the complex interplay between the satellite, the stripped
 material in the wake and the ambient plasma is revealed.

Computational limitations are also an issue. To properly resolve the
intricacies of gas halo mass loss a resolution of order
$0.1\units{kpc}$ is necessary \citep{Roediger2015}, whereas the
typical size of clusters is of order several $\unitstx{Mpc}$.  In
addition, exploring the relevant parameter space which affects RPS
will incur high computational costs.

By making use of analytic models one may sidestep this
issues. Analytic models are typically computationally cheap and
therefore can be used to explore many different settings and parameter
values. In our simulated clusters, galaxies are not properly resolved
and the use of analytic models is a necessity.

It is extremely challenging to construct an analytic model which will
capture the intricacies of the RPS process. To overcome this hurdle,
we employ a simple toy model, motivated by the insights gleaned from
numerical studies, to describe the satellite gas halo and its
interactions with the ICM. However, when making use of simple
toy-models to describe such complicated processes one must be well
aware of their limitations.

At the outset, it must be stressed that we do not intend to use these
models to attempt to perfectly reproduce the mass loss from the
satellite haloes, but rather to obtain a lower limit on the amount of
mass removed by RPS. In the model we will assume an upper limit for
the binding gravitational force, so that the stripping is
under-estimated. As we will show, we still find substantial stripping
in this limit and can be assured that almost all of the halo gas can
be removed, despite the simplicity of the model.

In what follows, we will first construct fully analytic models, i.e.\@
models for both cluster and satellite, since this allows a basic
understanding of the important properties of cluster and satellites
which set the efficiency of RPS. Next, we will use our suite of
simulated clusters to model the background gas density and use the
analytic satellite models to gauge the effects of RPS.

\subsection{Cluster and Satellite Models}\label{sec:models}
The gaseous halo of a satellite or cluster, i.e.\@ the ICM, resides
within a dark matter halo which sets the gravitational potential which
in turns sets the thermodynamic properties of the gas.  We make use of
two simple density profiles to model the dark matter distribution for
our clusters and satellites - the Isothermal Sphere profile and the
NFW profile \citep{Navarro1996,Navarro1997}. The former is used to
obtain a simple analytic form for the stripping radius which
highlights the dependence on the cluster and satellite masses. The
latter, which is accepted as a reliable approximation for the dark
matter distribution, is used to obtain a more precise numerical
estimate for the stripping. To obtain the gaseous distribution we make
the simple assumption that the gas is isothermal and in hydrostatic
equilibrium within the dark matter potential.

\subsubsection{The Isothermal Sphere}\label{sec:IsoSphere}
The density profile for the dark matter in an Isothermal Sphere model is given by 
\begin{equation}\label{eq:isoRho}
\rho_{\rm iso}(r)=\frac{\Mv}{4\mathrm{\pi}\Rv^3}\left(\frac{r}{\Rv}\right)^{-2},
\end{equation}
with the mass profile given by 
\begin{equation}\label{eq:isoMass}
M_{\rm iso}(<r)=\Mv\left(\frac{r}{\Rv}\right) .
\end{equation}

Under the assumption of a hydrostatic isothermal gaseous component,
the gas density profile, $\rho_{\rm g}$, is set by the equation
\begin{equation}\label{eq:hydrostatCond}
  \frac{\kboltz T_{\rm g}}{\mu m_{\rm p}}\frac{\diff \ln \rho_{\rm g}}{\diff r}=-\frac{G M(r)}{r^2},
\end{equation}
where $\kboltz$ is the Boltzmann constant, $T_{\rm g}$ is the gas
temperature and $\mu m_{\mathrm{p}}\simeq 0.59 m_{\mathrm{p}}$ is the
average particle mass ($m_{\mathrm{p}}$ being the proton mass). Under
the assumption that the gas temperature is the same as the virial
temperature of the halo, $T_{\rm g}=\Tv$, solving
\cref{eq:hydrostatCond} results in $\rho_{\rm g}\propto r^{-2}$,
i.e.\@ the gaseous distribution follows that of the dark matter and is
also described by an Isothermal Sphere model. We therefore set the
density and mass profiles for the gaseous distribution by $\rho_{\rm
  g}=\fg \rho_{\rm iso}$ and $M_{\rm g}=\fg M_{\rm iso}$ where $\fg$
is the gas fraction defined by the ratio of the total gaseous mass
within $\Rv$ to the virial mass in the halo $\fg=M_{\rm gas}/\Mv$.

The strength of the Isothermal Sphere model is its simplicity, which
often allows for analytic solutions to toy-models leading to important
insights as to the key-players of a given model. Of course, this is
also its weakness - it is a poor approximation of the actual
distribution of gas and dark matter seen in observations and
simulations. In particular, the model over-predicts the density in the
outskirts of clusters, beyond $\Rv$, which leads to an over-estimation
of the RPS in these regions.

\subsubsection{The NFW Model}\label{sec:nfwModel}
The NFW model \citep{Navarro1996,Navarro1997} has long been accepted
as a fair approximation for the dark matter distribution in haloes. The density profile is given by
\begin{equation}\label{eq:nfwDensity}
\rho_{\mathrm{NFW}}(r)=\rho_u \delta_s \left( \frac{r}{r_s} \right)^{-1} \left( 1+\frac{r}{r_s} \right)^{-2},
\end{equation}  
where $\rho_u$ is the mean density of the universe, $\delta_s$ is an
over-density parameter and $r_s$ is the scale radius of the model. The
over-density parameter is given by
\begin{equation}\label{eq:deltaNFW}
\delta_s=\frac{\delvir}{3} \frac{\cvir^3}{\afunc{1}{\cvir}}, 
\end{equation}
where $\cvir \equiv \Rv/r_s$ and we define
\begin{equation}\label{eq:nfwADef}
\afunc{x}{\cvir} \equiv \ln\left(1+\cvir x \right)-\frac{\cvir x}{1+\cvir x}.
\end{equation} 
Using \cref{eq:virialDef} one may recast the model using the
parameters $\Mv$, $\Rv$ and $\cvir$
\begin{equation}\label{eq:nfwDensity2}
\rho_{\mathrm{NFW}}(r)=\frac{\Mv}{4 \mathrm{\pi}\Rv^3}
    {\afunc{1}{\cvir}}^{-1} \left( \frac{r}{\Rv} \right)^{-1} \left(
    \cvir^{-1}+\frac{r}{\Rv} \right)^{-2}.
\end{equation}  
 The mass profile for the model is
\begin{equation}\label{eq:nfwMass}
M_{\mathrm{NFW}}(r)=\Mv \frac{\afunc{r/\Rv}{\cvir}}{\afunc{1}{\cvir}}.
\end{equation}

To use this model one must determine an additional parameter, namely
$\cvir$. Analysis of \nbody~simulations yields a power-law relation
between $\Mv$ and $\cvir$ as a function of redshift
\citep{Bullock2001a,Wechsler2002,Maccio2008}. We use the relation
found in \citet{Munoz-Cuartas2011} to determine $\cvir$ for a given
value of $\Mv$.

To set the gaseous distributions we solve the hydrostatic equation,
\cref{eq:hydrostatCond}, under the same assumptions as before. The
equation can be solved analytically
\citep[e.g.\@][]{Makino1998,Brueggen2008}, resulting in a distribution
which closely resembles an NFW profile for $0.1\Rv < r < \Rv$, but
possesses a density core for smaller radii. For radii of $r>\Rv$, the
gas density drops off less rapidly than the dark matter density, and
levels off at $r\gtrsim2\Rv$ a behavior which is \emph{not} observed
in real clusters \citep{Finoguenov2001}. In light of this, we opt
instead to assume that the gas distribution follows that of the dark
matter, i.e. $\rho_{\rm g}=\fg \rho_{\rm NFW}$ and $M_{\rm g}=\fg
M_{\rm NFW}$ where once again $\fg$ is the ratio of the gaseous mass
to virial mass $\fg=M_{\rm gas}/\Mv$. In this way the gas density in
the outskirts of the cluster is not over-estimated, and the density
profile of the satellite is similar to the results of the isothermal,
hydrostatic gas distribution.

The NFW model is a better fit for observed distributions of dark
matter and is expected to give better predictions for the RPS in the
cluster outskirts. However, employing this profile in toy-models often
results in implicit equations which entail a numerical solution. While
more precise, it is sometimes harder to glean the important aspects
from such a solution.

\subsection{Tidal Stripping versus Ram Pressure Stripping}\label{sec:haloTidal}
Before continuing with assessing the stripping due to ram pressure, we
must ensure that it is indeed the dominant stripping mechanism in the
region of interest, and that other stripping mechanisms, namely tidal
stripping, do not void the basic assumptions of the model.

Tidal stripping occurs when the tidal forces acting on a satellite
overcome the gravitational binding force of the satellite. Unlike RPS,
which is a hydrodynamical process that affects only the gas and leaves
the dark matter halo intact, tidal stripping is a dynamical process
which can remove the dark matter halo. If tidal stripping is effective
in the region of interest, then modelling the gas halo as an Isothermal
Sphere or by the NFW model may no longer be valid since the dark
matter potential well which determines the gas properties of the halo
may have been totally or partially removed (along with the gas).

To gauge the effectiveness of the tidal stripping we compare the tidal
force acting on a gas element in the satellite halo and the
gravitational force exerted by the satellite on said gas element. The
tidal force is maximal along the line connecting the centres of the
cluster and satellite. Perpendicular to that line, the tidal forces
actually \emph{compress} the satellite \citep{Dekel2003}. For a
satellite whose centre is at a distance $r$ from the cluster centre,
the tidal force on a gas element of mass $\diff m$ at a distance
$\ell$ from the satellite centre (along the line connecting the
centres) is
\begin{equation}\label{eq:tidalForce}
F_\mathrm{T}=-\frac{G M_{\rm clust}(r+\ell) \diff m}{(r+\ell)^2}+\frac{G M_{\rm clust}(r) \diff m}{r^2}
\end{equation}
and the gravitational force exerted by the satellite is
\begin{equation}\label{eq:satForce}
F_{\rm sat}(r)=-\frac{G M_{\rm sat}(|\ell|)\diff m}{|\ell|^3}\ell.
\end{equation}

Combining \cref{eq:tidalForce,eq:satForce} and assuming an Isothermal Sphere
profile for both cluster and satellite results in the following
equation for the tidal radius $\tl_t$ expressed in units of the
satellite virial radius
\begin{equation}\label{eq:lStripTid1}
\tl_t^2-\mu^{1/3}\tilde{r}\tl_t -\tilde{r}^2=0,
\end{equation}
where $\tl_t \equiv \ell_t / \Rsat$, $\tilde{r}\equiv r/\Rc$ and
$\mu=\Msat/\Mc$. Solving the equations yields the tidal radius of the
satellite as a function of the satellite position in the cluster
\begin{equation}\label{eq:lStripTid2}
\tl_t=\tilde{r}\frac{\mu^{1/3}}{2}\left(1+\sqrt{1+4\mu^{-2/3}}\right)\approx
  \tilde{r}\left(1+\frac{\mu^{1/3}}{2}\right),
\end{equation}
with the second, approximate relation valid when \mbox{$\mu \ll
  4^{3/2}=8$}, which is always the case since the mass of satellite
haloes which host single galaxies in the cluster is of order 1 per
cent of the cluster mass or lower.

We therefore find that the tidal radius of satellites found beyond the
virial radius of the cluster is always larger than the virial radius
of the satellite. We found this result to hold when both the satellite
and cluster are modeled with an NFW profile as well. We can therefore
safely conclude that tidal stripping is ineffective in the cluster
outskirts.

\begin{figure}
  \centering
  \includegraphics[width=8cm,keepaspectratio,bb=0 0 8.5in 11.0in,trim=0.63in 2.46in 0.91in 2.84in, clip]{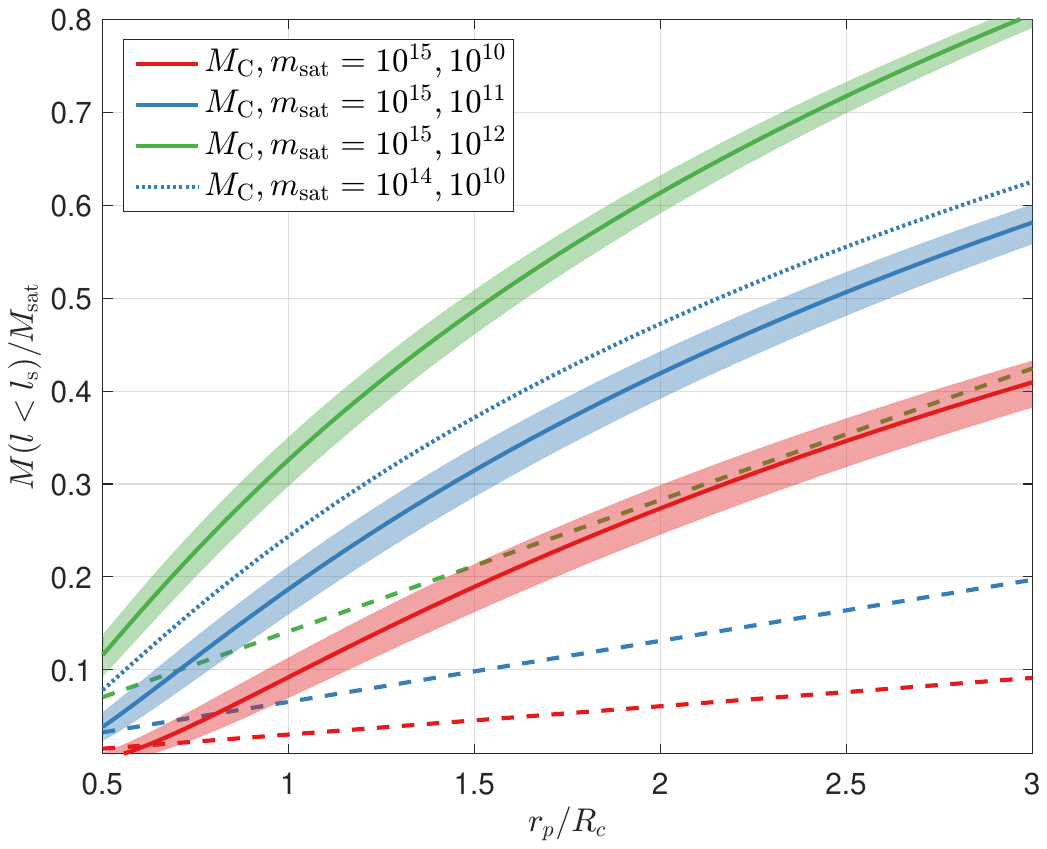}
  \caption{A simple toy-model for the amount of mass remaining in the
    gas halo of satellites undergoing RPS in the ICM of a
    $\Mc=10^{15}\msun$ cluster, as a function of the position in the
    cluster in units of the cluster virial radius $\Rc$. Solid lines
    represent the stripping for the NFW model, with the shaded regions
    corresponding to variations of up to $\pm 10$ per cent in the
    values of both $\cvc$ and $\cvs$. Dashed lines show the stripping,
    for the same values of the satellite-to-cluster mass ratios, for
    the Isothermal Sphere model. The RPS efficiency parameter is set
    to $\alfP= 0.5$. Satellites lose more than \perc{70} of their gas
    halo mass before reaching the virial radius, with low-mass
    satellites losing as much as \perc{90}. For comparison, the blue
    dotted line shows the case of stripping in a $10^{14}\msun$
    cluster with the same satellite-to-host mass ratio as the blue
    solid line ($10^{-4}$). The stripping in this case is reduced
    since both the cluster and satellite are more concentrated.}
  \label{fig:rpsSimple}
\end{figure}

\begin{figure*}
  \centering
  \subfloat[$r=2\Rc$]{\label{fig:rpsAlpha_2Rv}
    \includegraphics[width=8cm,keepaspectratio,bb=0 0 5.56in 4.35in]{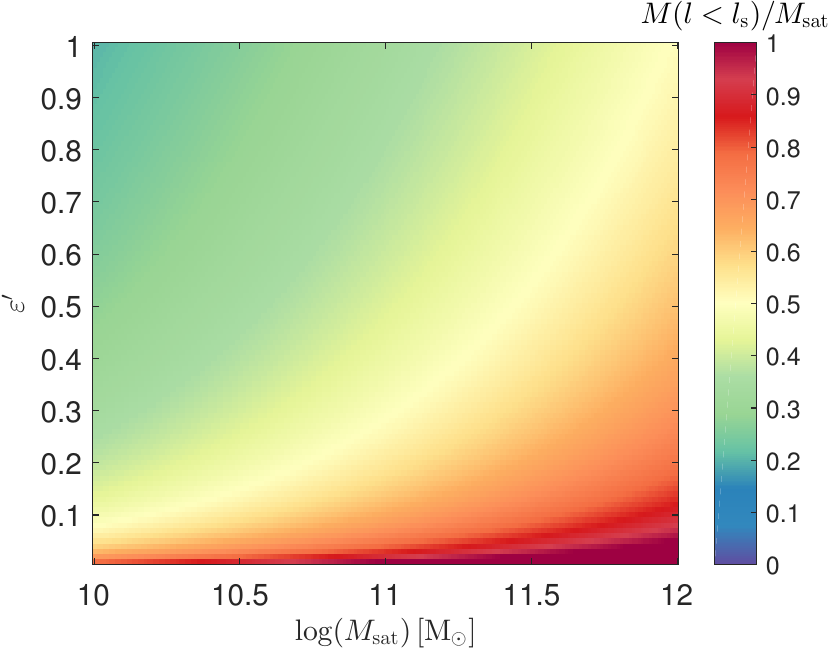}}
  \subfloat[$r=1\Rc$]{\label{fig:rpsAlpha_1Rv}
    \includegraphics[width=8cm,keepaspectratio,bb=0 0 5.56in 4.35in]{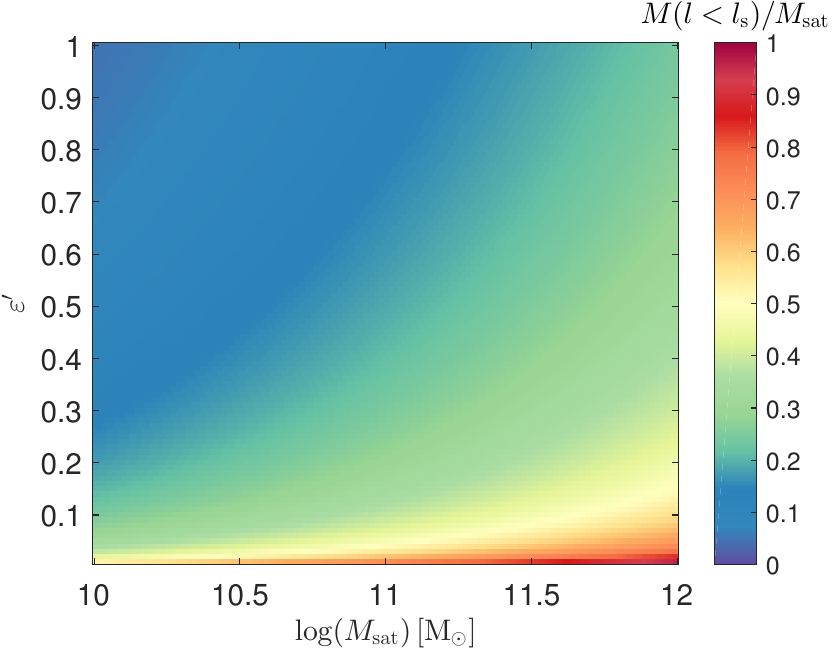}}
  \caption{We explore the dependence of the simple NFW RPS toy-model
    on the efficiency parameter $\alfP$, for a range of satellite
    masses within a $\Mc=10^{15}\msun$ cluster.  We examine the
    stripping for a satellite positioned at twice the cluster virial
    radius \subrfig{rpsAlpha_2Rv} and at the virial radius
    \subrfig{rpsAlpha_1Rv}. At $2\Rc$, low-mass satellites lose more
    than half the mass, even for low values of $\alfP \gtrsim 0.1$. At
    $1\Rc$, even high-mass satellites lose more than half their halo
    gas mass for low values of $\alfP$, and low-mass galaxies are
    stripped to well below \perc{20} for most values of $\alfP$.}
  \label{fig:rpsAlpha}
\end{figure*}

\subsection{Gauging the Mass Stripping}

\subsubsection{Ram Pressure}\label{sec:ramPressure}
A galaxy moving at a velocity of $\vec{v}_{\rm sat}$ within a gaseous
medium of density $\rho_{\mathrm{med}}$ will be subjected to a
ram pressure of
\begin{equation}\label{eq:rpsDef}
P_{\rm ram} = C_{\rm ram}\rho_{\mathrm{med}}\left(\vec{v}_{\rm sat}-\vec{v}_{\mathrm{med}}\right)^2,
\end{equation} 
where $\vec{v}_{\mathrm{med}}$ is the local velocity of the medium and
$C_{\rm ram}$ is a unit-less pre-factor.

For the motion of the satellite we will assume its speed is constant
and equal to the virial velocity of the cluster\footnotemark
\begin{equation}\label{eq:satVel}
  v_{\rm sat}=\sqrt{\frac{G\Mc}{\Rc}}\simeq
  \begin{cases}
    1290 M_{15}^{1/3}\units{km\,s^{-1}} & z=0 \\ 
    1520 M_{15}^{1/3}\units{km\,s^{-1}} & z=0.6
\end{cases},
\end{equation}
where $M_{15}\equiv \Mc /10^{15}\mathrm{M_\odot}$.  \footnotetext{Velocity
  profiles in the simulated cluster show that typical velocities, even in the
  radial range between $\Rv$ and $3\Rv$ differ from $\Vv$ by at most a factor
  of 2. }

In addition we assume that the ICM is at rest compared to the
satellite motion, ($v_{\mathrm{med}}=0$). While this is clearly not
the case at every given point in the cluster, since the gas in the ICM
is constantly in motion, it is true in an average sense if one takes
into account that the velocity field of the ICM is dominated by random
motions. Thus, as a satellite travels through the ICM, its velocity
will at times be aligned with that of its surroundings and
anti-aligned at other times (with the ram pressure shrinking and
growing accordingly), but on average the ICM velocity will be zero.

As can be seen in \cref{fig:CL103_rps}, this assumption appears to be
valid with one important exception -- motion along the gas streams
falling into the cluster, where the velocity field is flowing
coherently over large distances. We discuss the implications of this
exception in \rfsec{rpsSat_simClust}.

Inserting \cref{eq:satVel} into \cref{eq:rpsDef} we use \cref{eq:isoRho,eq:nfwDensity2} to 
find the ram pressure for the Isothermal Sphere,
\begin{equation}\label{eq:ramPressIso}
P_{\rm ram}(r)=\alf \fc \frac{G}{4 \mathrm{\pi}} \frac{\Mc^2}{{\Rc}^4} 
\tilde{r}^{-2},
\end{equation}
and the NFW Model,
\begin{equation}\label{eq:ramPressNFW}
P_{\rm ram}(r)=\alf \fc \frac{G}{4 \mathrm{\pi}} \frac{\Mc^2}{{\Rc}^4} 
{\left[{\afunc{1}{\cvc}}\tilde{r} \left(\cvc^{-1}+\tilde{r} \right)^{2}\right]}^{-1}, 
\end{equation}
where $\tilde{r}\equiv r/\Rc$. 
In the above expressions, $f_{\rm c}$ is the cluster gas fraction,
$\Mc$ and $\Rc$ are the virial mass and radius of the cluster
respectively, and $\alf$ is a fudge-factor, added to incorporate any
uncertainties in the assumptions we made in our model, e.g.\@ motion
with velocities different than the virial velocity, and which also
accounts for $C_{\rm ram}$.

\subsubsection{Gravitational Binding}\label{sec:gravForce}
We wish to compare the ram pressure to the binding gravitational force
per unit area acting on the satellite gas halo and determine the
amount of mass stripped from the satellite. While expressing the
ram pressure is relatively simple, as shown above, determining the
correct form of the gravitational binding force for the appropriate
mass elements affected is not so straightforward.

Ideally, one wishes to construct a simple spherical approximation for
the binding force at a radius $\ell$ in the satellite which will lead
to a definition of a stripping radius $\ell_s$.

One option is to find the gravitational force which binds a spherical
shell of width $\diff \ell$ at a given radius $\ell$, divided by the
cross section of that shell $\diff A=2\mathrm{\pi}\ell \diff \ell$
\begin{equation}\label{eq:onion}
\frac{F_{\rm grav}}{\diff A}=\frac{G M(\ell)}{\ell^2}\frac{\rho(\ell) 4\mathrm{\pi}
  \ell^2 \diff \ell}{2\mathrm{\pi} \ell \diff \ell}= 2\frac{G M(\ell)
  \rho(\ell)}{\ell},
\end{equation}
where $M(\ell)$ is the \emph{total} mass profile and $\rho(\ell)$ is
the gas density profile of the satellite. This approximation suggests
that the ram pressure `peels' off spherical shells, like an onion,
which is incongruous with the anisotropic nature of RPS. However, if
the satellite halo is rotating with respect to its direction of
motion, such that all parts of the outer shell experience the
ram pressure, this approach may be justified.

Another option is to examine a cylindrical shell of radius $\ell$ and
width $\diff \ell$ ($\diff A=2\mathrm{\pi}\ell \diff \ell$), oriented
in the direction of motion of the satellite, and thus the direction of
the ram pressure force. \citet{McCarthy2008} employed this method by
finding the maximal value of the restoring gravitational acceleration
in the direction of the motion (and, by definition, the cylinder axis)
as well as an upper limit of the projected surface density, resulting
in an overestimation of the binding force which ensures that one finds
an upper limit to the stripping radius.

For a power-law spherically symmetric density profile $\rho\propto
r^{-\alpha}$ the maximal restoring acceleration is\footnotemark
  \begin{equation}\label{eq:gzMax}
g_{\mathrm{max}}(\ell)=\frac{G M(\ell)}{\ell^2}
\sqrt{\frac{ (\alpha-1)^{\alpha-1}}{\alpha^\alpha}},
\end{equation}
and the surface density is
\begin{equation}\label{eq:SigmaCyl}
\Sigma(\ell)=2 \rho(\ell)\ell \int_0^{L}
(1+x^2)^{-\frac{\alpha}{2}}\diff x.
\end{equation}
The upper limit of the integral is formally
\mbox{$L=\sqrt{R_{\mathrm{vir}}/\ell-1}$} but can be taken to $L\to
\infty$ when the integral converges (or to some set scale otherwise)
to achieve an upper limit for the surface density. The gravitational
binding force is then
\begin{equation}\label{eq:fgravCyl}
\frac{F_{\rm grav}}{\diff A}=g_{\mathrm{max}}(\ell)\Sigma_{\mathrm{max}}(\ell)=\nu\frac{G M(\ell)\rho(\ell)}{\ell},
\end{equation} 
where $\nu$ is a factor of order unity 
\begin{equation}\label{eq:nuDef}
\nu=2 \sqrt{\frac{ (\alpha-1)^{\alpha-1}}{\alpha^\alpha}}
\int_0^{\infty} (1+x^2)^{-\frac{\alpha}{2}}\diff x.
\end{equation}
For an Isothermal Sphere profile we find $\nu=\mathrm{\pi}/2$.
\footnotetext{For a power-law density profile, $\rho\propto
  r^{-\alpha}$, the mass profile is $ M \propto
  r^{3-\alpha}(3-\alpha)^{-1}$, which is not defined for the case of
  $\alpha=3$. Thus, \cref{eq:gzMax} is strictly correct for the
  non-pathological cases of \mbox{$\alpha\ne 3$}. For $\alpha=3$ one
  may still define a mass profile, e.g.\@ by adding a small core
  radius \mbox{$\rho\propto (r+r_c)^{-\alpha}$}, to be set by boundary
  conditions, and equate the maximal acceleration with
  $GM(\ell)/\ell^2$ up to a numerical factor, which will depend on the
  core radius. }

Yet another option is to assume that the gas halo is in hydro-static
equilibrium within the gravitational potential of the dark matter halo
of the satellite, and use the gas pressure as a proxy for the binding
force per unit area. For a power-law density profile finding the
pressure by integration of the hydro-static equation results in
\begin{equation}\label{eq:presureProxy}
P(\ell)=(2\alpha-2)^{-1}\frac{G M(\ell)\rho(\ell)}{\ell}.
\end{equation}
For an Isothermal Sphere profile the pre-factor is $1/(2\alpha-2)=1/2$.

We find that employing several different methods of approximating the
gravitational binding force all lead to an expression of the same
form, namely
\begin{equation}\label{eq:gravForce1}
\frac{F_{\rm grav}}{\diff A}=\kappa \frac{G M(\ell) \rho(\ell)}{\ell} ,
\end{equation} 
with $\kappa$ being a factor of order unity.  While the above
expression has been derived for a simple power-law density profile, it
is reasonable to assume it is equally valid for more general density
profiles, such as the NFW profile. Naturally, if one were to guess an
approximation to the binding force per unit area based on a simple
dimensional analysis, the above expression would be the result.

For the satellite models used in this work (\rfsec{models}) the
gravitational binding force per unit area is 
\begin{equation}\label{eq:gravForceIso}
\frac{F_{\rm grav}}{\diff A}=\kappa \fg \frac{G}{4 \mathrm{\pi}}
\frac{\Msat^2}{{\Rsat}^4} \tl^{-2}
\end{equation} 
and 
\begin{equation}\label{eq:gravForceNFW}
\frac{F_{\rm grav}}{\diff A}
=\kappa \fg \frac{G}{4 \mathrm{\pi}} \frac{\Msat^2}{{\Rsat}^4} 
\frac{\afunc{\tl}{\cvs}}{ {\afunc{1}{\cvs}}^2} \left[ \tl \left(
    \cvs^{-1}+\tl\right)\right]^{-2}
\end{equation} 
for the Isothermal Sphere and NFW models respectively where once
again, $\tl\equiv \ell / \Rsat$. In the equations above, we have
assumed the gravitational potential is dominated by the dark matter
component and neglected the gaseous component. Taking this component
into account introduces an additional factor of $(1+\fg)$. The effect
of this additional factor on the stripping radius (\equnp{lstrip1})
for typical values of $\fg$ is only of order several per cent.

\begin{figure*}
  \subfloat[\zeq{0}]{\label{fig:rps_sat_simClust_a1}
    \includegraphics[width=\columnwidth,keepaspectratio,bb=0 0 5.22in 4.14in ]{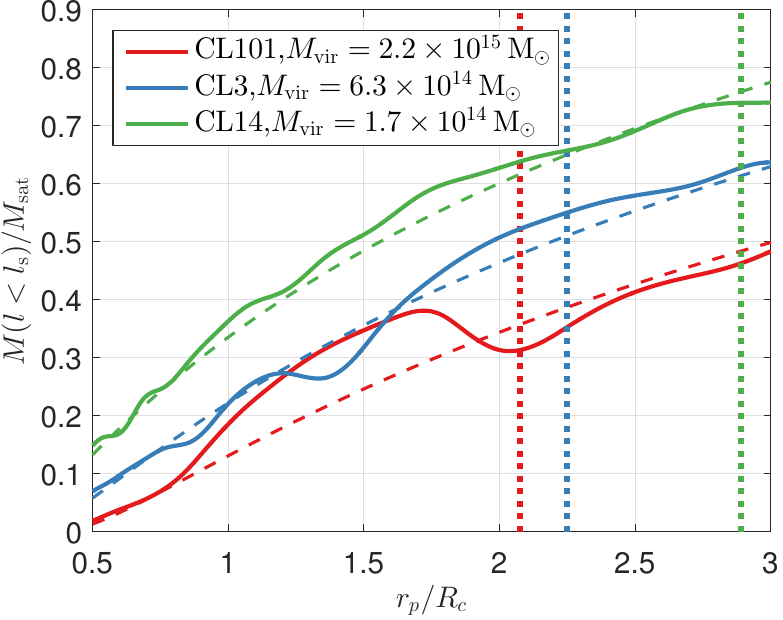}}
   \subfloat[\zeq{0.6}]{\label{fig:rps_sat_simClust_a06}
    \includegraphics[width=\columnwidth,keepaspectratio,bb=0 0 5.22in 4.14in ]{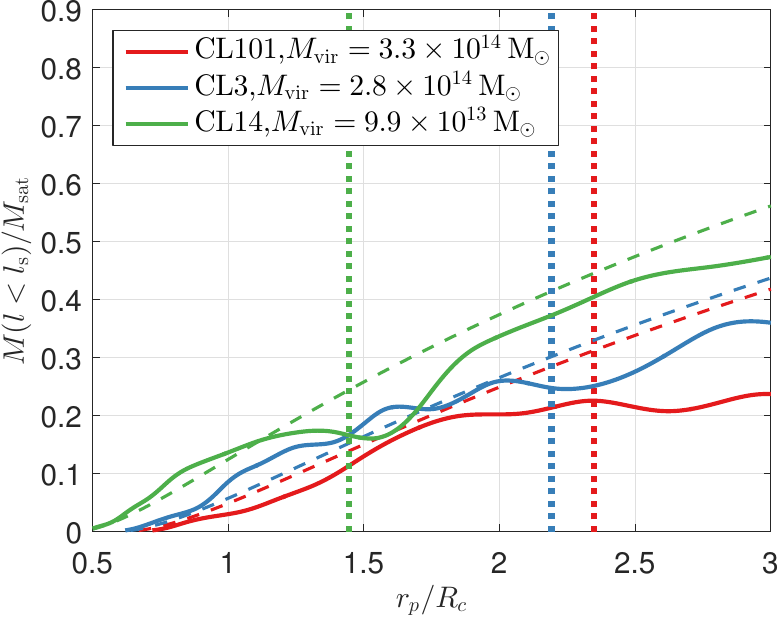}}
  \caption{The remaining gas mass of an $\Mv=10^{11}\msun$ satellite
    halo (modeled with an NFW profile) in 3 representative simulated
    clusters: CL101 ({blue}), CL3 ({red}) \& CL14 ({green}). The
    ram pressure of the cluster is calculated by using the spherically
    averaged gas density profile, and assuming the satellite travels
    at the virial velocity of the cluster. In this calculation we have
    assumed $\alfP=0.5$. The dashed lines show in comparison the
    remaining mass assuming an NFW model for the cluster gas (as in
    \cref{fig:rpsSimple}). We present the results for the 3 clusters
    at \zeq{0} \subrfig{rps_sat_simClust_a1} and \zeq{0.6}
    \subrfig{rps_sat_simClust_a06}. The dotted lines denote the lowest
    value of the shock edge as defined in \rfsec{edgeFind}
    (\cref{fig:edgeMvir}). The stripping in the simulated clusters is
    very similar to the NFW toy-model. The stripping is very effective
    in the high-mass clusters, removing more than \perc{80} of the gas
    before the satellites reach the virial radius. The stripping in
    the lower mass clusters is less effective but still more than
    \perc{60} of the gas is removed by the time the satellite reaches
    the virial radius of the cluster.}
  \label{fig:rps_sat_simClust}
\end{figure*}
\begin{figure*}
  \subfloat[\zeq{0}]{\label{fig:rps_sat_simClust_12rv_a1}
    \includegraphics[height=7cm,keepaspectratio,bb=0 0 5.13in 4.13in ]{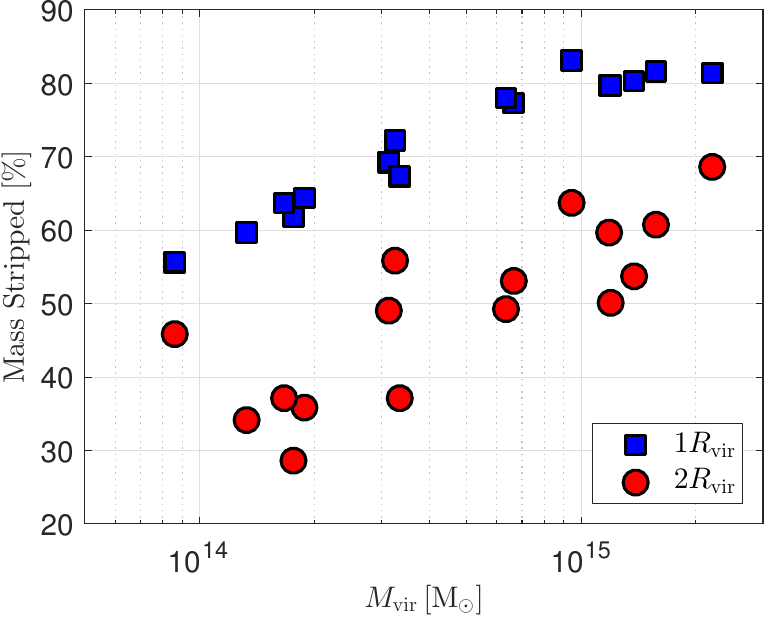}}
   \subfloat[\zeq{0.6}]{\label{fig:rps_sat_simClust_12rv_a06}
    \includegraphics[height=7cm,keepaspectratio,bb=0 0 5.46in 4.13in ]{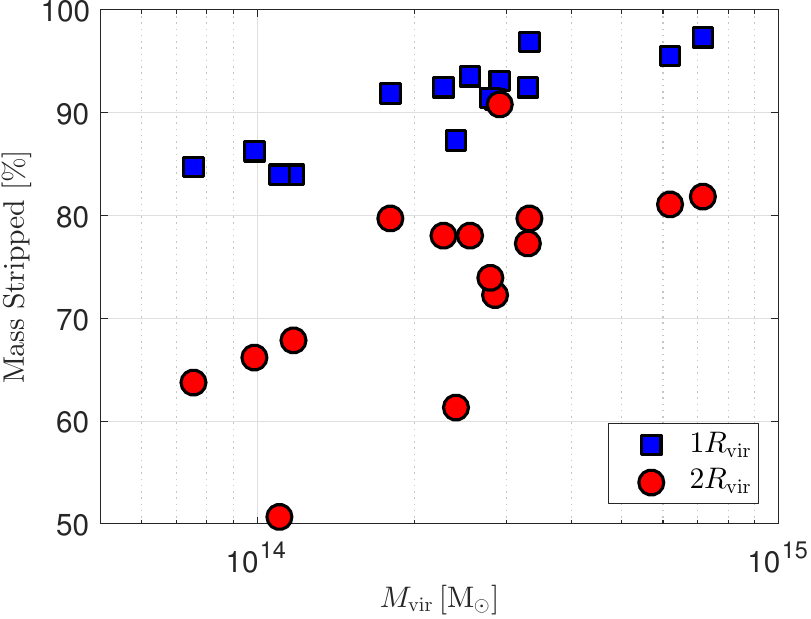}}
  \caption{The total amount of gas stripped from an NFW satellite halo
    of $\Msat=10^{11}\msun$ at twice the virial radius of the cluster
    ($2\Rc$, {red circles}) and at the virial radius of the cluster
    ($1\Rc$, {blue squares}) in all the clusters of the simulation
    suite at \zeq{0} \subrfig{rps_sat_simClust_12rv_a1} and \zeq{0.6}
    \subrfig{rps_sat_simClust_12rv_a06}. The effectiveness of the
    stripping in our simple model is dependent on the cluster mass,
    with the satellite losing about \perc{60} of its gas mass in the
    low-mass clusters and as much as \perc{80} for the high mass
    clusters. The stripping is more effective at \zeq{0.6}, due to the
    higher densities found at the virial radius and the lower
    concentrations of satellites at that epoch.}
  \label{fig:rps_sat_simCLust_rv12}
\end{figure*}

\subsubsection{Defining a Stripping Radius}
To find the stripping radius we equate the ram pressure, and the
binding force. For the Isothermal Sphere model we use
\cref{eq:ramPressIso,eq:gravForceIso} to find the stripping radius,
$\tl_s\equiv \ell_s/\Rsat$, as a function of the satellite position
within the cluster, $\tilde{r}_p\equiv r_p/\Rc$ 
\begin{equation}\label{eq:lstrip1}
\tl_s=\sqrt{\frac{\kappa}{\alf}\frac{f_g}{f_c}}\left(\frac{\Msat}{\Mc}\right)^{\frac{1}{3}}\tilde{r}_p =\frac{1}{\sqrt{\alfP}}\mu^{\frac{1}{3}}\tilde{r}_p,
\end{equation}
making use of the relation $\Rv\propto\Mv^{\frac{1}{3}}$ (\equnp{virialDef}). 
In this final form we combine the two fudge-factors
$\kappa$ and $\alf$ into a single parameter $\alfP=\alf/\kappa$
which incorporates all the uncertainties in the model. Low values of
$\alfP$ reduce the effectiveness of the RPS mechanism. In addition, we
hereafter make the reasonable assumption\footnotemark that $f_c\simeq
f_g$ and as before define the satellite to cluster mass ratio
$\mu\equiv\Msat/\Mc$. Scaling by typical values allows us to estimate
the value of the stripping radius
\begin{equation}\label{eq:lstrip3}
\tl_s \simeq 0.05 \left( \frac{\alfP}{0.5}\right)^{-\frac{1}{2}}  
\left(\frac{\Msat}{10^{11}} \right)^{\frac{1}{3}} 
\left(\frac{\Mc}{10^{15}}\right)^{-\frac{1}{3}} 
\tilde{r}_p.
\end{equation}
\footnotetext{The ratio $f_g/f_c$ is of
  order unity, and enters the expression for the stripping radius as a
  power of $1/2$. Any deviation from equality of the two gas fractions
  will have only a small effect on the stripping radius.}

Under the assumption of an Isothermal Sphere profile for the satellite
haloes, the value of $\tl_s$ also embodies the fraction of gas mass
stripped from the halo ( \equnp{isoMass}). \Cref{eq:lstrip3} shows us
that ram pressure stripping of the halo gas is very efficient with
$\gtrsim 95$ per cent of the gas having been stripped before the
satellite reaches the virial radius. 

The above result is hardly surprising due to the very simple gas model
employed which is completely defined by a single parameter, in this
case the virial mass of the halo, and for which the temperature and
virial velocity are constants. This implies that the relation be
governed completely by the mass ratio of the two bodies (and the free
parameter $\alfP$) and that the model is completely scale free.

The Isothermal Sphere model is useful for achieving an explicit
solution for the stripping, and shows that the halo gas in satellites
is stripped almost completely when the satellite is at $\Rv$. However,
as noted above, this is most likely an over-estimation of the
stripping since the gas in this model is denser in the outskirts than
more realistic profiles of the cluster density. A better approximation
can be reached using the NFW model. Equating the ram pressure with the
gravitational binding force for an NFW model,
\cref{eq:ramPressNFW,eq:gravForceNFW}, results in an implicit equation
which can be solved numerically. Once a stripping radius is found, the
mass remaining in the satellite can be calculated with
\cref{eq:nfwMass}.

In \cref{fig:rpsSimple} we show the amount of mass remaining in the
gas halo of satellites of virial mass \mbox{$10^{10},\,10^{11}$ and
  $10^{12}\msun$} within the ICM of a $10^{15}\msun$ cluster, both of
which are modeled with the NFW profile. The solid line shows the
result for an NFW profile where the concentration parameters of both
host and satellite were set by the \citet{Munoz-Cuartas2011} relation,
and the shaded region accounts for the results after varying the
concentration parameter by as much as $\pm 10$ per cent for both host
and halo. In addition, the results of the Isothermal Sphere model for
the corresponding mass ratios are also shown by the dashed lines.

We find that typical satellites in clusters will have already lost a
substantial amount of their gas even at $2\Rv$, as much as \perc{40}
for very massive satellites and \perc{70} (or more) for low mass
satellites. A satellite at the virial radius of the cluster will
retain only a small fraction of its initial gas halo, with values of
\perc{10} and less for low-mass satellites.  As expected, the
Isothermal Sphere model can be seen to over-estimate the mass
stripping, especially in the outer regions of the cluster.

An interesting feature can be seen when comparing the results for the
$10^{11}\msun$ satellite in a $10^{15}\msun$ host (solid blue line) to
the results for a $10^{10}\msun$ satellite in a $10^{14}\msun$ cluster
(dotted blue line). Though the satellite-host mass ratio is identical
for both cases, the stripping is markedly weaker for the lower mass
host. This is due to the fact that the concentration parameter is
higher for lower mass haloes. The relative drop
in density (at a given position in units of $\Rv$) is larger for a
more concentrated host halo, reducing the ram pressure. In addition, a
more concentrated satellite is more tightly bound and therefore harder
to strip.

In \cref{fig:rpsAlpha} we explore the dependence of the NFW model on
the stripping effectiveness parameter $\alfP$. For a cluster of
$\Mc=10^{15}\msun$ we plot the mass remaining within the gas halo at
the virial radius of the cluster and at twice the distance, for a
range of satellite masses and $\alfP$ values. We can see that even for
very weak stripping, i.e.\@ low values of $\alfP\simeq 0.1$, all but
the most massive satellites lose more than \perc{60} of their mass at
the virial radius.

Since we find that the stripping is effective even for low values of
$\alfP$ we can rest assured that the various approximations made in
our simple model, especially with respect to the gravitational binding
force (\rfsec{gravForce}), do not change the conclusion that the gas
haloes lose all but a small fraction of their initial mass
\emph{before} they reach the virial radius of the cluster for the
first time.

\begin{figure*}
  \subfloat[Stripping of $10^{11}\msun$ Satellite at \zeq{0}]{\label{fig:CL103_rpsM_a1}
    \includegraphics[width=\columnwidth,keepaspectratio,bb=0 0 4.83in 4.35in ]{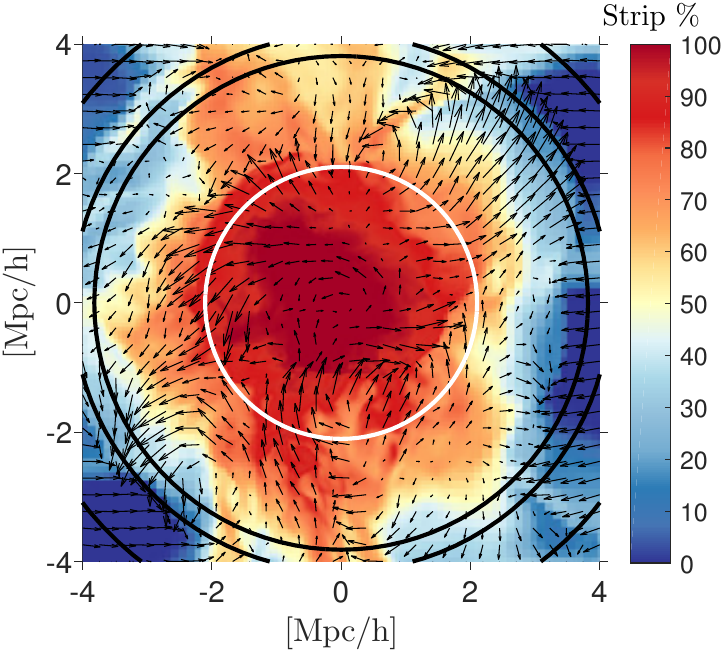}}
  \subfloat[Stripping of $10^{11}\msun$ Satellite at \zeq{0.6}]{\label{fig:CL103_rpsM_a06}
    \includegraphics[width=\columnwidth,keepaspectratio,bb=0 0 5.01in 4.35in]{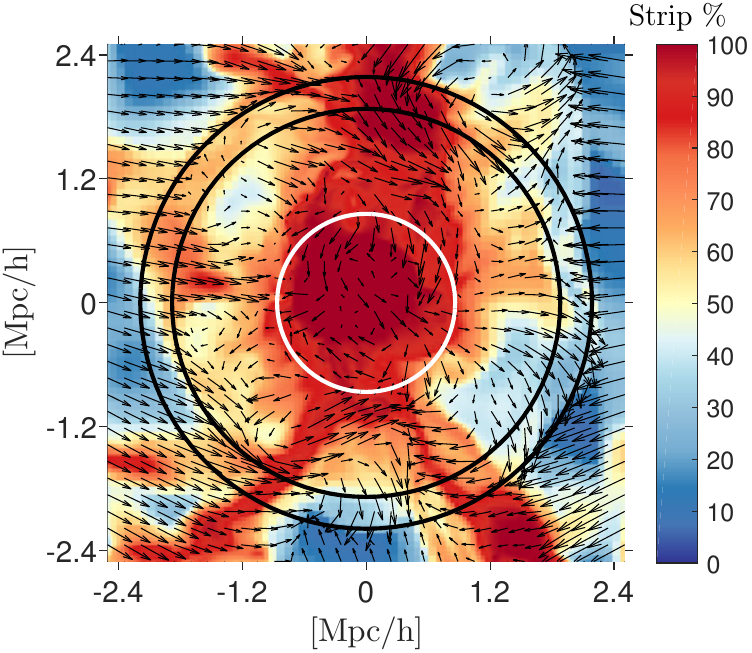}}
  \caption{Stripping of the gaseous halo of a satellite calculated
    based on the simulated gas density in the ICM of the cluster CL103
    at \zeq{0} (left) and \zeq{0.6} (right). The RPS effectiveness
    parameter is assumed to be $\alfP=0.5$. The figures show the
    percentage of gas mass stripped from an $\Mv = 10^{11}\msun$
    satellite at each point in the cluster, travelling at the virial
    velocity of the cluster, $\Vv=1497\units{km\,sec^{-1}}$ at \zeq{0}
    and $\Vv=1009\units{km\,sec^{-1}}$ at \zeq{0.6}. We see that
    already in the vicinity of the shock front (marked by black
    circles) the satellite will be stripped of at least \perc{30} of
    its mass. At the virial radius of the cluster (white circle) the
    satellite has lost between $70$ to \perc{90} of its halo gas
    mass. The black arrows show the velocity field which highlights
    the inflow along the high density streams.}
  \label{fig:CL103_rps}
\end{figure*}

\begin{figure*}
  \subfloat[Ram Pressure \zeq{0}]{\label{fig:CL103_rp_a1}
    \includegraphics[width=\columnwidth,keepaspectratio,bb=0 0 4.92in 4.37in ]{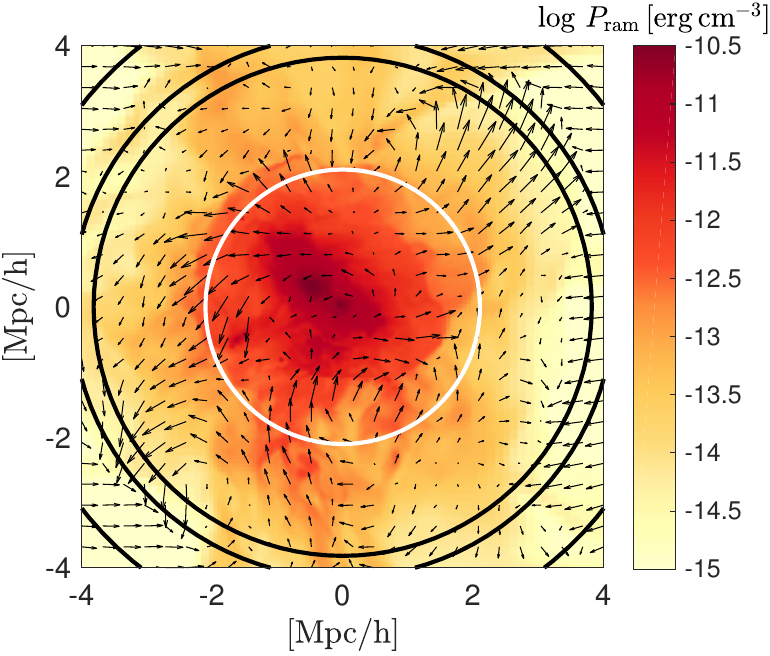}}
  \subfloat[Ram Pressure at \zeq{0.6}]{\label{fig:CL103_rp_a06}
    \includegraphics[width=\columnwidth,keepaspectratio,bb=0 0 5.1in 4.37in]{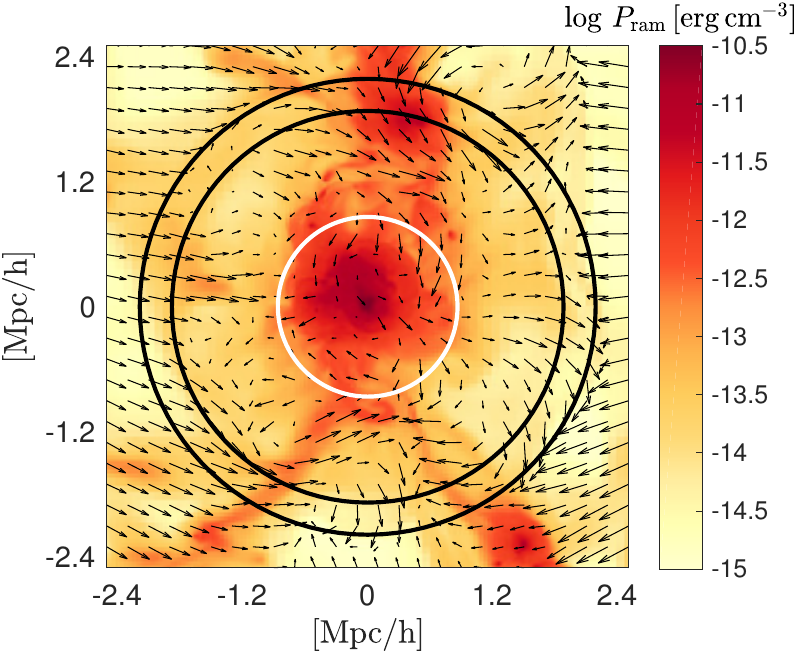}}
  \caption{The ram pressure experienced by a satellite within the
    cluster CL103 at \zeq{0} (left) and \zeq{0.6} (right). The
    satellite velocity is equal to the virial velocity of the cluster,
    $\Vv=1497\units{km\,sec^{-1}}$ at \zeq{0} and
    $\Vv=1009\units{km\,sec^{-1}}$ at \zeq{0.6}. The RPS effectiveness
    parameter is assumed to be $\alfP=0.5$.  The black arrows show the
    velocity field which highlights the inflow along the high density
    streams.}
  \label{fig:CL103_rp}
\end{figure*}

\subsubsection{Ram Pressure Stripping in Simulated Clusters}\label{sec:rpsSat_simClust}
We now make use of our simulated cluster systems to gauge the
stripping of the satellite gas halo in a more realistic setting. The
satellite is still modeled with an NFW profile, moving at the virial
velocity, but the gas density of the ICM is taken directly from the
simulations.

In \cref{fig:rps_sat_simClust} we show representative examples for a
$10^{11}\msun$ satellite halo in 3 simulated clusters (CL101,CL3,CL14)
at \zeq{0} and \zeq{0.6}. The ram pressure exerted by the cluster gas,
\cref{eq:rpsDef}, is calculated using the spherically averaged density
profile from the simulation. The lowest value of the shock edge, as
defined in \rfsec{edgeFind}, is also shown. For comparison, the
stripping due to a cluster with an NFW profile of the same virial mass
is also shown. The concentration parameter $\cvir$ is defined here by
the $\cvir$--$\Mv$ relation and \emph{not} calculated from the dark
matter distribution in the simulation.

The stripping in the simulated clusters is very similar to the results of the
simple NFW toy-model. This indicates that the NFW model is indeed a very good
approximation for the gas density profile of the cluster, at least for these
simulated clusters. The stripping is more effective for higher mass clusters,
since we assume the stripping is proportional to $\Vv^2$ and higher mass
clusters have higher virial velocities \cref{eq:satVel}. In high mass cluster
the satellite retains less than \perc{20} of its gas mass at $\Rc$, but even
for the lower-mass clusters, the satellite loses more than \perc{60} of its gas
mass. The stripping at the location of the shock edge is already strong enough
to strip over \perc{60} of the halo gas, at least for the high-mass clusters.

In \cref{fig:rps_sat_simCLust_rv12} we show the stripped mass of a
\mbox{$\Msat = 10^{11}\msun$} satellite calculated at a position within the
cluster of $2\Rc$ and $1\Rc$ for all the clusters in the simulation suite at
\zeq{0} and \zeq{0.6}. The mass dependence of the effectiveness of the
stripping is evident here. Even at $2\Rc$ the satellite halo in the high mass
halo loses over half of the gas and at $1\Rc$, only \perc{20} per cent of the
gas still remains. For the low-mass haloes, more than half the halo gas has
been removed at $\Rc$. At first glance it would seem surprising that the
stripping seems less effective at \zeq{0} than at \zeq{0.6} but one must keep
in mind that the comparisons in
\cref{fig:rps_sat_simClust,fig:rps_sat_simCLust_rv12} are performed at the
virial radius, which changes by a factor of 2 between the two epochs. The
density at the virial radius at \zeq{0.6} is higher by a factor of 2 than at
\zeq{0}, thus the stripping is greater, but in terms of absolute distance, at
\zeq{0} the stripping occurs at twice the distance. In addition, the
concentration of both clusters and satellites is lower at \zeq{0.6} which also
leads to more effective stripping.

In \cref{fig:CL103_rps} we see the effect of the gas stripping within
a representative cluster (CL103), as well as a representation of the
velocity field in the cluster. The left and right panels show the
cluster at \zeq{0} and \zeq{0.6} respectively. In
\cref{fig:CL103_rpsM_a1,fig:CL103_rpsM_a06} we show the amount of mass
stripped from a $\Msat = 10^{11}\msun$ halo at every point within the
halo. The satellite is assumed to be traveling at a speed equal to the
virial velocity of the cluster, $\Vv=1497\units{km\,sec^{-1}}$ at
\zeq{0} and $\Vv=1009\units{km\,sec^{-1}}$ at \zeq{0.6}.  As can be
seen, the halo is stripped to between $70$ to \perc{90} of its
original mass at the virial radius. In fact, simply entering the
shocked ICM leads to stripping of $30$ to \perc{50} of the mass.

It is important to note that in one respect the results shown in
\cref{fig:CL103_rps} may be misleading -- stripping along the gas
streams flowing into the clusters. This is especially pertinent since
most galaxies are expected to accrete onto the cluster via these
streams. The density in the streams is high compared to the
surrounding ICM, so the stripping appears enhanced in the
streams. However, the ram pressure is a result of the motion of the
satellite relative to the surrounding medium, \cref{eq:rpsDef}. Within
the ICM, which is characterized by random motions, one does not expect
the satellite motion to be correlated to its surrounding.  Clearly
this is not the case within the streams, which exhibit highly ordered
flows. This is especially apparent in
\cref{fig:CL103_rpsM_a06}.

If a satellite travels along a stream (an example of this can be seen
in the bottom-right quadrant of
\cref{fig:cl106_S_a1,fig:cl106_T_a1,fig:cl106_N_a1}), and its velocity
is similar to the surrounding stream velocity, the RPS would be
attenuated and the satellite could reach the virial radius or even the
cluster centre, while suffering only minor mass loss. Thus, along the
streams the RPS may be stronger or weaker depending on whether or not
the drop in ram pressure due to the satellite co-moving with the
streams is compensated by the increased density in the stream.

The results of our model are in agreement with those of
\citet{Bahe2013} who employ a cosmological SPH simulations in which
individual satellites can be followed and their gas content
monitored. Their results show that the hot gas haloes are removed by
the time the satellites reach the cluster virial radius. In these
simulations, the effect of stripping seems to be enhanced in the
filaments, with the increased density within the stream outweighing
the lower contribution of the relative velocity. 

To complete the picture, we show in \cref{fig:CL103_rp} the magnitude
of the ram pressure experienced by a satellite in the cluster CL103
moving at a speed equal to the virial velocity of the cluster at
\zeq{0} and \zeq{0.6}.

\section{Ram Pressure Stripping of the Galactic Disc}\label{sec:rpsDisc}
In the previous section we demonstrated that ram pressure can be a
very efficient mechanism for stripping the gas haloes of satellites
well outside the virial radius, thus removing the gas reservoir which
can cool on to the galaxy. We now address the possibility of removing
the gas held within a galactic disc by ram pressure stripping.

To do so, we wish to compare the force exerted by the ram pressure on
a gas element \emph{within} the galaxy disc with the gravitational
binding force exerted by the galaxy, as was formulated in
\cref{eq:rpsCond}. Once again we wish to construct a simple toy-model to
evaluate the effectiveness of RPS on the gas in the disc.

Here too we must stress that we intend to use the simple galaxy model
to obtain a limit on the RPS, and not to precisely reproduce the
stripping, a feat which is beyond such simple models. In this case, we
wish to show that RPS is \emph{not} effective in removing the ISM gas
from galaxies found in the outskirts of clusters beyond the virial
radius. To do so, we will employ several assumptions intended to
maximize the effect of the ram pressure and show that even then the
amount of gas removed from the galactic disc is negligible. Once again
we begin with fully analytic models, and afterwards cast our analytic
galaxies in into the more complex and detailed ICM afforded by our
simulation suite.

In our simple toy-model we make the following assumptions:
\begin{enumerate}[label=\emph{\alph*})]
\item The galaxy is modeled as a thin exponential disc\footnotemark of
  stars and gas and a spherical stellar bulge.  \footnotetext{In order
    to gauge the stability of the disc we extend the stellar disc
    model to account for the vertical structure of the disc. However, we do
    not consider the vertical structure when calculating the RPS.}
\item The centre of mass velocity of the galaxy relative to the ICM is
  perpendicular to the galactic disc plane, thus the effect of the
  ram pressure is maximal. Several numerical studies
  \citep{Quilis2000,Roediger2006,Jachym2009} found that except when
  the disc is nearly edge-on compared to its direction of motion, RPS
  is largely insensitive to the inclination angle with respect to the
  direction of motion.
\end{enumerate}

We now describe the galaxy model employed to calculate the
gravitational binding force per unit area within the disc plane.

\subsection{Disc Galaxy Model}\label{sec:discModel}
Our galaxy model is comprised of three galactic components: a stellar
disc and a gaseous disc, both modeled as thin exponential discs, and a
stellar bulge component, all of which are embedded within a dark
matter halo. In the following sections we present the main
characteristics of the model. 

\subsubsection{Stellar Disc}\label{sec:stellarDisc}
We model the stellar disc as an exponential disc, with surface density
and mass of
\begin{align}
\Sigma_{\mathrm{stars}}(R)&=\sigstar e^{-\frac{R}{\Rd}} \label{eq:starRho}\\
M_{\mathrm{stars}}(R)&=\Ms\left[1-e^{-\frac{R}{\Rd}} \left(1+\frac{R}{\Rd} \right)\right], \label{eq:starMass}
\end{align}
where $R$ denotes the distance from the disc centre in the plane of
the disc, $\Ms$ is the stellar mass in the disc and $\Rd$ is the
exponential scale radius. The surface density parameter $\sigstar$ is
related to the other parameters via the relation
\mbox{$\sigstar=\Ms/(2\mathrm{\pi}\Rd^2)$}.

The gravitational acceleration in the plane of a thin, exponential
disc is \citep[][Ch.\@ 2.6]{Binney2008}
\begin{equation}\label{eq:gdisc1}
\begin{split}
g(\tR)&=-\frac{\diff \Phi}{\diff R}\\
&=-\mathrm{\pi} G \sigstar \tR\times \\
& \left[I_0\left(\frac{\tR}{2}\right) K_0\left(\frac{\tR}{2}  \right)-                   
 I_1\left(\frac{\tR}{2}\right) K_1\left(\frac{\tR}{2}  \right)\right],
\end{split}
\end{equation}
where $I_\nu$ and $K_\nu$ are the modified Bessel functions of the
first and second kind and we have defined $ \tR=R/\Rd$. In the
interest of convenience we introduce the following function
\begin{equation}\label{eq:bfuncDef}
\bfunc{\nu}{x}=I_0\left(\nu\frac{x}{2}\right) K_0\left(\nu\frac{x}{2}\right)-I_1\left(\nu\frac{x}{2}\right) K_1\left(\nu\frac{x}{2}\right),
\end{equation}
such that \cref{eq:gdisc1} is now
\begin{equation}\label{eq:gdisc2}
g(\tR)=-\mathrm{\pi} G \sigstar \tR \bfunc{1}{\tR}.
\end{equation}

\subsubsection{Gaseous Disc}\label{sec:gasDisc} 
The gaseous disc is also modeled as an exponential disc
\citep{Bigiel2012} with a mass of $\Mg$ and a scale radius $\Rg$. We
present these parameters in units of the stellar disc parameters as
$\fgs=\Mg/\Ms$ and $\bet=\Rd/\Rg$. By this definition, values of
$\bet>1$ correspond to a gaseous component which is more
\emph{compact} than the stellar component and values of $\bet<1$
result in a gaseous disc which is more \emph{extended} than the stars.

The surface density of the gas disc is therefore (see
\equnp{starRho})
\begin{equation}\label{eq:gasRho}
\Sigma_{\mathrm{gas}}(\tR)=\Sigma_g e^{-\frac{R}{\Rg}}=\sigstar\fgs\bet^2e^{-\bet \tR},
\end{equation}
from which the mass profile is found to be
\begin{equation}\label{eq:gasMass}
\Mg(<\tR)=\Ms\fgs\left(1-e^{-\bet\tR}\left(1+\bet\tR \right)\right).
\end{equation}
The gas disc contribution to the gravitational acceleration is
\begin{equation}\label{eq:gGas}
g_{\rm gas}(\tR)=-\mathrm{\pi} G \sigstar \fgs \bet^3 \tR \bfunc{\bet}{\tR}.
\end{equation} 

\subsubsection{Stellar Bulge}\label{sec:bulge}
To model the stellar bulge component we use the Hernquist profile
\citep{Hernquist1990} for which the density profile is
\begin{equation}\label{eq:hern}
\rho(r)=\frac{\Mb}{2\mathrm{\pi}}\frac{\Rb}{r}\frac{1}{\left(\Rb+r\right)^3},
\end{equation}
where the stellar mass of the bulge is $\Mb$ and the scale radius is
$\Rb$. As before, we recast the bulge model parameters in terms of the
stellar disc parameters, $\fbs=\Mb/\Ms$ and $\xii=\Rd/\Rb$, noting
$\fbs$ is essentially the `Bulge-to-Disc' stellar mass ratio. In many
cases, observed galaxies are classified according to the
`Bulge-to-Total' stellar mass ratio. The relationship between the two
is
\begin{equation}\label{eq:BT}
\frac{B}{T}\equiv\frac{\Mb}{\Ms+\Mb}=\frac{\fbs}{\fbs+1}.
\end{equation}
The gravitational acceleration exerted by the bulge (in the disc
plane) is thus
\begin{equation}\label{eq:gBulge}
g_{\rm bulge}(\tR)=-\frac{G\Mb}{\left(\Rb+r\right)^2}=-\frac{2\mathrm{\pi} G \sigstar\fbs\xii^2}{\left(1+\xii\tR\right)^2}.
\end{equation}

\subsubsection{Dark Matter Halo}\label{sec:halo}
For a spherically symmetric dark matter halo, for which the mass
profile is $M_{\rm H}(r)$ the contribution to the gravitational acceleration
will be
\begin{equation}\label{eq:haloAccel}
g_{\rm H}(r)=\frac{G M_{\rm H}(r) }{r^2} \to g_{\rm H}(\tR)=2\mathrm{\pi} G \sigstar \frac{M_{\rm H}(\tR)}{\Ms} \tR^{-2}.
\end{equation}
We assume an NFW model for the dark matter halo as described in detail
in \rfsec{nfwModel}.

\begin{figure}
\includegraphics[width=8cm,keepaspectratio,bb=0 0 5.24in 4.19in ]{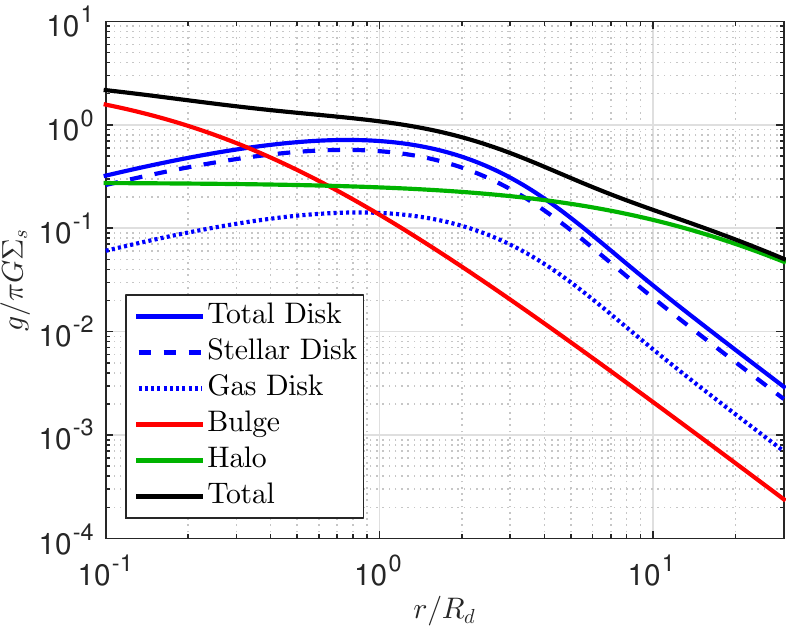}
\caption{Decomposition of the gravitational acceleration into the
  contributions from the various components in a
  $\Ms=10^{10}\msun$ galaxy. The total disc contribution
  ({solid blue}) is divided into the stellar disc contribution
  ({dashed blue}) and the gaseous disc ($\fgs\simeq
  0.2,\bet\simeq 0.9$) contribution ({dotted blue}). The bulge
  component ($\fbs\simeq 0.11,\, \xii=3.6$) can be seen to dominate
  the acceleration in the centre of the galaxy ({red}). The
  galaxy is embedded in an $\Mv=5.7\times 10^{11} \msun, \,
  c_\mathrm{vir}\simeq 9.5$ NFW dark matter halo ({green}). The
  total profile ({black}) is also shown. The halo becomes the
  dominant component at $r \gtrsim 5\Rd$, where it becomes an
  important contribution to the disc stability.}
\label{fig:gDecompose}
\end{figure} 

\subsubsection{Full Model}\label{sec:fullModel}
For the fully assembled galaxy, the total gravitational
acceleration is the sum of the contributions of all three
components
\begin{equation}\label{eq:gFull}
\begin{split}
g_{\rm gal}(\tR)=&-\mathrm{\pi} G \sigstar \tR \times \\
&\left[  \bfunc{1}{\tR} + \fgs\bet^3\bfunc{\bet}{\tR} +\frac{2\fbs\xii^2}{\left(1+\xii \tR\right)^2}\right].
\end{split}
\end{equation}
The full acceleration is obtained by adding the contribution of the
halo (\equnp{haloAccel}) to the above expression.  As one would
expect, at very large distances the acceleration drops as
\mbox{$g_{\rm gal} \propto r^{-2}$}.

In \cref{fig:gDecompose} we show a decomposition of the contributions
of the various components of the galaxy and halo to the gravitational
acceleration of a representative $\Ms=10^{10}\msun$
galaxy. The gaseous component is defined by a gas fraction of
$\fgs\simeq 0.3$ and with $\bet\simeq0.9$. The bulge-to-disc mass
ratio is $\fbs\simeq 0.2$ with $\xii=3.6$, and thus can be seen to
dominate the acceleration in the centre of the galaxy. In addition,
the galaxy is embedded in a dark matter NFW model halo with
$\Mv=4.8\times 10^{11} \msun$ and $
c_\mathrm{vir}\simeq10$. The halo contribution is comparable to that
of the disc in the inner parts of the galaxy $r\lesssim \Rd$ but
becomes the dominant component at $r\gtrsim 5\Rd$. This has important
consequences on the stability of the disc.

The force per unit area affecting the gas is
\mbox{$F=(g_{\rm gal}+g_{\rm H})\Sigma_{\rm gas}$}. The total force per
$\unitstx{kpc^{−2}}$ is therefore
\begin{equation}\label{eq:fullForce}
\begin{split}
&F(\tR)=-\mathrm{\pi} G \sigstar^2 \fgs \bet^2  e^{-\bet \tR}  \tR \times \\
&\left[ \bfunc{1}{\tR}+\fgs\bet^3\bfunc{\bet}{\tR}+\frac{2\fbs\xii^2}{\tR\left(1+\xii\tR\right)^2}+\frac{2 M_{\rm H}(\tR)}{\Ms\tR^3}\right],
\end{split}
\end{equation}
which we separate into a unit-less function $\ftilde(\tR)$ describing
the spatial dependence of the force and a constant factor comprised of
the relevant galaxy parameters
\begin{equation}\label{eq:ftildeDef}
\ftilde(\tR)=\frac{F(\tR)}{\mathrm{\pi} G \sigstar^2 \fgs} .
\end{equation}

As noted earlier, the true complexities of the RPS process are beyond
the scope of a simple analytic model. In reality, the gas will be
pushed out and the disc will be warped by the ram pressure
\citep[e.g.\@][]{Vollmer2008,Vollmer2008a}.

\begin{figure}
\centering
\includegraphics[width=8cm,keepaspectratio,bb=0 0 5.38in 4.32in ]{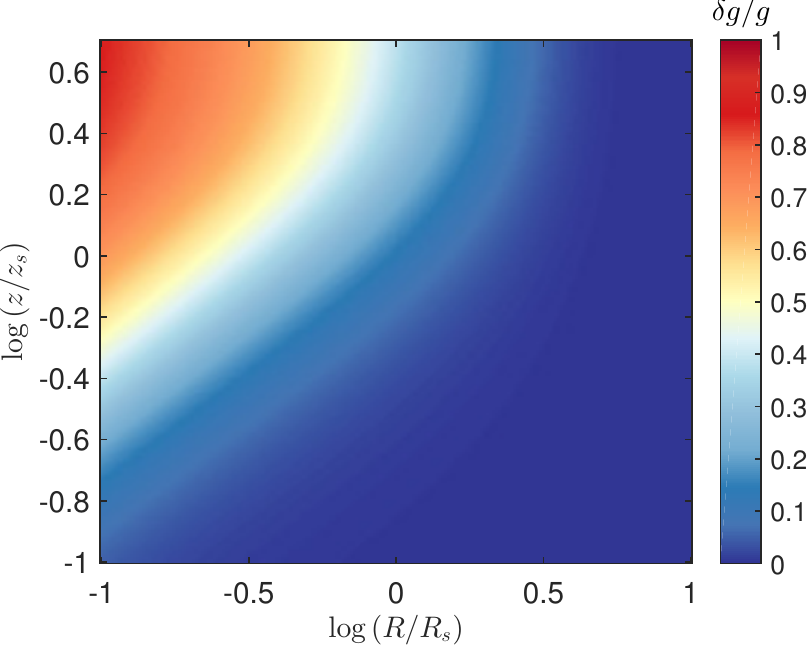}
\caption{The fractional error incurred by neglecting the $z$ component
  of the gravitational acceleration as a function of \mbox{$\tR=R/\Rd$} and
  $z/\zd$. We find that beyond $\Rd$ we may approximate the
  gravitational binding force using the component within the disc
  plane, neglecting the $z$ component which accounts for less than
  \perc{10}.}
\label{fig:gFracError}
\end{figure}

We make a simplifying assumption that the total gravitational binding
force is represented by the force acting on a gas element in the plane
of the disc. 
In order to estimate the error implicit in this assumption we consider
a thick disc model to find the acceleration in the $z$ direction.
In a more realistic thick disc, the gravitational binding
force on a gas element $\diff m$ at a position $(R,z)$, where $R$ is
the distance from the centre in the plane of the disc and $z$ is the
perpendicular distance from the disc plane will be
\begin{equation}\label{eq:gravF}
  F_g=\sqrt{g_{\rm R}^2+g_{\rm z}^2}\diff m,
\end{equation}
where $g_{\rm R}$ and $g_{\rm z}$ are the gravitational acceleration
components parallel and perpendicular the disc.

In order to estimate the error of neglecting the acceleration in the
$z$ direction, we employ the Isothermal Sheet model which assumes that
the vertical distribution of the stellar disc is locally isothermal
such that the distribution function of stars in the $z$ direction at
every $R$ is Maxwellian \citep{Spitzer1942,Mo2010}. The resulting
gravitational acceleration perpendicular to the disc in this model is
given by
\begin{equation}\label{eq:gsDisc}
g_{\rm z}(R,z)=\frac{ \Ms}{\Rd^2}e^{-\frac{R}{\Rd}}\tanh\left(\frac{z}{2\zd}\right),
\end{equation} 
where $\zd$ is the scale height of the disc. 

In \cref{fig:gFracError} we examine the fractional error in the
gravitational acceleration incurred by neglecting the $z$-component
defined as
\begin{equation}\label{eq:gFracError}
\frac{\delta g}{g}=1-\frac{g_{\rm R}}{\sqrt{g_{\rm R}^2+g_{\rm z}^2}}.
\end{equation} 
As can be seen, in the regime relevant for stripping $R\gtrsim
\Rd,\,z<\zd $, ignoring the vertical component of the acceleration
leads to negligible errors in estimating the gravitational binding,
and the total acceleration can be approximated by the acceleration in
the disc plane.

\subsection{Constructing Mock Galaxy Catalogs}\label{sec:mockCatalog}
The disc model presented above is fully described by 2 primary
parameters $\Ms,\Rd$ of units $\unitstx{\msun}$ and $\unitstx{kpc}$
respectively, while the rest of the parameters, namely
$\fgs,\bet,\fbs$ and $\xii$, are unit-less ratios with respect to
these parameters.

All together, this results in a rather large parameter space which we
need to explore to find the effectiveness of RPS in removing gas from
the galactic disc. To contend with this challenge, we generate a large
catalog of $\gtrsim 10^4$ mock galaxies, in which the different
parameters have been chosen randomly, thus allowing us to investigate
the effects of RPS on galaxies of many different attributes.

In the interest of achieving a physically motivated choice for these
parameters, we employ the model presented in \citet{Mo1998} to relate
the scale radius $\Rd$ to the stellar mass of the disc and the
properties of the host halo. The model assumes that the gas which
formed the stellar disc conserves its specific angular momentum during
the formation process, and more importantly, that the specific angular
momentum was equal to that of the dark matter of the host halo in
which the galaxy formed. \citet{Dekel2013} and \citet{Danovich2015}
find that the spin of a galaxy and its host dark matter halo are
indeed very similar (up to a factor of 2), despite the very different
history of angular momentum buildup by gas and dark
matter. Observations of the relation between $\Rd$ of the galaxy and
$\Rv$ of the halo are in general agreement with this model
\citep{Kravtsov2013}. While the original model assumed that all the
gas ended up as stars in a stellar disc and bulge, we have extended it
to allow some of the gas to form a gaseous disc component.

We relate the properties of the galaxy with the dark matter halo by
expressing the angular momentum of the disc $J_\mathrm{d}$ as a
fraction $\jd$ of the total angular momentum of the dark
matter halo $J$, such that \mbox{$ J_\mathrm{d}= \jd J$}. The
angular momentum of a dark matter halo can be expressed by the
unit-less spin parameter as defined by \citealt{Bullock2001}
\begin{equation}\label{eq:lambdaDef}
\lambda^\prime=\frac{J}{\sqrt{2}\Mv\Vv\Rv}.
\end{equation}

The angular momentum of the disc can be calculated by\footnotemark 
\begin{equation}\label{eq:discAMDef}
\begin{split}
J_\mathrm{d}&=\int_0^{\Rv} \Sigma_{\mathrm{T}}(R) R V_c(R) 2\mathrm{\pi} R \diff R\\
&=2 \mathrm{\pi} \sigstar \Rd^3 \int_0^{\frac{\Rv}{\Rd}} x^2 \left( e^{-x}+\fgs \bet^2 e^{-\bet x}\right) \Vc(x\Rd)  \diff x \\
&= \Ms \Rd \Vv \tau^{-1},
\end{split}
\end{equation} 
where we have used the total disc surface density
\mbox{$\Sigma_{\mathrm{T}}=\Sigma_{\mathrm{stars}}+\Sigma_{\mathrm{gas}}$}
(\equnpTwo{starRho}{gasRho}), and the circular velocity of the disc is given by $\Vc$. We define 
\begin{equation}\label{eq:tauDef}
\tau^{-1}= \int_0^{\infty} x^2 \left(e^{-x}+\fgs\bet^2 e^{-\bet x}\right) \frac{V_c(x\Rd)}{\Vv} \diff x, 
\end{equation}
setting the upper limit for the integral to $\infty$ since the
integrand drops exponentially, and the ratio $\Rv/\Rd$ is usually of
order $\sim 10\textrm{--}100$. \footnotetext{We
  assume the stellar bulge has negligible angular momentum.}

The circular velocity needed to calculate \cref{eq:tauDef} is
determined by the \emph{entire} system, i.e.\@ stars, gas bulge and
dark matter. In calculating it one must take into account that the
sinking of baryons to the centre of the halo to form the galaxy leads
to a contraction of the dark matter in the centre of the halo. We
assume that the halo response to the assembly of the disc is adiabatic
\citep{Blumenthal1986,Flores1993,Dalcanton1997}, and therefore the
angular momentum of individual dark matter particles is conserved.

Combining \cref{eq:lambdaDef,eq:discAMDef} the scale radius of the disc can be expressed as
\begin{equation}\label{eq:rScale}
\Rd=\sqrt{2} \lambda^\prime \left(\frac{\jd}{\md}\right)\left(1+\fgs\right)\tau  \Rv,
\end{equation} 
where $\md$ is the fraction of total mass initially in the halo
found in the galactic disc
\begin{equation}\label{eq:mdDef}
\md =\frac{\Ms +\Mg}{\Mv}=\frac{\Ms}{\Mv}\left(1+\fgs\right),
\end{equation}
and \mbox{$\tau=\tau(\Ms,\Rd,\fgs,\bet,\fbs,\xii,m_d,\cvir)$} is a
function of the model parameters. The implicit equation
\cref{eq:rScale} can be solved iteratively once the values of the
disc, bulge and halo parameters have been determined.

To implement this procedure, we must select the properties of the host
dark matter halo for each galaxy ($\Mv$,$\cvir$ and $\lambda^\prime$),
the values of the parameters relating the halo to the galaxy ($\jd$
and $\md$) and of course the galaxy properties
($\Ms,\,\fgs,\,\bet,\,\fbs$ and $\xii$). The values of these
parameters are randomly chosen for each galaxy based on physical or
observational considerations as well as wisdom gleaned from
simulations.

\subsubsection{Setting the Galaxy Parameters}
To achieve a comprehensive collection of model galaxies, we generate
mock catalog by sampling the parameter space of the disc model in the
following fashion:
\begin{enumerate}[label=\emph{\alph*})]
\item $\Ms$ is randomly drawn from a uniform logarithmic distribution
  in the range $9 < \log(\Ms) < 11.5$, allowing us to gauge the
  effectiveness of RPS as a function of mass.
\item We assume a gas content which can comprise up to \perc{50} of
  the total disc mass. Accordingly, $\fgs$ is randomly selected from a
  uniform distribution in the range $[0.05, 1]$.
\item We assume a gas distribution which is either identical or
  extended up to $2.5$ times as the stellar distribution.  This
  entails selecting $\bet$ from a uniform distribution in the range
  $[0.4,1]$.
\item Bulge-to-Total ratios are in the range $[0,0.5]$, thus $\fbs$ is
  selected from a uniform distribution in the range $[0,1]$.
\item We allow the bulge scale radius to lie between \perc{25} of
  $\Rd$ to $2\Rd$, which is achieved by drawing $\xii$ from a uniform
  distribution in the range $[0.5,4]$.
\end{enumerate}

\subsubsection{Setting Halo Parameters}
The host haloes are initially modeled using the NFW profile
\citep{Navarro1996}. The properties of the host haloes of the galaxies
are selected at random according to the following prescription
\begin{enumerate}[label=\emph{\alph*})]
\item The virial mass of the host halo is set as a factor of the total
  stellar mass of the galaxy. For a given stellar mass ($\Ms+\Mb$), we
  randomly select the ratio of stellar to dark matter mass based on
  the relation (and scatter) found in \citet{Moster2010a}. 
\item Analysis of \nbody~simulations yields a power-law relation
  between $\Mv$ and $\cvir$ as a function of redshift
  \citep{Bullock2001a,Wechsler2002,Maccio2008}. The concentration
  parameter $\cvir$ is determined according to the relation and
  scatter between $\Mv$ and $\cvir$ found in
  \citet{Munoz-Cuartas2011}, with the scatter added randomly to the
  mean relation.
\item Several studies and observations
  \citep{Syer1999,Bullock2001,Maccio2008} have found that the spin
  parameter $\lambda^\prime$ follows a log-normal distribution 
\begin{equation}\label{eq:lambdaDistribution}
P(\lambda^\prime)=\frac{1}{\lambda^\prime\sqrt{2\mathrm{\pi}}\sigma}\exp\left(-\frac{\ln^2(\lambda^\prime/\lambda^\prime_0)}{2\sigma^2}
\right), 
\end{equation}
which is independent of halo mass and appears to have a very weak
dependence on redshift. We draw the value of the spin parameter of the
host halo based on the above distribution with
$\lambda^\prime_0=0.031,\,\sigma=0.57$ based on
\citet{Munoz-Cuartas2011}.  We limit our sample to haloes with
$\lambda > 0.01$ due to numerical convergence considerations.  Systems
that do not meet this criterion account for $\sim 3.5$ per cent of the
population.
\end{enumerate}

\subsubsection{Selecting $\md$ and $\jd$}
The relationship between $\jd$, the fraction of total angular momentum
found in the disc, and $\md$, the fraction of mass in the disc, can be
set by physical considerations. It is often assumed that the baryons,
who initially share the density distribution and angular momentum of
the halo, conserve their specific angular momentum as the disc is
formed. This leads to the following relation
\begin{equation}\label{eq:jdmd}
\frac{J}{\Mv}=\frac{J_d}{\Ms+\Mg}  \Rightarrow \md=\jd  .
\end{equation}  
\cite{Mo1998} find that the above relation succeeds in fitting the
sizes of \zeq{0} discs and we adopt this assumption unless noted
otherwise\footnotemark. The above relations assume that no angular
momentum was transferred to the halo. If the baryons do lose angular
momentum to the halo then \cref{eq:jdmd} should be considered as an
upper limit.\footnotetext{In the above, we have assumed the bulge has
  no angular momentum, and whatever angular momentum lost during the
  bulge formation was transferred entirely to disc. Relaxing this
  assumption, namely that the entire baryonic component conserves its
  specific angular momentum leads to $\jd=\md+m_{\mathrm{b}}$, with
  $m_{\mathrm{b}}=\Mb/\Mv$.}

The value of $\md$ is set by \cref{eq:mdDef} once the values of $\Ms$,
$\fgs$ and $\Mv$ have been set.

\subsubsection{Disc Stability}\label{sec:discStability}
Since the values of the parameters are selected randomly and
independently of each other, the resulting galaxies generated for our
catalog may not be dynamically stable. In order to weed out such
galaxies we employ the Toomre stability analysis
\citep{Toomre1964,Dekel2009a}, in which a thin stellar disc is deemed
unstable when the local gravity overcomes the combined stabilizing
effects of rotation and pressure due to turbulent or thermal
motions. This criterion is expressed by the Toomre $Q$ parameter
having a lower value than a stability threshold $Q_c=1$
\begin{equation}\label{eq:ToomreDef}
Q=\frac{\sigma_{\rm R} \kappa}{3.36 G \Sigma(R)}<1,
\end{equation}
where $\sigma_{\rm R}$ is the radial velocity dispersion of the stars,
$\kappa$ is the epicyclic frequency and $\Sigma(R)$ is the surface
density of the stellar disc.

The parameter $\sigma_{\rm R}$ is not defined in our model since we do not
account for the orbital properties of the stars but treat the disc as
a single entity. To solve this shortcoming, we consider the case of a
thick galactic disc. For a stable, relaxed system one may assume that
the velocity dispersion components are proportional to each
other. Thus $\sigma_{\rm R}\propto \sigma_{\rm z}$ where $z$ is the direction
perpendicular to the disc plane. By expanding our disc model to
account for the disc thickness we can estimate the value of $\sigma_{\rm z}$
and by proxy obtain a value for $\sigma_{\rm R}$. We once again turn to the
Isothermal sheet model, in which the velocity dispersion in the $z$
direction is related to the velocity dispersion in the $z$ direction
by
\begin{equation}\label{eq:sigZ}
\sigma_{\rm z}^2(R)=2 \mathrm{\pi} G \Sigma(R) \zd.
\end{equation}
Observations show that that the scale height $\zd$ is generally
independent of $R$ and, to a good approximation, a constant fraction
of the scale radius $\Rd$
\citep{vanderKruit1981,vanderKruit1981a,vanderKruit1982}, such that
$\zd=\zeta\Rd$. This allows us to randomly select appropriate scale
heights for the galaxies based on observed distributions
\citep{Bizyaev2002}, and thus determine $\sigma_{\rm z}$.

By further assuming that $\sigma_{\rm R}\propto \sigma_{\rm z}$
\citep{Kregel2005}, the $Q$ parameter can be calculated. We discard
galaxies for which the value of $Q$ for any point on the disc drops
below $1$. Using this criteria we find that $\sim 1$ per cent of the discs
generated are unstable.  Based on the assumptions detailed above, we
have generated a mock catalog of order $1.9\times 10^4$ stable
galaxies, which will serve for testing the ram pressure stripping of
galactic gas.

\begin{figure}
\centering 
  \includegraphics[width=8cm,keepaspectratio,bb=0 0 5.74in 4.51in ]{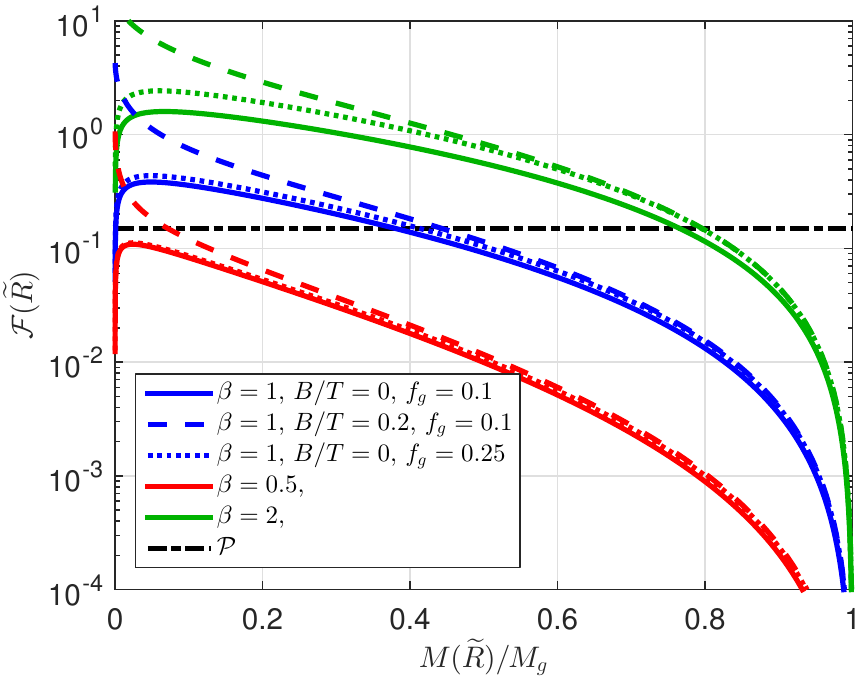}
  \caption{The binding gravitational force in a $\Ms = 10^{10}
    \msun$ disc versus the enclosed gas mass for various disc
    models. Solid lines denote bulge-less disc galaxies of gas
    fraction $\fgs = 0.1$ for $\bet=1$ ({blue}), $\bet= 0.5$
    ({red}) and $\bet= 2$ ({green}). Dashed lines denote
    galaxies with $B/T = 0.2$, $\xii= 3$ and $\fgs=0.1$ and dotted
    lines denote bulge-less galaxies of $\fgs = 0.25$. The galaxy is
    embedded in $M=4.8\times 10^{11} \msun$, $\cvir = 10$ NFW
    dark matter halo. The black horizontal line demarks a value of
    $\ptilde$, the L.H.S of the stripping condition
    \cref{eq:rpsDisc2}, which is typical for a satellite at $0.5\Rc$
    of a $10^{15}\msun$ cluster. The intersection point of
    that line with the $\ftilde$ curves denoting the amount of gas
    being stripped in each case. As a galaxy moves closer to the
    cluster centre, the $\ptilde$ line will rise in value. It is clear
    that varying the gas fraction has little effect on the binding
    force, and the bulge has a significant effect only in the central
    areas.}
  \label{fig:ftilde}
\end{figure}

\begin{figure*}
  \subfloat[$r=2\Rc$]{\label{fig:rpsCat_2rv}
    \includegraphics[width=\columnwidth,keepaspectratio,bb=0 0 5.53in 4.36in]{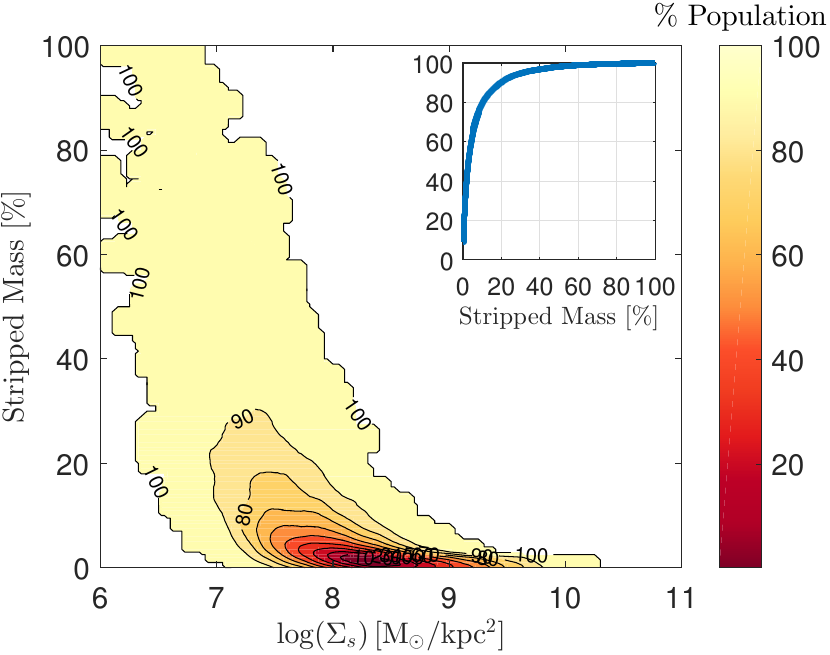}}
  \subfloat[$r=\Rc$]{\label{fig:rpsCat_1rv}
    \includegraphics[width=\columnwidth,keepaspectratio,bb=0 0 5.53in 4.36in]{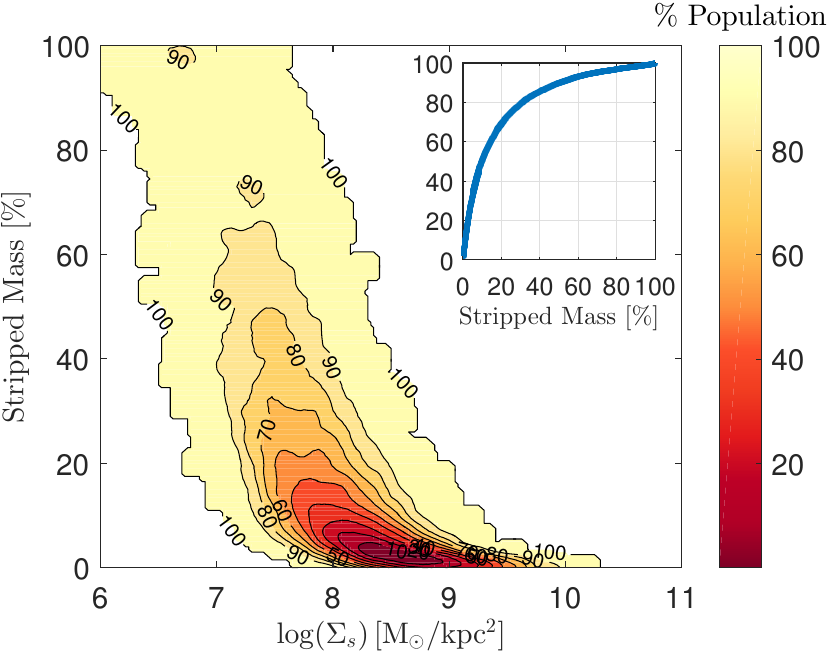}}\\
  \subfloat[$r=0.5\Rc$]{\label{fig:rpsCat_05rv}
    \includegraphics[width=\columnwidth,keepaspectratio,bb=0 0 5.53in 4.36in]{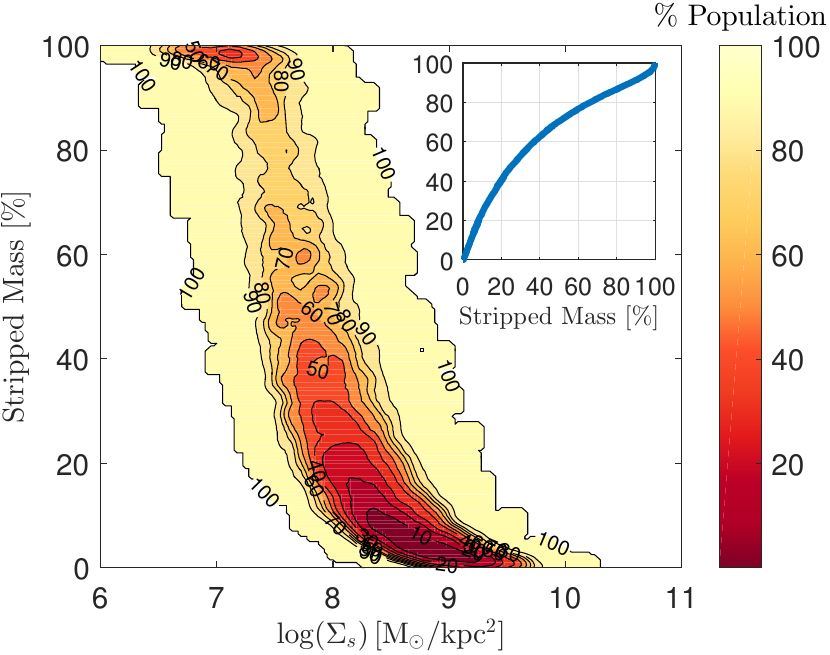}}
  \subfloat[$r=0.2\Rc$]{\label{fig:rpsCat_02rv}
    \includegraphics[width=\columnwidth,keepaspectratio,bb=0 0 5.53in 4.36in]{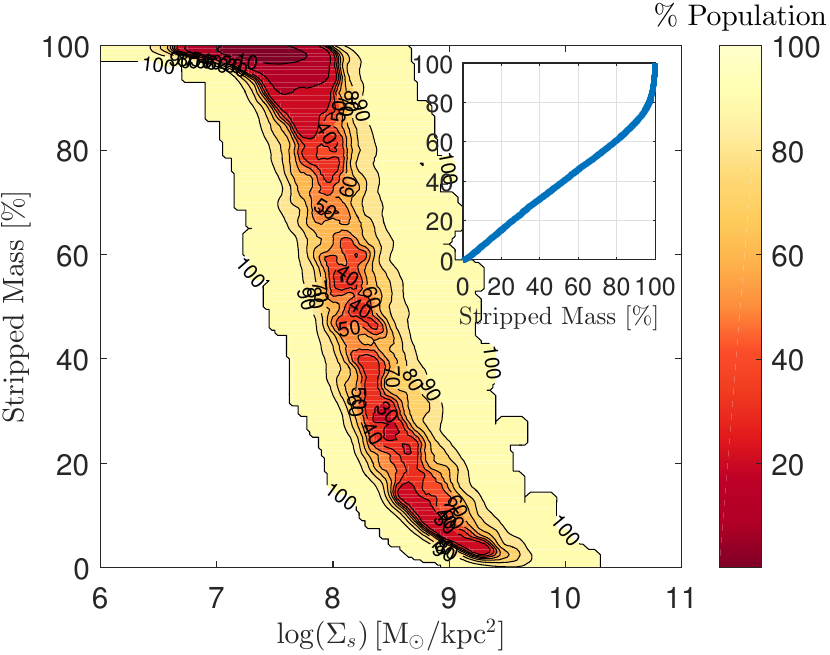}}
  \caption{ The distribution of satellite galaxies in the mock catalog
    on a plane relating the parameter $\sigstar$ and the percentage of
    stripped mass at a given position in a $\Mc= 10^{15}\msun$ NFW
    cluster and a gas fraction of $f_c = 0.15$. Coloured contours
    correspond to the percentage of the galaxy population enclosed
    within the contours. The four panels correspond to the stripping
    at different positions in the cluster: $2\Rc$
    \subrfig{rpsCat_2rv}, $\Rc$ \subrfig{rpsCat_1rv}, $0.5\Rc$
    \subrfig{rpsCat_05rv} and $0.2\Rc$ \subrfig{rpsCat_02rv}. The
    inlay shows the cumulative distribution of the stripped mass in
    the catalog galaxies. At $1\Rc$ we find that \perc{50} of the
    population has lost less than \perc{20} of the mass and less than
    \perc{10} of the galaxies have lost more than \perc{60} of the
    gas. The stripping can be seen to be ineffective at $1\Rc$ and
    beyond, but within the virial radius a large part of the
    population has lost a significant amount of gas. The percentage of
    galaxies who lost more than half their gas is $\lesssim5$ per cent
    at $2\Rc$, \perc{15} at $1\Rc$, \perc{30} at $0.5\Rc$ and
    \perc{60} at $0.2\Rc$.}
  \label{fig:rpsCat}
\end{figure*}

\begin{figure*}
  \subfloat[CL101, \zeq{0}]{\label{fig:rpsDiscSimClust_cl101_a1}
    \includegraphics[height=2.3in,keepaspectratio,bb=0 0 8.5in 11.0in,trim=1.375in 3.25in 1.75in  3.6in, clip]{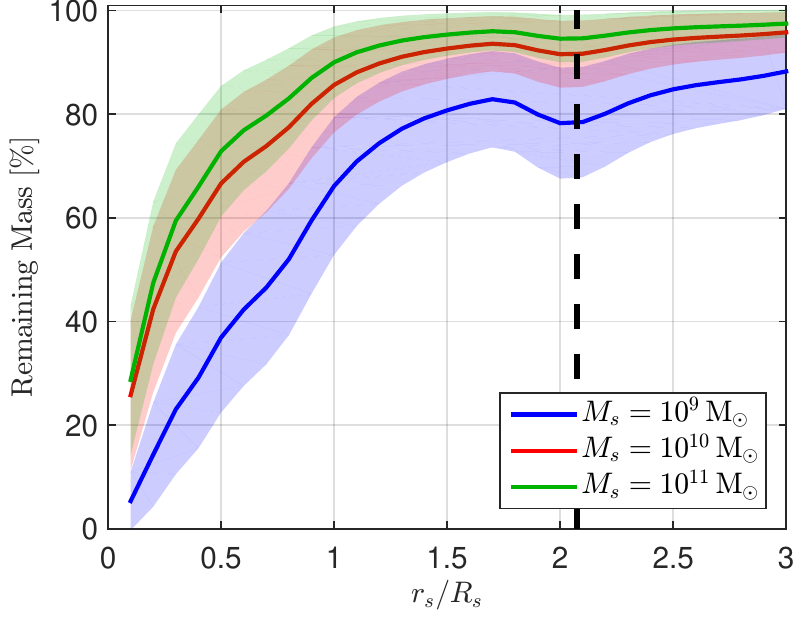}}
  \subfloat[CL101, \zeq{0.6}]{\label{fig:rpsDiscSimClust_cl101_a06}
    \includegraphics[height=2.3in,keepaspectratio,bb=0 0 8.5in 11.0in,trim=1.375in 3.25in 1.75in  3.6in, clip]{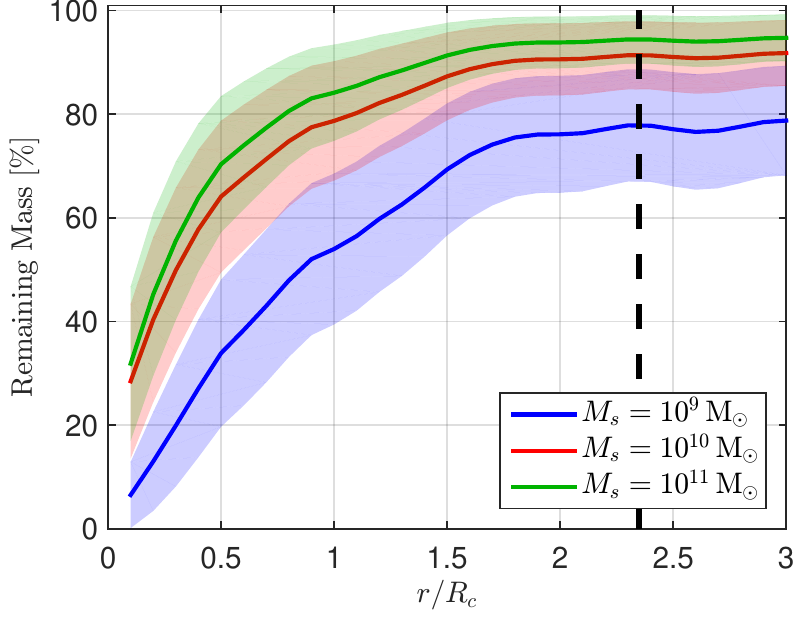}}\\
 \subfloat[CL3, \zeq{0}]{\label{fig:rpsDiscSimClust_cl3_a1}
    \includegraphics[height=2.3in,keepaspectratio,bb=0 0 8.5in 11.0in,trim=1.375in 3.25in 1.75in  3.6in, clip]{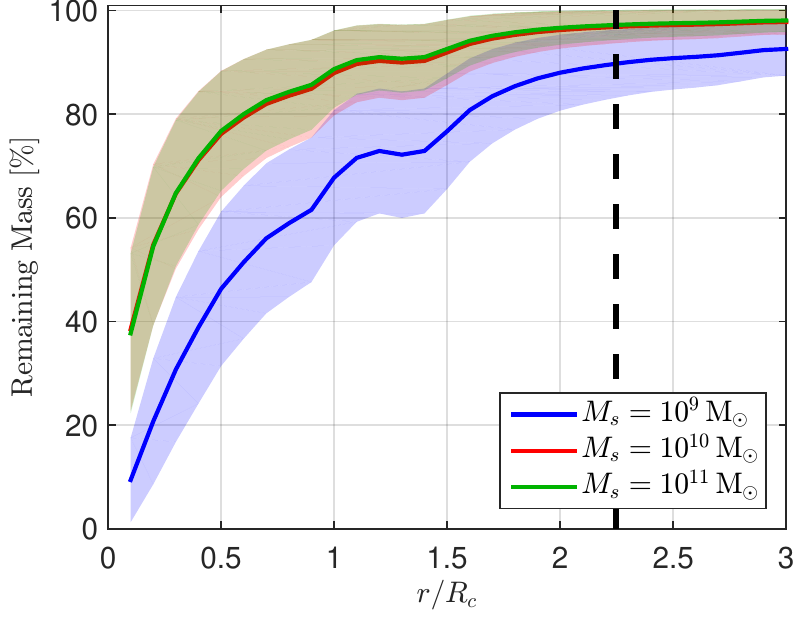}}
 \subfloat[CL3, \zeq{0.6}]{\label{fig:rpsDiscSimClust_cl3_a06}
    \includegraphics[height=2.3in,keepaspectratio,bb=0 0 8.5in 11.0in,trim=1.375in 3.25in 1.75in  3.6in, clip]{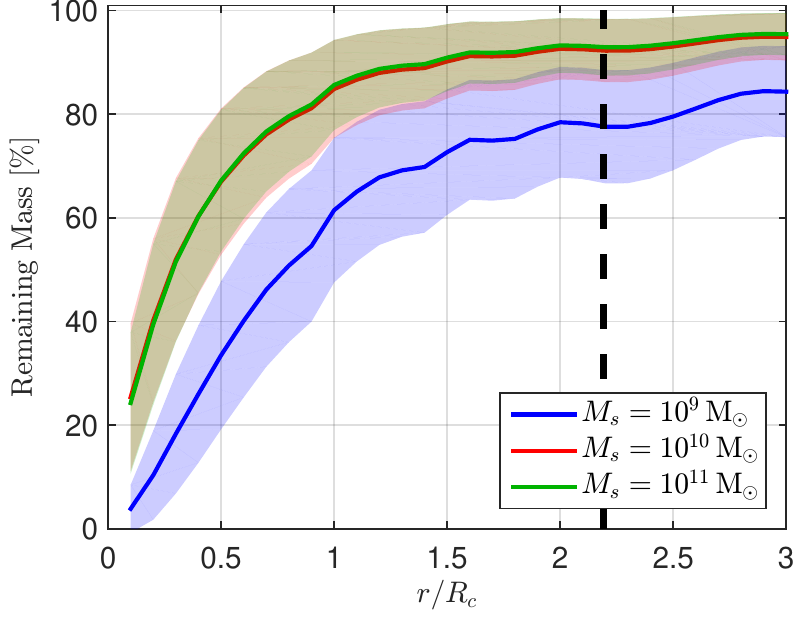}}\\
 \subfloat[CL14, \zeq{0}]{\label{fig:rpsDiscSimClust_cl14_a1}
    \includegraphics[height=2.3in,keepaspectratio,bb=0 0 8.5in 11.0in,trim=1.375in 3.25in 1.75in  3.6in, clip]{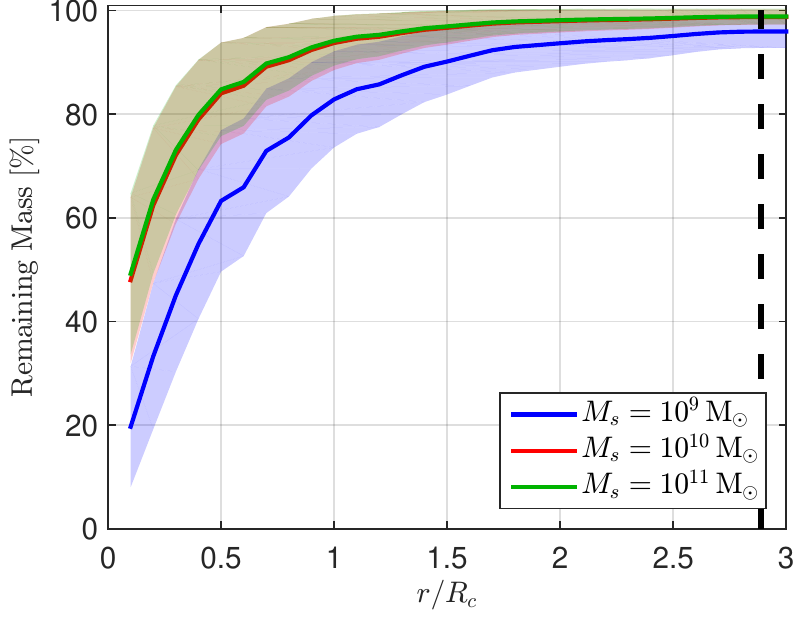}}
  \subfloat[CL14, \zeq{0.6}]{\label{fig:rpsDiscSimClust_cl14_a06}
    \includegraphics[height=2.3in,keepaspectratio,bb=0 0 8.5in 11.0in,trim=1.375in 3.25in 1.75in  3.6in, clip]{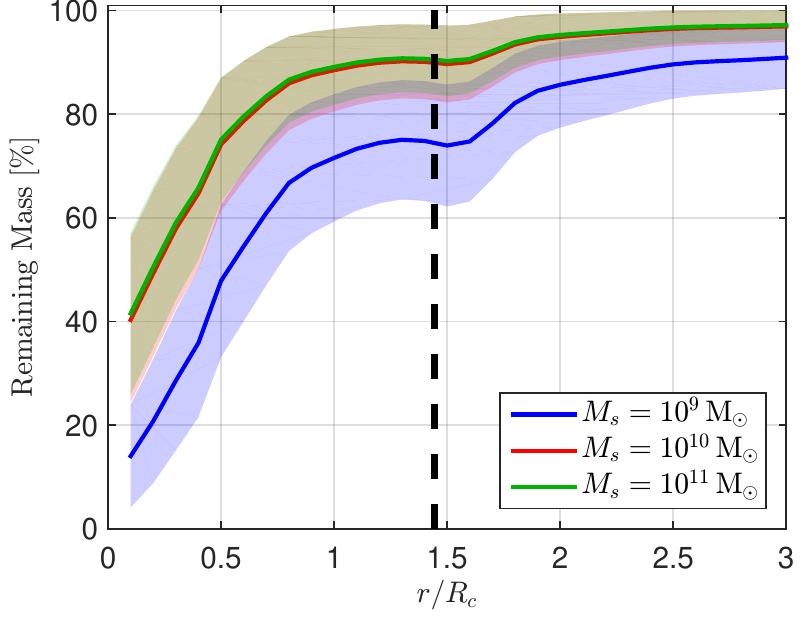}}
  \caption{RPS of galaxies in the simulated clusters. The ram pressure
    is calculated from the gas density profiles of 3 simulated
    clusters: CL101 \subrfig{rpsDiscSimClust_cl101_a1} \&
    \subrfig{rpsDiscSimClust_cl101_a06}, CL3
    \subrfig{rpsDiscSimClust_cl3_a1} \&
    \subrfig{rpsDiscSimClust_cl3_a06} and CL14
    \subrfig{rpsDiscSimClust_cl14_a1} \&
    \subrfig{rpsDiscSimClust_cl14_a06}, at \zeq{0} ({left}) and
    \zeq{0.6} ({right}). Mass stripping is calculated for three
    mock catalogs of 1000 galaxies each for which the stellar disc
    mass is constant: $10^{9}\msun$ ({blue}),
    $10^{10}\msun$ ({red}) and $10^{11}\msun$
    \emph({green}).  The mean value of the remaining gas mass for each
    of the catalogs is shown (solid line) with the shaded region
    corresponding to a range of $1\sigma$ about the mean. Also shown
    is the minimal location of the shock edge as defined in
    \rfsec{edgeFind} ({black dashed}). The value of the
    fudge-factor for RPS is $\alf=1$. We find that the stripping
    becomes pronounced within the virial radius of the cluster. Beyond
    the virial radius, satellites of $\Ms>10^{10}\msun$
    undergo very little stripping while the low-mass satellites still
    retain more than \perc{60} of their gas mass. Satellites in higher
    mass clusters naturally experience more stripping. }
  \label{fig:rpsDiscSimClust}
\end{figure*}

\subsection{Employing the Stripping Condition in an NFW Model Cluster}\label{sec:rpsDiscCond}
With a catalog of galaxies covering the relevant parameter
space in hand we can now assess the effectiveness
of RPS on the gas in the galaxy, embodied by \cref{eq:rpsCond}.

As before, we begin by modelling the ICM with an NFW model (\rfsec{nfwModel}) and assume the galaxy speed is
identical to the virial velocity (\rfsec{ramPressure}). Under these
assumptions, the ram pressure, \cref{eq:ramPressNFW} and binding
gravitational force \cref{eq:ftildeDef} can be inserted into the
stripping condition \cref{eq:rpsCond}
\begin{equation}\label{eq:rpsCondDisc}
\alf \fc \frac{G}{4 \mathrm{\pi}} \frac{\Mv^2}{{\Rc}^4} 
{\left[{\afunc{1}{\cvir}}\tilde{r} \left(\cvir^{-1}+\tilde{r} \right)^{2}\right]}^{-1}
\ge
\mathrm{\pi} G \sigstar^2 \fgs \ftilde(\tR).
\end{equation}
For the sake of convenience, we define a new quantity encapsulating
the cluster parameters $\sigclust\equiv \Mv/(2\mathrm{\pi} \Rc^2)$. Scaling for
typical values we find, making use of \cref{eq:virialDef},
\begin{equation}\label{eq:sigclustScale}
  \sigclust \simeq 
  \begin{cases}
2.4\times10^7\left(\frac{\Mv}{10^{15}} \right)^{\frac{1}{3}}\units{\msun\,kpc^{-2}} & z=0\\ 
4.6\times10^7\left(\frac{\Mv}{10^{15}} \right)^{\frac{1}{3}}\units{\msun\,kpc^{-2}} & z=0.6
\end{cases}.
\end{equation}

The stripping condition now takes the form
\begin{equation}\label{eq:rpsDisc2}
\ptilde(\tilde{r}_p)\equiv \alf
\frac{f_c}{\fgs}\left(\frac{\sigclust}{\sigstar}\right)^2 {\left[{\afunc{1}{\cvir}}\tilde{r} \left(\cvir^{-1}+\tilde{r} \right)^{2}\right]}^{-1}     \ge
\ftilde(\tR).
\end{equation}
We recall that $\ftilde(\tR)$ describes how the binding force changes
as a function of distance from the galaxy
centre. $\ptilde(\tilde{r}_p)$ embodies the strength of the
ram pressure at a given position, normalized to the properties of a
specific galaxy. As before, $\alf$ is a fudge-factor of order unity
which encapsulates any uncertainties in the model.

To find the effect of the ram pressure we relate the gravitational
binding force acting on a gas element at given radius with the gas
mass enclosed within that radius. A galaxy at a given position within
the cluster will be characterized by a value of $\ptilde$ which can be
equated with $\ftilde$, which in turn determines how much mass is
unaffected by the ram pressure.

In principle, stripping of mass from the galaxy decreases the
self-gravity. We ignore this effect since we are only stripping the
gas, which accounts for a small contribution to the gravitational
force, while the main contributors to the force, the stellar disc and
dark matter halo, are unaffected by RPS (and as shown in
\rfsec{haloTidal}, tidal stripping which can affect the stellar and
dark matter components is not relevant in the cluster outskirts). We
assume the rest of the galaxy is unaffected by the ram pressure, an
assumption which is supported by simulations \citep{Quilis2000}.

In \cref{fig:ftilde} we show the $\ftilde(\tR)-M(\tR)$ relation for
several illustrative examples of a $\Ms = 10^{10}\msun$ galaxy
with a gas-to-stellar mass ratio of $\fgs=0.1$ embedded in an $\Mv=
4.8 \times 10^{11}\msun$, $\cvir = 10$ NFW dark matter
halo. The different coloured solid lines in the plot correspond to
different gas distributions: (a) A distribution identical to the
stellar component ($\bet=1$, blue), (b) A distribution which is twice
as extended, ($\bet=0.5$, red), (c) A distribution which is twice as
compact ($\bet=2$, green), all in a galaxy without a bulge
component. Gas rich galaxy models with an enhanced gas fraction of
$\fgs=0.25$ are plotted with dotted lines. In addition, galaxy models
with a bulge of $B/T = 0.2$ and disc-to-bulge scale radius ratio of
$\xii = 3$ (dashed lines), are also shown.

We see in the figure that a value of $\ptilde=0.15$, typical for a
satellite at $0.5\Rc$ of a $10^{15}\msun$ cluster (shown as a
horizontal black line), corresponds to a stripping of $\sim 20$ per cent of
the mass for the compact distribution, about \perc{40} in the case of
an identical distribution and in the case of an extended gas
distribution all of the gas is removed in the bulge-less model,
whereas \perc{10} or so of the gas remain if a bulge is present. Thus
we can see that the gas distribution is very important in setting the
effectiveness of RPS. The bulge can be seen to have a strong influence
in the central areas of the galaxy, while increasing the gas fraction
in the galaxy does very little to affect the binding force. As a
galaxy moves closer to the cluster centre, $\ptilde\propto
\tilde{r}_p^{-2} $ increases and the horizontal line will rise,
indicating more stripping of the gas.

In our catalog, we find a mean value\footnotemark of
\mbox{$\langle\log\sigstar\rangle = 10^8$} which means that at the
virial radius $r/\Rc = 1$, the parameter $\ptilde$ will be of order
$0.05$ for a typical cluster, using
\cref{eq:sigclustScale,eq:rpsDisc2} (assuming $\alf=1$). Thus, a
galaxy of \mbox{$\Ms = 10^{10}\msun$}, as shown in
\cref{fig:ftilde}, will lose at most $\sim 50$ per cent of its gas if its gas
distribution is extended compared to the stars. If the gaseous
component is not extended the mass loss will be much smaller. This
indicates that RPS is most likely not effective in removing the gas
from galaxies when they are beyond the virial radius of
clusters.\footnotetext{The distribution of $\sigstar$ values in our
  catalog is Gaussian in $\log(\sigstar)$ with a mean value of $8.03$
  and $\sigma= 0.57$.}

We examine the stripping for the entire galaxy catalog in
\cref{fig:rpsCat}, where the galaxies are assumed to be in an NFW model
$10^{15}\msun$ cluster with a gas fraction of $f_c=0.15$, assuming
$\alf=1$. We find the amount of stripped mass in each galaxy and show
the distribution of the galaxy population according to the stripped
mass and $\sigstar$ for 4 different positions within the cluster:
$2\Rc$, $1\Rc$, $0.5\Rc$ and $0.2\Rc$. The contours correspond to the
percentage of the population enclosed within them, and the cumulative
distribution of the stripped mass is also shown in the inlay.

At twice the virial radius (\cref{fig:rpsCat_2rv}) the stripping is
practically negligible, with \perc{50} of the population retaining
more than \perc{95} of their gas, and \perc{90} of the population
retaining more than \perc{65} of their gas. At the virial radius
(\cref{fig:rpsCat_1rv}), the stripping is still largely ineffective,
with about \perc{80} of the galaxies retaining more than \perc{60} of
their gas, although there are already some galaxies ($\sim 3$ per cent)
who have been stripped of more than \perc{90} of their gas. These
levels of stripping were achieved with the assumption of maximal RPS
effectiveness $\alf=1$.

Within the virial radius (\cref{fig:rpsCat_05rv,fig:rpsCat_02rv}), RPS
becomes an important process, affecting nearly all the galaxies to
some extent. In the central regions ($\sim 0.2\Rc$) over half of the
galaxies have lost more than \perc{60} of their gas, and of these most
of lost virtually all their gas.

\subsection{RPS of the Galactic Disc in Simulated Clusters}\label{sec:rpsDiscSimClust}
We examine the effectiveness of the stripping in a more realistic
setting in \cref{fig:rpsDiscSimClust} by using the gas density
profiles from $3$ of our simulated clusters: CL101, CL3 and CL14 at
\zeq{0} and \zeq{0.6}, in a similar fashion to
\cref{fig:rps_sat_simClust}. As before, the ram pressure is calculated
based on the gas density profile and under the assumption that the
speed of the galaxy is equal to the virial velocity of the
cluster. The fudge factor $\alf$ is also employed to account for
uncertainties in the model, and is set to $\alf=1$ to ensure maximal
stripping.

As test cases we generated mock catalogs of 1000 galaxies each, in
which the stellar disc mass was identical for all
galaxies. Three such catalogs were generated with stellar
disc masses of $10^{9},\,10^{10}$ and $10^{11}\msun$.

As one would expect, galaxies travelling in clusters of higher mass
experience more stripping. One can also see that galaxies of lower
masses are more susceptible to RPS. Here too the RPS becomes effective
when galaxies are within the virial radius of the cluster. As noted in
relation to \cref{fig:rps_sat_simCLust_rv12}, while the stripping may
seem more effective at \zeq{0.6}, one must bear in mind that the
virial radius at \zeq{0.6} is lower by a factor of $\sim 2$. In
essence, one is examining stripping at a much smaller radius in
comparison to the \zeq{0} case, since the cluster gas density profile does not
change significantly between these times at a given position.

\begin{figure}
\centering 
  \includegraphics[width=8cm,keepaspectratio,bb=0 0 5.33in 4.15in ]{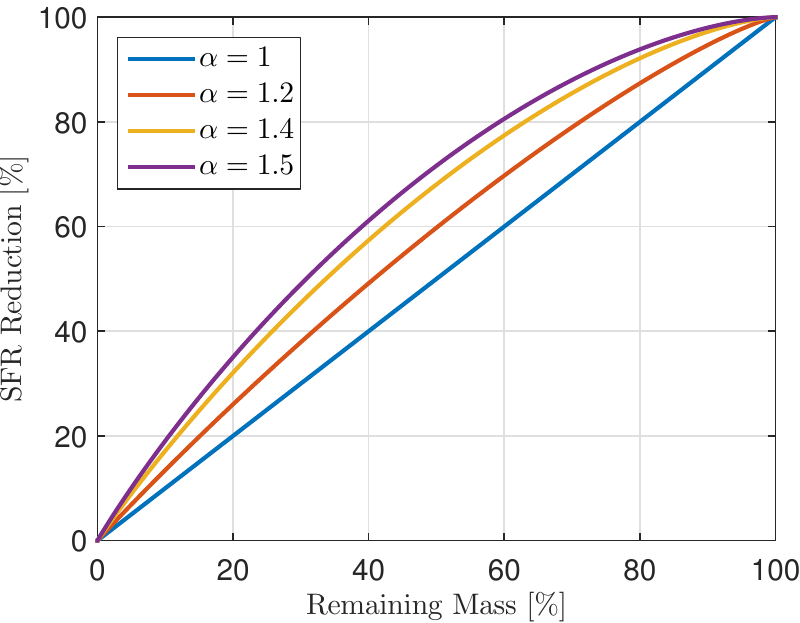}
  \caption{The relation between the remaining gas mass in a galaxy and
    the reduction in SFR as a result of the mass loss. The different
    lines correspond to different values of $\alpha$, the exponent in
    the Kennicutt--Schmidt law
    $\Sigma_{\mathrm{SFR}}\propto\Sigma_{\rm gas}^\alpha$. The SFR of a
    galaxy with only \perc{40} of its gas remaining will be $\sim
    60$ per cent of its original value, for values of $\alpha\simeq 1.4$. To
    reduce the SFR of a galaxy to half its initial value one must
    remove at least \perc{70} of the gas.}
  \label{fig:mstripSfr}
\end{figure}

\subsection{Relating the Mass Stripping to Star Formation Quenching}\label{sec:stripQuench}
In the previous sections we explored the amount of gas removed from
the disc of satellite galaxies as a result of RPS. Clearly, removal of
the gas from the galaxy should result in a drop in the star formation,
but the relation between the two is not necessarily linear.

The observational Kennicutt--Schmidt relation
\citep{Schmidt1959,Kennicutt1998} relates the surface density of the
gas in a galaxy with the star formation rate per unit area,
$\Sigma_{\mathrm{SFR}}\propto \Sigma_{\rm gas}^\alpha$, with an exponent
in the range of $\alpha=1-1.5$. This relation is seen to hold locally
within a galaxy, as well as on average over the entire disc.

Thus, the central areas where the density is higher (and the
gravitational binding is stronger) will account for a larger part of
the star formation than the outskirts of the disc, even though they
contain more mass. RPS, on the other hand, preferentially removes gas
from the outskirts. As a result, removing half the gas in the galaxy
will not result in a \perc{50} drop in star formation, but rather in a
smaller reduction.

For the exponential disc model we employed in this paper (see
\rfsec{gasDisc}) one can approximate the SFR using the
Kennicutt--Schmidt law for a given galaxy model. For a galaxy which has
undergone stripping, one can relate the amount of gas remaining in the
galaxy to the reduction in the total SFR, compared to its value before
the stripping took place.

In \cref{fig:mstripSfr} we show this relation for several representative
values of $\alpha$, the exponent in the Kennicutt--Schmidt law. We see
that for $\alpha=1$, the fraction of remaining mass is equivalent to
the reduction in SFR. For $\alpha>1$, the reduction in SFR will always
be smaller than the reduction in mass. The SFR of a galaxy stripped of
half its gas will be $\sim 70$ per cent of its original value for
$\alpha\simeq1.4$. Conversely, to reduce the SFR of a galaxy to half
its original value, one must strip roughly \perc{70} of its gas.

This means that the results for the stripped mass in galaxies shown in
\cref{fig:rpsCat,fig:rpsDiscSimClust} should be considered as
upper-limits to the amount of star-formation quenching in those
galaxies. This only enhances the conclusion that RPS is not an
effective mechanism for inducing star formation quenching in the
outskirts of clusters. 

\section{Star Formation Quenching}\label{sec:quench}
The quenching of star formation in galaxies can be achieved by removing enough
gas from the galaxy such that there is insufficient fuel for forming new
stars. It is common to distinguish between two modes of quenching.  If the gas
is removed from the galactic disc, new stars cannot be formed and the
shut-down of star formation is very rapid.  However, if the gas in the galaxy
remains intact, and only the gas from the surrounding halo is removed, star
formation can still continue within the galaxy for some time before exhausting
the gas in the disc. Since the gas in the disc cannot be replenished the star
formation in the galaxy will eventually cease, albeit much more slowly. This
is often referred to as quenching by `starvation'.

In \rfsec{rpsHalo} and \rfsec{rpsDisc} we presented two toy models which
allowed us to assess the effectiveness of removing the gas, via RPS, from the
gas haloes surrounding satellite galaxies as well as from the within the
galactic disc. The two main results of these toy models with respect
to star formation quenching in the cluster outskirts ($r\gtrsim\Rv$) are that
RPS is \emph{not} effective at removing the gas from within the galactic disc
in these regions and thus, rapid quenching cannot take place in the outskirts
(assuming no other quenching processes take place) and that RPS is \emph{very}
effective in stripping the halo gas surrounding galaxies, removing as much as
\perc{80} of the gas by the time the satellites reach the virial radius of the
cluster.

We therefore find that star formation quenching in the outer regions
can occur by `starvation'-- the removal of the gas reservoir in the
halo and subsequent decline of star formation over time as the gas
within the galaxy is depleted. In contrast, within the virial radius
and especially in the inner regions of clusters, ram pressure can lead
to the complete removal of gas from the galaxy leading to very rapid
quenching.

In light of these results we wish to determine whether galaxies, after
crossing the accretion shock at the edge of the system and losing
their gas reservoir will be quenched by the time they reach the virial
radius. To do so we must compare the time it takes a satellite to
travel from the accretion shock to the virial radius $(\ttrav)$, to the
depletion time, the time it takes to use up all the gas within a
galaxy $(\tdepl)$.

\begin{figure} 
     \includegraphics[height=6.5cm,keepaspectratio,bb=0 0 5.17in 4.19in]{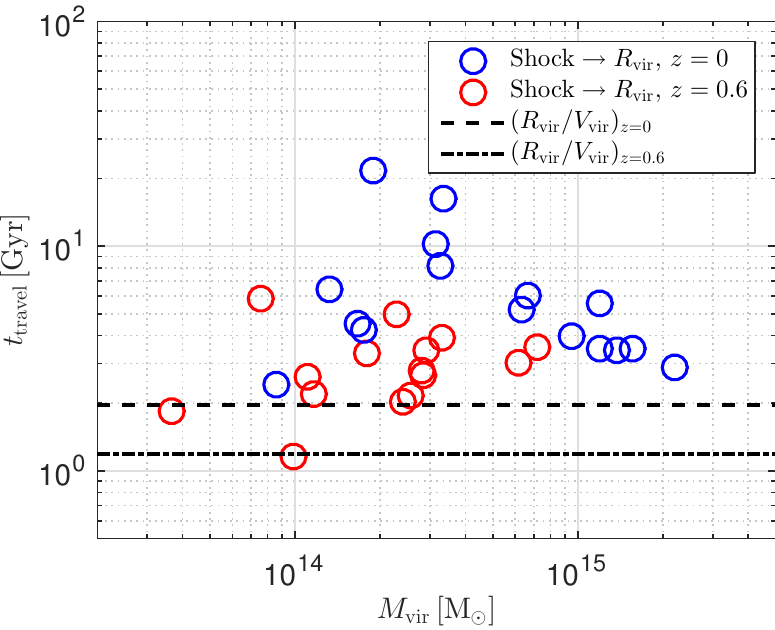}
    \caption{Time of travel for a satellite galaxy, $\ttrav$, from
      $R_{\mathrm{shock}} \to \Rv$ in the suite of simulated clusters
      is shown for \zeq{0} ({blue}) and \zeq{0.6}
      ({red}). The values for $t_{2\to1\Rv}$ at \zeq{0} and
      \zeq{0.6} are also shown ({dashed} and {dot-dashed},
      respectively).}
  \label{fig:tTravel}
\end{figure}

\subsection{Travel Time}\label{sec:travel}
As we have seen in \rfsec{edgeFind}, the shock front demarking the
edge of the ICM is typically found at $\gtrsim 2\Rv$. As a first
approximation, we assume galaxies travel at the virial velocity and
find the time it takes to travel from $2\Rv$ to $1\Rv$, based on the
virial relation \cref{eq:virialDef}
\begin{equation}\label{eq:tTravelVir}
\begin{split}
t_{\rm travel,\mathrm{vir}}&=\frac{\Rv}{\Vv}= \frac{\Rv}{\sqrt{G\Mv/\Rv}} \\ 
&= 1.96
\left(\frac{\delvir(z)}{337}\frac{\rho_{\mathrm{ref}}(z)}{\rho_{\mathrm{mean}}(0)}\right)^{-\frac{1}{2}}
\units{Gyr}.
\end{split}
\end{equation}
This results in a time-scale which, at any given time, is independent
of the virial parameters of the system. At \zeq{0.6} we find
$t_{2\to1\Rv} = 1.2\units{Gyr}$.

For a more detailed result we turn to our suite of simulated
clusters. To assess the typical velocity of incoming objects we
created a mass-weighted inflowing radial velocity profile by averaging
only over simulation cells in which the radial velocity of the gas is
inflowing, i.e.\@ $v_r < 0$,
\begin{equation}\label{eq:vrProfDef}
\langle v_r \rangle=\frac{\iint v_r(r,\theta,\varphi) W(r,\theta,\varphi)\diff \theta \diff \varphi}
{\iint  W(r,\theta,\varphi)\diff \theta \diff \varphi},
\end{equation}
where
\begin{equation}\label{eq:windowDef}
 W(r,\theta,\varphi)=
\begin{cases}  
\rho(r,\theta,\varphi) & v_r<0 \\
0 & v_r\ge 0
\end{cases}.
\end{equation}
The travel time can now be calculated 
\begin{equation}\label{eq:tTravel}
\ttrav=\int_{R_{\rm out}}^{R_{\rm in}}\frac{\diff r}{\langle v_r \rangle}.
\end{equation}
In \cref{fig:tTravel} we show the calculated travel times for
travelling from the shock edge (lowest value, see \rfsec{edgeFind}) to
$1\Rv$, for the clusters at \zeq{0} and \zeq{0.6}. As can be seen, the
results are of order several giga years and greater than the
virial approximation $t_{2\to1\Rv}$ by a factor of a few.

\begin{figure} 
     \includegraphics[height=6.5cm,keepaspectratio,bb=0 0 5.32in 4.06in]{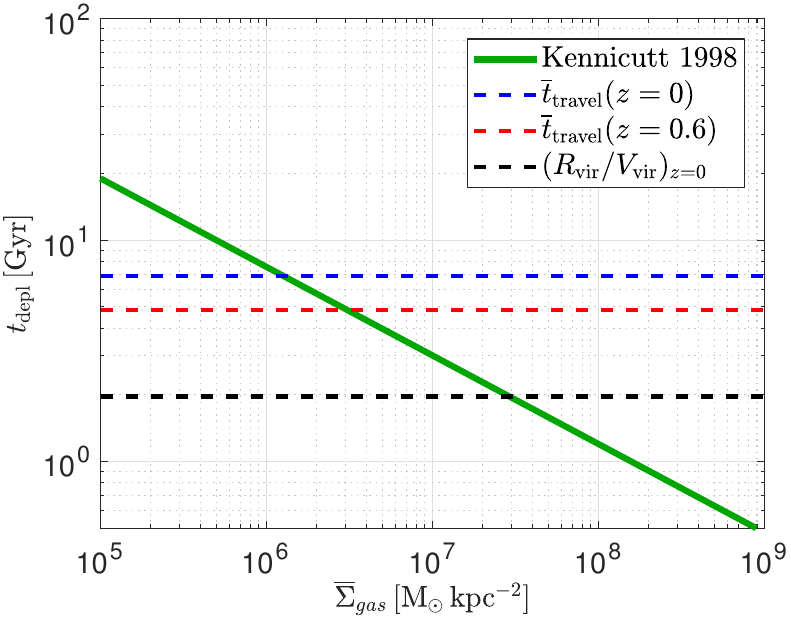}
    \caption{The depletion time inferred from the Kennicutt--Schmidt
      law as a function of average gas surface density
      ({green}). Also shown are the median value for the
      distribution of travel times in the simulated clusters at
      \zeq{0} and \zeq{0.6} ({blue} \& {red}, respectively)
      as shown in \cref{fig:tTravel}, as well as $t_{2\to1\Rv}$
      ({black}). The depletion times for typical galaxies are of
      the order of the travel times showing showing that galaxies may
      quench by `starvation' even before reaching the virial radius of
      the cluster.}
  \label{fig:tDepl}
\end{figure}

\subsection{Gas Depletion Time}\label{sec:deplete}
To find the gas depletion time we use the Kennicutt--Schmidt law  
\begin{equation}\label{qe:kenLaw}
\olSig_{\mathrm{SFR}}=A\left( \frac{\olSig_{\rm gas}}{\units{\msun\,pc^{-2}}}\right)^\alpha
\end{equation}
for the average surface densities of the gas and
SFR. \citet{Kennicutt1998} found this relation to hold over several
orders of magnitudes in surface density and found the values of the
parameters based on observations to be $\alpha=1.4 \pm 0.15$ and $
A=(2.5\pm 0.7)\times10^{-4}\units{\msun\, kpc^{-2}\,
  yr^{-1}}$. Additional studies
\citep[e.g.\@][]{Kennicutt2007,Leroy2008,Tacconi2013} found different
values for the exponent in the range $\alpha = 1.0\textrm{--}1.5$, and
also found that the relation holds over a large range of redshifts
\citep{Tacconi2013}. The relation is also seen to apply locally within
a single galaxy, although the parameters of the relation usually vary
between different galaxies \citep{Bigiel2008}.

The relation can be theoretically motivated if one assumes that
$\mathrm{SFR}\propto \rho_{\rm gas}/\tau$, where $\tau$ is a relevant time
scale. If $\tau$ is taken to be the free-fall time $t_{\mathrm{ff}}
\propto \rho_{\rm gas}^{-1/2}$ we naturally arrive at the
Kennicutt--Schmidt law with an exponent of $\alpha = 1.5$
\citep{Madore1977}. More detailed analysis derived the relation from
the physics of molecular clouds, where the actual star-formation
occurs \citep{Krumholz2005}, or by the properties of the density
distribution function of the ISM \citep{Kravtsov2003}.

Gas can be depleted either through star formation, or through feedback
process, such as supernova feedback or radiative feedback from massive
main-sequence stars, which drive outflows from the galaxies. These
processes can be assumed to scale with the star formation rate such
that $\olSig_{\mathrm{outflow}}=\tau\olSig_{\mathrm{SFR}}$ where
$\tau\approx 1$ \citep{Dekel2014}. The average gas density then
follows the equation
\begin{equation}\label{eq:sigasDiffEq}
\dot{\olSig}_{\rm gas}=-\olSig_{\mathrm{SFR}}(1+\tau)=-\widetilde{A}(1+\tau)\olSig_{\rm gas}^\alpha,
\end{equation}
where $\widetilde{A}\equiv A / 10^{6\alpha} $ due to unit conversion. The
solution to this equation is \footnote{In the case of $\alpha= 1$, the
  solution to \cref{eq:sigasDiffEq} is an exponential decay with a
  decay time-scale of $\left(\widetilde{A}(1 + \tau)\right)^{-1}.$}
\begin{equation}\label{eq:sigasSol}
\olSig_{\rm gas}(t)=\left(\olSig_0^{1-\alpha}+(\alpha-1)(1+\tau)\widetilde{A}t\right)^{\frac{1}{1-\alpha}},
\end{equation}
where $\olSig_0\equiv \olSig_{\rm gas}|_{t=0}$. As one would expect, the
gas surface density goes to $0$ asymptotically.

We define the depletion time as the time in which the average surface
density drops to a given fractional value $\epsilon$
\begin{equation}\label{tdeplDef}
\begin{split}
\tdepl&=\left(\epsilon^{1-\alpha}-1\right)\frac{\olSig_0^{1-\alpha}}{(\alpha-1)(1+\tau)\widetilde{A}}\\ &=3.01\left(\frac{\olSig_{\rm
    gas}}{10^7\msun\units{kpc^{-2}}}\right)^{-0.4}\units{Gyr}.
\end{split}
\end{equation}
The typical depletion value given above is defined for representative
values of $\epsilon=0.1$ and $\alpha=1.4,\,\widetilde{A}=10^{-14}$ and
is consistent with observations
\citep{Kong2004,Bigiel2008,Leroy2008,Pflamm-Altenburg2009,Bauermeister2013}.

In \cref{fig:tDepl} we show the depletion time as a function of
$\olSig_{\rm gas}$ as well as the median of the travel time distribution
in our simulated clusters. As can be seen, the depletion time and
travel times are of the same order for ordinary disc galaxies
($\olSig_{\rm gas} \sim 10^6\textrm{--}10^7 \units{\msun\,kpc^{−2}}$). 

The values of $\tdepl$ found above should be treated as an upper limit
to the star formation quenching time.  Observations and theoretical
studies point to a threshold gas density necessary for star formation
\citep{Krumholz2009}. This threshold may be reached before the gas is
completely depleted. Another factor to consider is that the
ram pressure exerted by the ICM can also enhance the star formation
rate of a satellite galaxy due to the additional pressure exerted on
the ISM of the satellite galaxy \citep{Bekki2003,Kronberger2008}. If
so, the actual depletion time may be even shorter.

We see therefore that the depletion time (and by extension the star
formation quenching time) and the travel time are of the same order
and that galaxies beyond the virial radius in clusters may quench due
to gas depletion in the galaxy after the loss of their halo gas
reservoir to RPS, even before crossing the virial radius for the first
time.

While we have shown that quenching via `starvation' is feasible for
galaxies beyond the virial radius of the cluster, we note that not all
galaxies will necessarily be quenched. As seen in \cref{fig:tTravel},
the typical travelling time varies between clusters, and galaxies with
long depletion times may reach the virial radius before quenching if
the travel time is sufficiently low. This can account for the observed
star-forming galaxies found within the virial radius.

\section{Discussion: Validity of the RPS Model}\label{sec:discuss}
Of the various processes occurring in the extended ICM which can lead
to gas depletion and star formation shutdown we focus here on
ram pressure stripping. This hydrodynamic process is known to occur in
clusters \citep{Cayatte1994,Kenney2014,Abramson2014} and has been
studied analytically \citep{Gunn1972,Gisler1976} and
numerically (\citealt{Gisler1976,Abadi1999}, also see
\citealt{Roediger2009} and references therein).

To study the effectiveness of RPS we employed a simple analytic
toy-model which we apply to stripping of the hot halo gas surrounding
satellite galaxies and to the stripping of gas within the galactic
disc itself. Our model, while easy to implement, is based on a set of
simplifying assumptions. The model is instantaneous rather than
dynamical-- we assess the mass loss for a given halo/galaxy at a given
position in the cluster while disregarding its history and ignoring
the subsequent evolution. We assume mass loss occurs rapidly and do
not address the dynamical response of the system to the mass loss
\citep{Smith2012}.

In modelling the gas distribution of the gas halo of the satellites,
we have assumed that the gas is isothermal and in hydrostatic
equilibrium within the dark matter potential. Gas distributions which
assume a more general polytropic equation of state for the gas have
been suggested \citep[e.g.\@][]{Komatsu2001}, and we intend to extend
our RPS model to include these profiles in a future work. In addition,
we are ignoring the multiphase nature of the gas in the disc, an
omission which over-estimates the ability of RPS to remove the clumpy
cold gas component \citep{Tonnesen2009}.

In reality, studies have shown that the gas removal from a galaxy
undergoing stripping is a multi-stage process in which the gas located
beyond the stripping radius is first displaced from the disc over a
time-scale of \mbox{$\sim 10\units{Myr}$} and is subsequently
completely unbound over longer time-scales of \mbox{$\sim 100
  \units{Myr}$}
\citep{Schulz2001,Marcolini2003,Roediger2005,Roediger2007}. In the
interim, some of the displaced gas can be re-accreted to the galaxy. A
third phase of prolonged viscous stripping \citep{Nulsen1982} follows.

Despite the inaccuracies of the simple model employed in this paper,
we find it is a very useful tool whose results regarding the stripping
in the cluster outskirts can be trusted, if wielded properly. In both
the halo gas and disc gas stripping scenarios we used the model to
obtain a lower or upper limit for the stripping effectiveness to
overcome the shortcomings of the model.

In the halo gas stripping scenario we found that the stripping was
very effective for typical satellite masses even when the effect of
RPS was reduced by hand (via the fudge-factor $\alfP$,
\cref{fig:rpsAlpha}). Since RPS only affects the gas and not the dark
matter in the cluster outskirts (\rfsec{haloTidal}) the potential well
of the satellite will remain largely unchanged, and since the RPS is
so strong, the assumption of rapid gas removal is justified.

Due to the shortcomings of the model which we listed above, one may
argue that for the scenario of stripping gas from the disc, the model
over-estimates the stripping by assuming rapid and total gas
removal. However, we find that the stripping is largely ineffective,
even when ensuring a maximal effect of the ram pressure.  This is done
by assuming a relatively high cluster mass ($10^{15}\msun$)
and taking a maximal value of the fudge-factor $\alf=1$). We can thus
be assured that in the outskirts of clusters RPS is not an effective
form of disc gas stripping.

In focusing on RPS we have also neglected other processes which may
lead to gas depletion. Two relevant dynamical process are tidal
stripping by the cluster and interactions with other satellites in
close encounters, the combined effect of which is known as `galaxy
harassment' \citep{Moore1996,Gnedin2003a}. Tidal stripping is strongly dependent
on the cluster-centric distance \citep{Dekel2003} and as shown in
\rfsec{haloTidal}, can be ignored in the outskirts of clusters (see
also the Appendix of \citealt{McCarthy2008}). Due to the high relative
velocities between satellite galaxies, perturbations induced by close
encounters are expected to be small \citep{Boselli2006}.

Two additional relevant \emph{hydrodynamical} processes are viscous
stripping \citep{Nulsen1982} and thermal evaporation
\citep{Cowie1977}. Viscous stripping occurs when the gas in the outer
layers of a galaxy travelling through the ICM experiences a momentum
transfer due to viscosity which can lead to gas removal. The
effectiveness of the stripping depends on whether the flux is laminar
or turbulent, which in turn depends on the size of the galaxy. Thermal
evaporation occurs at the interface between the hot ICM and the much
cooler ISM. The mass loss rate is sensitive to the temperature of the
ICM $\dot{M}\propto T_{\mathrm{ICM}}^{3/2}$, and the presence of
magnetic fields which can reduce the efficiency of this process.

The effect of these two processes on the disc are qualitatively similar to
that of RPS, in that only the gas in the satellite is affected (and not the
stellar or dark matter components) and that the processes work on the gas from
the outside in. In terms of relative importance, in the central regions of
clusters both RPS and viscous stripping are expected to be the dominant mass
loss channels in galaxies which are extended and/or characterized by high
orbital velocities, while thermal evaporation becomes important for smaller
(dwarf) galaxies \citep{Boselli2006}. \citet{Roediger2008} and
\citet{Roediger2009} find that viscous stripping has a minor effect compared
with RPS and its impact is felt only over much longer time-scales.

While thermal evaporation has been shown to be equally as important as
RPS in the central regions of cluster \citep{Nulsen1982}, since it is
strongly dependent on the temperature, its effect is greatly reduced
in the outskirts of clusters. As seen in \cref{fig:tempProfs}, the
temperature in the outskirts can drop to as much as \perc{20} of the
temperature within $\Rv$ which entails a mass loss rate due to
evaporation which is \perc{90} smaller than in the central regions of
the cluster.

Another aspect of the interaction we have neglected is the effect of
magnetic fields on the mass loss due to RPS. Interactions between the
magnetic fields of the ISM and ICM may somewhat suppress mass loss,
depending on the orientation of the ICM magnetic field to the
direction of motion of the galaxy, with a parallel orientation leading
to stronger suppression of mass loss.  The overall effect however is
found to be mild \citep{Shin2014,Tonnesen2014,Ruszkowski2014}.

For the case of halo gas stripping, one must consider the confining
effect produced by the ICM pressure. The ICM pressure may lead to a
contraction of the gas halo and reduce the efficiency of
RPS. \citet{Bahe2012} find that the ram pressure is generally dominant
over confinement pressure.

\section{Conclusion and Discussion}\label{sec:summary}
It is common practice to treat the virial radius, defined by the mean
over density of the dark matter halo with respect to the mean
cosmological density, as the outer edge of a collapsed cosmological
object. Although this definition is physically motivated by the
process of energy-conserving virialization (embodied in the spherical
collapse `Top Hat' model), in the case of cluster-sized systems
\mbox{$\gtrsim 10^{14}\msun$} the true extent of the cluster
hot gas is far greater.

We examined 16 high-resolution simulated cluster systems and found
that, at least from \zeq{0.6} and onwards, the virial accretion shock
extends to well beyond the virial radius and is usually found at a
distance of $\sim 2.5 \Rv$, an estimate based on the peak of the
entropy profile of the cluster and the maximal negative entropy
gradient along the profile.

The virial accretion shock is comprised of several shock fronts
(`lobes') which extend between the large scale filaments and merge
seamlessly with the cylindrical shocks surrounding the filaments that
feed the cluster. Though very rough, our edge detection method manages
to capture the main features of the shock edge, and gives a reliable
estimate for the shock edge fronts in the simulated clusters.

The relative positions of the accretion shock and $\Rv$ show that the
standard virial radius encloses $\lesssim 10$ per cent of the ICM
volume. $\Rv$ should be treated as a useful ball-park scale of the
cluster size, but not necessarily its edge. As a result it should come
as no surprise that galaxies at cluster-centric distances of several
times $\Rv$ are affected by the cluster environment \citep{Park2009}.

\citet{Lau2015} have shown that the profiles of cluster properties are
self-similar if normalized by the mass accretion rate into the
cluster. As a result, they find that the location of the accretion
shock, when cast in units of $R_{200}$, is strongly dependent on the
mass accretion parameter. Clusters with a high accretion rates have
smaller accretion shocks. We have confirmed this behavior in our
simulated cluster suite as well.

Another measure of the `edge' of a dark matter halo is the
`splashback' radius \citep{Diemer2014,Adhikari2014} which is defined
by a sharp drop in the density profile of the halo. \citealt{More2015}
find that the splashback radius is typically larger than the virial
radius in high mass haloes, in general agreement with our findings.

Observational confirmation of the shock location should be
possible. The virial shock front can accelerate charged particles to
very high energies of $\gtrsim \unitstx{TeV}$ which then Compton-scatter
CMB photons into $\gamma$-rays \citep{Loeb2000,Keshet2003}. The
$\gamma$-ray signal should be accompanied by a signal in hard $X$-rays
\citep{Kushnir2010} and a synchrotron emission
\citep{Loeb2000,Keshet2004}, which may be detected in radio
observations. The resultant signal is expected to resemble an
elongated ring due to the filamentary structure surrounding the
cluster (see \rfsec{edgeMaps}).

\citet{Keshet2017a} have reported the discovery of an elongated
$\gamma$-ray ring structure surrounding the Coma cluster at a radius
of \mbox{$\sim 5 \units{Mpc}$} which they argue is a signal of the virial
accretion shock. The virial radius of the Coma cluster is
$R_{200}\simeq 2.3 \units{Mpc}$, so that the shock signal is detected
at $\sim 2R_{200}$, in complete agreement with our findings.  We
expect that future observations will discover more evidence of the
existence and position of the virial accretion shock surrounding
clusters.

Having established that the environment of the cluster extends out to
$\sim (2\textrm{--}3) \Rv$ we examined how the extended environment
can affect satellite galaxies in the outskirts of the cluster. To that
end we employed simple analytic models to gauge the effect of RPS both
on the gas within a satellite galaxy as well as the gas in the halo
that surrounds it, and extended the analysis by examining the RPS
expected on analytic galaxy models within our simulated cluster suite.

The results of our RPS models lead us to the following conclusions in
terms of gas depletion and star formation quenching
\begin{enumerate}[label=\emph{\alph*})]
\item RPS in the outer regions of clusters, especially high-mass clusters, is
  very effective in removing the halo gas surrounding individual galaxies,
  between 40 and \perc{70} at $2\Rv$. When the satellites reach $\Rv$ less
  than \perc{30} of the gas remains in high-mass satellites, with low-mass
  satellites retaining less than \perc{10} of their initial gas haloes. RPS is
  \emph{not} an effective channel for removing the gas from within galactic
  discs in the outskirts of clusters.
\item Once a satellite crosses $\Rv$ and as it falls towards the
  centre of the cluster, RPS becomes an increasingly important
  mechanism for removing gas from the galactic disc.
\end{enumerate}

We address here quenching of satellite galaxies by the ICM of a host
cluster in the regions of $\gtrsim 1\Rv$. The quenching scenario we
envision is that of `starvation' -- the gas in the galaxy is depleted
from the galaxy via star formation and feedback processes over a
time-scale of several giga years and no new gas can be accreted from
the local gas reservoir since it has been removed by RPS. The gas
cannot be replenished from the hot ICM since its cooling time is very
long and the typical satellite velocity is higher than the escape
velocity of the satellite.

Once the gas density in the galaxy drops below a critical threshold,
star formation is quenched abruptly. This is consistent with the
`delayed-then-rapid' quenching scenario inferred by recent studies
\citep{Trinh2013,Mok2013,Wetzel2013}, as well as the findings of  
\citet{Tal2014} that the onset of quenching in satellites in galaxy
groups is delayed in comparison with the quenching of the central
galaxy \citep[see also][]{Woo2015}.

By comparing the typical times, i.e.\@ the time it takes a galaxy to
traverse the distance from the accretion shock to $\Rv$ (`travel
time'), and the gas depletion times we find that it is quite plausible
for a galaxy reaching the virial radius of the cluster for the first
time to be quenched due to `starvation'. This provides a natural
explanation for the quenched population of galaxies, and especially
disc galaxies, found in the outskirts of clusters \citep{Park2009}.

The scenario of satellite quenching in the environment of the cluster
which extends to $\sim (2\textrm{--}3) \Rv$ is consistent with the
observed `galactic conformity', i.e.\@ large scale correlations
between quenched satellites and quenched centrals over large distance
scales of several $\unitstx{Mpc}$
\citep{Weinmann2006,Ann2008,Kauffmann2010}.

The `starvation' quenching mode has been considered before
\citep{Larson1980,Balogh2000,McCarthy2008,Bekki2009,Vijayaraghavan2015} but usually under
the assumption that the virial radius marked the edge of the
system. As a result, the time-scales involved led to the conclusion
that the process was not fast enough and that to match the
observations of the quenched population, galaxies must have been
quenched in the centres of smaller groups before becoming part of the
cluster, a scenario known as `pre-processing'
\citep{Balogh2000,Jaffe2012,Wetzel2013,Fang2016}. However, if the edge 
of the ICM is much more extended and the gas reservoir in the haloes
of individual galaxies is removed much earlier, one need not invoke
`pre-processed' or `splashback' galaxies \citep{Wetzel2014} to account
for the quenched population in clusters.

\citet{Bahe2013} reached a similar conclusion based on the results of
SPH simulations of satellites in clusters \citep[see
  also][]{Cen2014a,Jaffe2015}. In this paper we have presented a more
comprehensive analytic framework to explore the RPS of the halo gas
and the gas within the galaxy. By making use of an analytic modelling
of the satellite galaxies we are in full control of the parameters and
attributes of the galaxies. This allows us to gauge the effectiveness
of RPS on a wide range of galaxy types, beyond those which are
reliably produced in a simulation, resulting in a better understanding
of what makes a galaxy more (or less) susceptible to stripping. The
model afforded the chance to explore the effectiveness of RPS over a
large and exhaustive parameter space thus enabling a mapping of the
limits of RPS as a quenching mechanism in satellite galaxies.

The ram pressure which acts to remove the gas halo reservoir of the
galaxy may also affect the depletion time indirectly by enhancing the
star formation rate due to the additional pressure exerted on the
ISM. Evidence for this has been seen in numerical studies by
\citet{Bekki2003,Kronberger2008,Kapferer2008,Kapferer2009}. \citet{Tonnesen2012}
however find no signs of star formation enhancement, perhaps due to
differences in the numerical scheme.

We note that star formation quenching may occur before the typical
depletion time has passed due to other processes which lead to more
rapid quenching. One such example is the `morphological quenching' --
the gas depletion leads to a drop in surface density of the gaseous
disc leading to an increased disc stability, \cref{eq:ToomreDef},
which may lead to a shutdown of star formation
\citep{Martig2009,Martig2013}. Another example is the compaction of
the gas in the disc which leads to increased star formation and
quenching \citep{Zolotov2015}.

Though we show that by removing the halo gas surrounding a galaxy,
star formation quenching via `starvation' can occur, it must be
stressed that not all the satellites will necessarily be quenched
completely, and indeed star forming galaxies are regularly observed
within clusters. In addition, \citet{Sun2007} and \citet{Jeltema2008}
find evidence of gas haloes surrounding satellite galaxies in the
central regions of clusters demonstrating that not all satellites are
stripped of their halo gas.

As we have pointed out in \citet{Zinger2016}, gas streams in clusters
can penetrate into the very centre of the cluster (e.g.\@ see
\cref{fig:CL103_rps}). Satellites which travel along the inflowing
streams may suffer very little stripping, if they are co-flowing with
the surrounding medium, and hypothetically could reach the centre of
the cluster with their gas content intact. Conversely, if the gas in
the filaments is hot enough, as shown in \citet{Zinger2016}, the
galaxies in the filaments may not be able to accrete new gas and
quench by `starvation', albeit over longer timescales, once the gas in
the galaxy and in the gaseous halo has been depleted. Further research
is needed to confirm the validity of this scenario and implications,
and we plan to address this issue in a future study.

In addition, as seen in \cref{fig:tDepl} galaxies with sufficiently
long depletion times may reach $\Rv$ before being quenched. Since the
typical travel times for reaching $\Rv$ vary between different
clusters (\cref{fig:tTravel}), one may expect to find clusters in
which star-forming galaxies can be found even well within the virial
radius.

In this paper we have explored in depth one aspect of the
environmental effect of the ICM on satellite galaxies found beyond the
virial radius, namely RPS. The understanding that the cluster
influence extends out to distances of several mega-parsecs should
prompt further observational, theoretical and numerical studies of the
ways the ICM affects galaxies in these regions.

\section*{Acknowledgments}
We acknowledge stimulating discussions with Yuval Birnboim, Joanna Woo
and Sandro Tacchella. We thank the anonymous referee for comments and
advice which improved the paper. This work was supported by ISF grant
24/12, by GIF grant G-1052-104.7/2009, by a DIP grant, by the I-CORE
Program of the PBC, by ISF grant 1829/12, and by NSF grants
AST-1010033, AST-1405962 and AST-1412768.

\bibliographystyle{mnras}
\bibliography{zinger_edge_rps_bib}

\label{lastpage}

\end{document}